\documentclass[a4paper,11pt]{article}
\pdfoutput=1 
\usepackage{jheppub}

\usepackage{braket}
\usepackage{array}

\newcommand{\C}{\mathbb{C}}
\newcommand{\R}{\mathbb{R}}
\newcommand{\Z}{\mathbb{Z}}

\makeatletter
\def\widebar{\accentset{{\cc@style\underline{\mskip10mu}}}}
\makeatother
\makeatletter
\def\wideubar{\underaccent{{\cc@style\underline{\mskip10mu}}}}
\makeatother

\makeatletter
\renewcommand{\theequation}{%
\arabic{section}.\arabic{equation}}
\@addtoreset{equation}{section}
\makeatother

\title{Matrix product states and equivariant topological field theories for bosonic symmetry-protected topological phases in (1+1) dimensions}

\author{Ken Shiozaki$^a$ and Shinsei Ryu$^b$}
\affiliation{$^a$Department of Physics, University of Illinois at Urbana Champaign, 1110 West Green Street, Urbana, IL 61801, U.S.A.
\\
$^b$James Franck Institute and Kadanoff Center for Theoretical Physics, University of Chicago, 5640 South Ellis Ave, Chicago, IL, 60637, U.S.A.}

\abstract{Matrix Product States (MPSs) provide a powerful framework to study and classify gapped quantum phases --symmetry-protected topological (SPT) phases in particular--defined in one dimensional lattices. On the other hand, it is natural to expect that gapped quantum phases in the limit of zero correlation length are described by topological quantum field theories (TFTs or TQFTs). In this paper, for (1+1)-dimensional bosonic SPT phases protected by symmetry $G$, we bridge their descriptions in terms of MPSs, and those in terms of $G$-equivariant TFTs. 
In particular, for various topological invariants (SPT invariants) constructed previously using MPSs, we provide derivations from the point of view of (1+1) TFTs. We also discuss the connection between boundary degrees of freedom, which appear when one introduces a physical boundary in SPT phases, and ``open'' TFTs, which are TFTs defined on spacetimes with boundaries.}

\begin{document} 
\maketitle
\flushbottom

\section{Introduction}
Symmetry-protected topological (SPT) phases of matter 
are gapped phases of quantum many-body systems with short-range entanglement. 
They are topologically distinct from topologically trivial states in the presence of symmetries.
In other words, SPT phases are separated from topologically trivial phases by 
quantum critical points. 
Here, ``short-range entanglement'' means in particular the absence of topological order,  
and hence the uniqueness of the ground state even when 
the system is put on an arbitrary spatial manifold. 
(This property is often called ``invertible'', and 
hence SPT phases are said to have an invertible topological order.) 

Bosonic SPT phases in (1+1) dimensions are 
known to be classified by the second group cohomology $H^2(G,U(1))$.  
\cite{ChenGuLiuWen2013}
For quantum many-body systems defined on one-dimensional lattices,
this can be most easily seen from the matrix product state (MPS) representations of
the quantum ground states of SPT phases. 
\cite{Chen2011, Schuch2011, PollmannBergTurner2012,PollmannTurner2012}

On the other hand, 
deep inside a gapped phase where 
the correlation length is very short (order of a few lattice constant), 
one could expect that the universal properties of the system 
can be described in terms of a topological quantum field theory (TQFT or TFT).
The canonical examples include 
Chern-Simons theories,
which describe various fractional quantum Hall liquids,
and 
the BF theory, 
which describes the topological limit of the $\mathbb{Z}_n$ lattice gauge theory.
\cite{wen2004quantum, fradkin2013field}

In this paper, we will undertake the task of bridging 
the descriptions of (1+1)d bosonic SPT phases using MPSs,
and those using (1+1)d TFTs. 
For (1+1)d bosonic SPT phases protected by symmetry $G$,  
where $G$ is a symmetry group, 
the relevant TFTs are $G$-equivariant TFTs discussed by 
Turaev and Moore-Segal. 
\cite{Turaev, turaev2010homotopy, Moore_lecture, Moore-Segal, Kapustin-Turzillo}
We in particular address the following two issues.

The first issue is about topological invariants (SPT invariants) of (1+1)d bosonic SPT phases.
These are quantities (numbers) which one can compute for a given
quantum ground state of a gapped (1+1)d system, 
and take the same value anywhere in a given gapped phase.
I.e., they are stable and remain unchanged against adiabatic deformations
of Hamiltonians so far as one stays within a given gapped phase. 
On the one hand, several topological invariants for bosonic 
SPT phases have been constructed so far by using MPSs.
\cite{PollmannTurner2012}
In this paper, we will rederive these invariants by using 
$G$-equivariant TFTs in (1+1)d.
The topological invariants are nothing but the partition functions of TFTs. 

Second,
the hallmark of (1+1)d SPT phases 
is the presence of boundary degrees of freedom that appear 
when the SPT phases are terminated by boundaries.
The canonical example is the spin 1/2 that appears at the end of
the spin 1 Haldane spin chain. 
In terms of MPSs, 
these physical boundary degrees of freedom are captured by 
degrees of freedom living in the auxiliary bond (entanglement) Hilbert space. 
On the TFT side, 
a natural framework to discuss the boundary degrees of freedom 
is an ``open'' TFT.
\cite{lazaroiu2001structure, Moore-Segal}
Open TFTs are TFTs defined on the (1+1)d spacetime which has (1+0)d boundaries.
In this paper, 
we will make an attempt to 
make a dictionary between 
MPSs with boundaries and open TFTs. 

The rest of the paper is organized as follows:
In Sec.\ \ref{Classification and topological invariants of MPSs},
we introduce the descriptions of bosnic SPT phases in  (1+1)d using MPSs. 
In particular, we review the known construction of
various topological invariants for (1+1)d SPT phases built out of MPSs.
By using the fixed point MPSs, 
we confirm that these topological invariants 
characterizes the elements of the group cohomology $H^2(G,U(1))$,
and describe the procedure to extract these from a given quantum ground state. 
(All results in Sec.\ \ref{Classification and topological invariants of MPSs} are 
known in the literature, 
so readers who are familiar with the MPS descriptions of (1+1)d bosonic SPT phases
and their topological invariants can skip this section.)

In Sec.\ \ref{$G$-equivariant Topological field theory},
we introduce $G$-equivariant TFTs following Moore and Segal. 
\cite{Moore-Segal}
We discuss both closed and open TFTs;
In closed TFTs we consider the (1+1)d spacetime which has no 
boundary, 
whereas in open TFTs the (1+1)d spacetime has (1+0)d boundaries.

In Sec.\ \ref{Fukuma-Hosono-Kawai state sum construction}, 
we will derive
the topological invariants from the point of view of (1+1)d TFTs.
To this end, we evaluate the partition functions of 
(1+1)d TFTs by using the so-called state sum construction~\cite{FHK}.
Introducing an orientation reversing operation on Frobenius 
algebras enables us to define partition function on the real 
projective plane $\R P^2$~\cite{Karimipour-Mostafazadeh}. 

Finally, Appendices are devoted to an introduction to 
the group cohomology, 
and 
projective representations, 
the relation to orbifolded theories ($(1+1)$d Dijkgraaf-Witten theories), 
and 
the derivations of algebraic relations in $G$-equivariant open and closed TFTs.

\section{Classification and topological invariants of SPT phases using MPSs}
\label{Classification and topological invariants of MPSs}

In this section, 
we briefly review the topological classification of bosonic SPT phases 
in (1+1)d~\cite{PollmannBergTurner2012, ChenGuLiuWen2013, Schuch2011}, 
and their topological invariants~\cite{PollmannTurner2012}. 
From the field theoretical point of view,
bosonic SPT phases are described by $G$-equivariant TFTs~\cite{Moore-Segal}, 
which will be introduced in the next section. 

\subsection{Symmetry and group cohomology classification}

Let us consider a short range entangled pure state $\ket{\Psi}$ on a closed chain of length $L$, 
which is represented by a MPS
\begin{align}
\ket{\Psi} 
&= \sum_{\{m_i \}} {\rm Tr}\, (A_{m_1} \cdots A_{m_L} ) \ket{m_1\cdots  m_L}, 
\quad 
\ket{m_1\cdots  m_L} = \ket{m_1}_1 \otimes \cdots \otimes \ket{m_L}_L, 
\end{align}
where 
$\ket{m}_j$ represents a state in the {\it physical} Hilbert space at the $j$-th site and $A_{m_j}$ is a $\chi \times \chi$ matrix which acts on the 
{\it auxiliary} Hilbert space or 
the {\it entanglement} Hilbert space living on the bonds;
The trace is taken over the $\chi\times \chi$ dimensional auxiliary Hilbert space. 
Here and henceforth, we assume the translational symmetry for simplicity.

Let $G$ be a symmetry group. 
The symmetry group $G$ possibly includes orientation reversing symmetries 
(time-reversal or inversion symmetry, say). 
For the purpose of specifying the orientation reversing symmetries,
let us introduce a homomorphism $\phi: G \to G_0$, 
where $G_0$ is a group consisting of the orientation preserving symmetries. 
The symmetry action $g \in G$ is defined on the basis $\ket{m}_j$ 
by a linear representation of $G$, 
\begin{align}
&\hat g (\ket{m}_j) = \ket{n}_j [U_g]_{nm}, && U_g U_h = U_{gh}, 
&& (\mbox{$g$ is on-site unitary symmetry}), \\
&\hat T (\ket{m}_j) = \ket{n}_{j} [U_T]_{nm}, && U_T U_g^* = U_{Tg}, 
&& (\mbox{$T$ is time-reversal symmetry}), \\
&\hat P (\ket{m}_j) = \ket{n}_{L-j} [U_P]_{nm}, && U_P U_g = U_{Pg}, 
&& (\mbox{$P$ is inversion symmetry}),
\end{align}
for any $h \in G$.

\begin{figure}[!]
 \begin{center}
  \includegraphics[width=0.6\linewidth, trim=0cm 0cm 0cm 0cm]{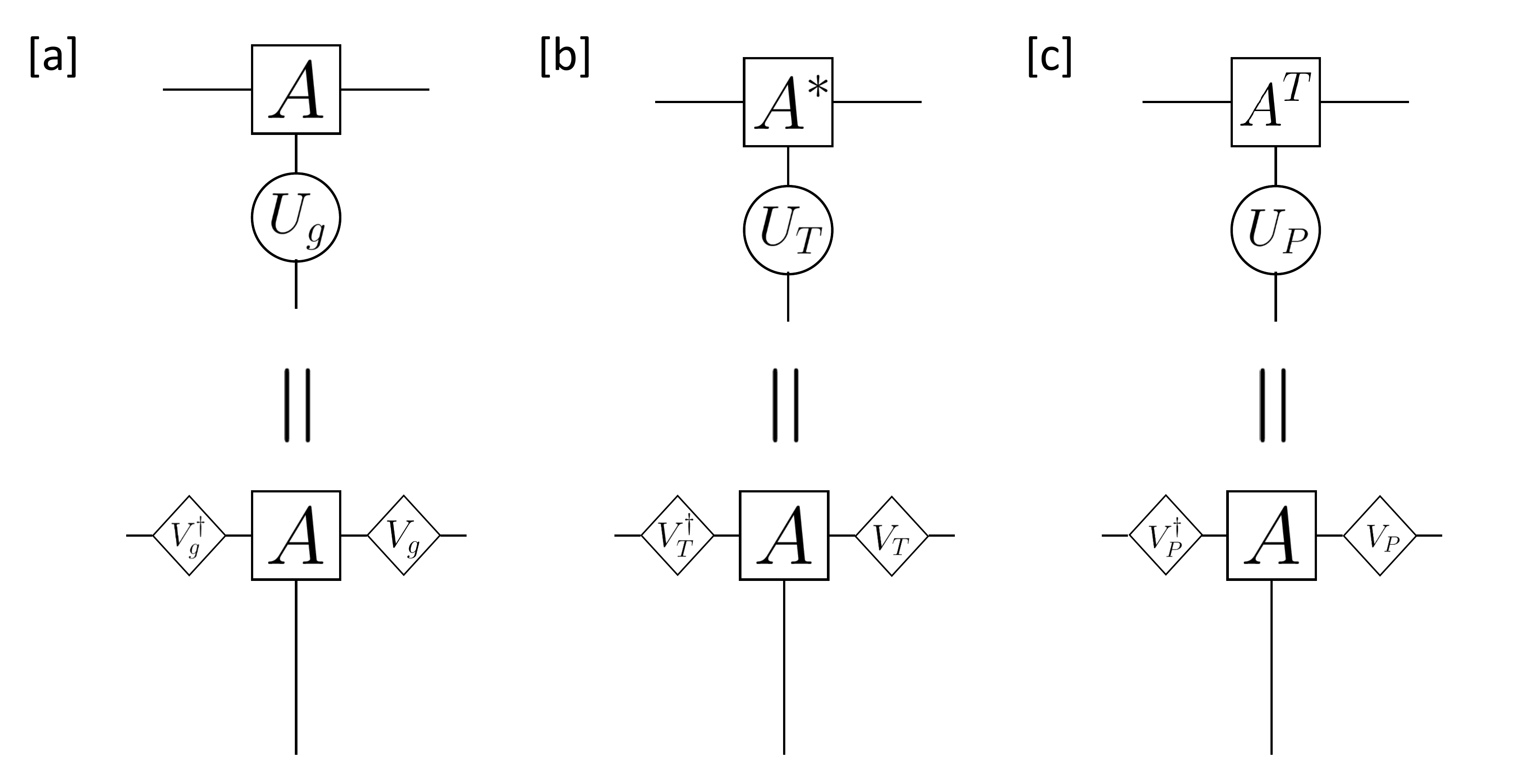}
 \end{center}
 \caption{The symmetry property of $A_m$ 
 under [a] on-site unitary, [b] time-reversal, and [c] inversion symmetries, respectively.}
 \label{Fig:Symmetry_MPS}
\end{figure}

By choosing different $A$, one can construct the ground states of gapped phases in (1+1) dimensions.   
One can consider to classify these gapped phase topologically, 
in the presence of a prescribed symmetry $G$. 
The topological classification of (1+1)d SPT phases of bosons is given by the classification of the symmetry action on $A_{m}$. 
Under the assumption that $\ket{\Psi}$ is a pure state, one can show~\cite{Cirac2008}
\begin{align}
&[U_g]_{mn} A_n = e^{i \theta_g} V_g^{\dag} A_m V_g, 
&& (\mbox{$g$ is on-site unitary symmetry}), \\
&[U_T]_{mn} A^*_n = e^{i \theta_T} V_T^{\dag} A_m V_T, 
&& (\mbox{$T$ is time-reversal symmetry}), \label{Eq:Sym_A_TRS} \\
&[U_P]_{mn} A_n^T = e^{i \theta_P} V_P^{\dag} A_m V_P, 
&& (\mbox{$P$ is inversion symmetry}), 
\end{align}
where $e^{i \theta_g}, e^{i \theta_T}$, and $e^{i \theta_P}$ are 1-dimensional linear representations of $G$, 
and $V_g, V_T$, and $V_P$ act on the entanglement Hilbert space,
and obey, for any $h\in G$,  
\begin{align}
\left\{ 
\begin{array}{ll}
V_g V_h = b(g,h) V_{gh}  & (\mbox{$g$ is on-site unitary symmetry}) 
\\
\\
V_g V_h^* = b(g,h) V_{gh} & (\mbox{$g$ is time-reversal or inversion symmetry}) \\
\end{array} \right. 
\label{Eq:ProjRep_V}
\end{align}
with a $U(1)$ phase $b(g,h) \in U(1)$ (2-cocycle). 
These symmetry actions on $A_m$ are diagrammatically represented 
in Fig.\ \ref{Fig:Symmetry_MPS}. 
(In the figure, we neglect the 1-dimensional representation $e^{i \theta_g}$.)
From the associativity condition of $V_g$, 
it follows that $b(g,h)$ is a representative of 
$\phi$-twisted second group cohomology $H^2(G,U(1)_{\phi})$. 
(Here, ``$\phi$-twisted'' means 
the $g \notin G_0$ action on $U(1)$ group is defined by complex conjugate. 
See Appendix \ref{Group cohomology}.) 
The factor group $[b(g,h)] \in H^2(G,U(1)_{\phi})$ classifies how symmetry $G$ acts on the 
short-range entangled pure state $\ket{\Psi}$ on the 1-dimensional closed chain. 

\subsection{Edge degrees of freedom}

In a non-trivial SPT phase 
(specified by a nontrivial group cohomology 
$[b(g,h)] \in H^2(G, U(1)_{\phi})$, $[b(g,h)] \neq 0$) 
on an open chain $I = \{1, \dots, L\}$, 
there emerge edge degrees of freedom where 
the symmetry action of $G$ is ``fractionalized''. 
The boundary degrees of freedom can be discussed 
by using the MPS of the open chain:
\begin{align}
\Psi(\bra{v_L} \otimes \ket{v_R}) = \sum_{\{m_i\}} 
v_L^{\dag} A_{m_1} \cdots A_{m_L} v_R \ket{m_1 \cdots m_L}, 
\qquad
\bra{v_L} \in V^*, 
\quad 
\ket{v_R} \in V, 
\end{align}
where $\bra{v_L}$ and $\ket{v_R}$ specify boundary conditions 
and belong to the edge Hilbert space $V^*$ and $V$, respectively. 
$V$ is a $b(g,h)$-projective representation defined by (\ref{Eq:ProjRep_V}) 
and $V^*$ is the its conjugate representation. 
Here, the symmetry fractionalization is realized in the following sense: 
A symmetry action $g \in G_0$ on the MPS on the open chain is given by
\begin{align}
\hat g \Psi(\bra{v_L} \otimes \ket{v_R}) = \sum_{\{m_i\}} v_L^{\dag} V_g^{\dag} A_{m_1} \cdots A_{m_L} V_g v_R \ket{m_1 \cdots m_L}
= \Psi(\bra{v_L} V_g^{\dag} \otimes V_g \ket{v_R}), 
\end{align}
where a symmetry operation $g \in G_0$ is projectively represented at the edges,
as opposed to the $g$ action on the bulk physical degrees of freedom, which is 
a linear representation.

For an on-site unitary symmetry $g \in G_0$, we can introduce 
a $g$-twisted MPS 
\begin{align}
\ket{\Psi_g} := \sum_{\{m_i\}} 
{\rm Tr}\, (A_{m_1} \cdots A_{m_L} V_g) \ket{m_1\cdots  m_L}, \quad (g \in G_0). 
\end{align}
From the perspective of Hamiltonians, 
$\ket{\Psi_g}$ is a ground state of a Hamiltonian with a $g \in G_0$ symmetry defect. 

For our purpose to make a comparison between MPSs and TFTs,  
it is useful to introduce an open to closed map $\imath^{g}$ 
and a closed to open map $\imath_{g}$~\cite{Moore-Segal}
as
\begin{align}
&\imath^{g} \big( \Psi(\bra{v_L} \otimes \ket{v_R}) \big) := \braket{v_L| V_g^{\dag} |v_R} \ket{\Psi_g}, 
\quad (g \in G_0), \\
&\imath_{g}(\ket{\Psi_g}) := \sum_a \Psi(\bra{a} \otimes V_g \ket{a}), 
\quad  (g \in G_0). 
\end{align}
Here $\{ \ket{a} \}_{a=1}^{{\rm dim} V}$ is a basis of $V$. 

Finally, we also introduce a formal ``gluing'' operation of two open MPSs by 
\begin{align}
\Psi(\bra{v_L} \otimes \ket{v_R}) \cdot \Psi(\bra{w_L} \otimes \ket{w_R}) 
:= \braket{w_L | v_R} \Psi(\bra{v_L} \otimes \ket{w_R}). 
\end{align}
This will be also useful when we make a comparison between MPSs and TFTs. 


\subsection{Simple and fixed point MPSs}
\label{Fixed point MPSs}

To study ground states deep inside a gapped phase,
or to study SPT phases in general, 
it is useful and convenient to introduce a simple and fixed point MPS. 
``Simple'' here means that the transfer matrix 
$T_{ab,cd} = \sum_{m} [A_m]_{ab} [A^*_m]_{cd}$
has an only one eigenstate with unit magnitude of eigenvalue $|\nu|=1$, 
i.e., unique ground state.~\cite{fidkowski2011topological}
``Fixed point'' means that we are in the limit of zero correlation length. 
It is in this limit where we expect
SPT phases and the corresponding MPSs
are faithfully described by TQFTs. 
In the following, we will construct fixed point MPSs
$\ket{\Psi}$ 
with a nontrivial group cohomology $H^2(G,U(1)_{\phi})$.

Let $G$ be a symmetry group with nontrivial group cohomology $H^2(G,U(1)_{\phi}) \neq 0$. 
We fix a nontrivial 2-cocycle $b(g,h) \in Z^2(G,U(1)_{\phi}), [b(g,h)] \neq 0$. 
We choose two $b(g,h)$-projective representations 
$V$ which satisfy (\ref{Eq:ProjRep_V}). 
We use the tensor product representation 
$V^* \otimes V$ as a 
physical Hilbert space, where 
$V^*$ is the complex representation of 
$V$. 
Note that in the product representation, 
the effect of the 2-cocycle cancels, $b^*(g,h) b(g,h)=1$. 
For each site $j$, the basis is given by $\{ \ket{a}^L_j \otimes \ket{b}^R_j \}$,
where $\ket{a}^L_j$ ($\ket{b}^R_j$) is the basis of $V^*$ ($V$) 
and transformed as 
\begin{align}
&\left\{\begin{array}{l}
\hat g (\ket{a}^L_j \otimes \ket{b}^R_j) = \ket{c}^L_j \otimes \ket{d}^R_j {[V^*_g]}_{ca} {[V_g]}_{db}, \\ 
{[U_g]}_{ab,cd} = {[V^*_g]}_{ac} {[V_g]}_{bd}, \\
\end{array}\right. && \mbox{($g$ is on-site unitary symmetry)},\\
&\left\{\begin{array}{l}
\hat T (\ket{a}^L_j \otimes \ket{b}^R_j) = \ket{c}^L_j \otimes \ket{d}^R_j {[V^*_T]}_{ca} {[V_T]}_{db}, \ \ \hat T i \hat T^{-1} = -i , \\
{[U_T]}_{ab,cd} = {[V^*_T]}_{ac} {[V_T]}_{bd}, \\
\end{array}\right. && \mbox{($T$ is time-reversal symmetry)}, \\
&\left\{\begin{array}{l}
\hat P (\ket{a}^L_j \otimes \ket{b}^R_j) = \ket{c}^L_{L-j} \otimes \ket{d}^R_{L-j} {[V^*_P]}_{cb} {[V_P]}_{da}, \\
{[U_P]}_{ab,cd} = {[V^*_P]}_{ad} {[V_P]}_{bc}, \\
\end{array}\right. && \mbox{($P$ is inversion symmetry)}. 
\end{align}
Note that $V$ and $V^*$ representations are exchanged under the inversion transformation. 

\begin{figure}[!]
 \begin{center}
  \includegraphics[width=\linewidth, trim=0cm 0cm 0cm 0cm]{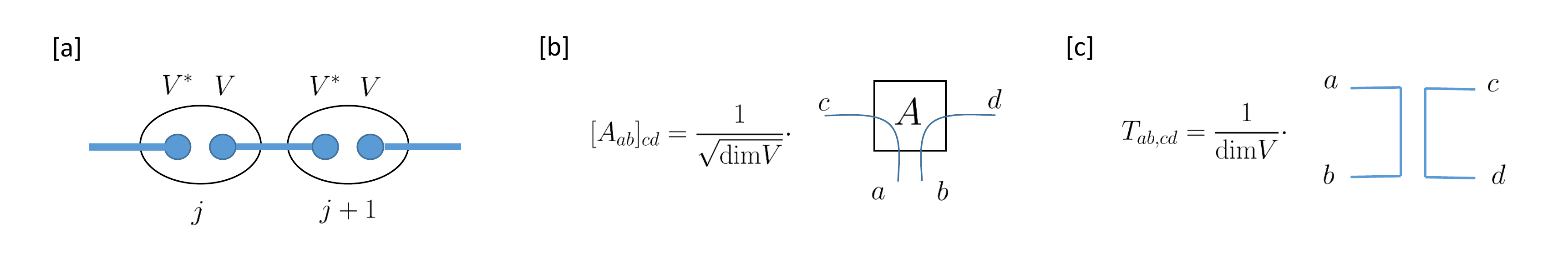}
 \end{center}
 \caption{[a] A fixed point MPS. The blue bond represents the singlet representation in $V^* \otimes V$. [b] $A$ matrix. [c] The transfer matrix.}
 \label{Fig:MPS_Fixed_Point}
\end{figure}

To write down ground state wave functions,
we make use of a singlet included in the decomposition of the product representations, 
$
V^* \otimes V = 1 \oplus \cdots. 
$
The fixed point MPS can be constructed as the product state of singlet bonds
(Fig.\ \ref{Fig:MPS_Fixed_Point} [a]) as 
\begin{align}
\ket{\Psi} 
= \cdots \otimes (\sum_{a} \ket{a}^R_j \otimes \ket{a}^L_{j+1}) \otimes (\sum_{b} \ket{b}^R_{j+1} \otimes \ket{b}^L_{j+2}) \otimes \cdots. 
\label{eq:fixed_point_MPS_def}
\end{align}
This can be written in the MPS form as 
\begin{align}
\ket{\Psi} = \sum \cdots [A_{b_j c_j}]_{a_j a_{j+1}} [A_{b_{j+1} c_{j+1}}]_{a_{j+1} a_{j+2}} \cdots 
\ket{ \cdots (b_j c_j) (b_{j+1} c_{j+1}) \cdots}
\end{align}
with 
\begin{align}
[A_{ab}]_{cd} = \frac{1}{\sqrt{{\rm dim} V}} \delta_{ac} \delta_{bd}. 
\end{align}
(See Fig.\ \ref{Fig:MPS_Fixed_Point} [b]. )
Here $(ab)$ is the physical index whereas $(cd)$ is the entanglement index. 
We abbreviated $\ket{b_j}^L_j \otimes \ket{c_j}^R_j$ by $\ket{(b_jc_j)}$. 
The prefactor $\frac{1}{\sqrt{{\rm dim} V}}$ is the normalization constant. 
The MPS $\ket{\Psi}$ is the AKLT state \cite{affleck1988valence} 
without any projection on the site degrees of freedom.~\cite{chen2012symmetry}

For the fixed point MPSs, the transfer matrix $T_{ab, a'b'}$ (Fig.\ \ref{Fig:MPS_Fixed_Point} [c]) is given by 
\begin{align}
T_{ab,cd}
= \sum_{ef} [A_{ef}]_{ab} [A^*_{ef}]_{cd}
= \frac{1}{{\rm dim} V} \delta_{ac} \delta_{bd}. 
\end{align}
In any symmetries, we have the following group action on $[A_{ab}]_{cd}$ for the fixed point MPS, 
\begin{align}
&[U_g]_{ab,ef} [A_{ef}]_{cd}
= \frac{1}{\sqrt{{\rm dim} V}} [V_g^*]_{ac} [V_g]_{bd}, 
\quad (\mbox{$g$ is on-site unitary symmetry}), \\
&[U_T]_{ab,ef} [A^*_{ef}]_{cd}
= \frac{1}{\sqrt{{\rm dim} V}} [V_T^*]_{ac} [V_T]_{bd}, 
\quad (\mbox{$T$ is time-reversal symmetry}), \\
&[U_P]_{ab,ef} [A^T_{ef}]_{cd} 
= \frac{1}{\sqrt{{\rm dim} V}} [V_P^*]_{ac} [V_P]_{bd}, 
\quad (\mbox{$P$ is inversion symmetry}). 
\end{align}

Finally, it is worth pointing out that in fixed point MPSs 
the length of the MPS chain is irrelevant because of the zero correlation length. 
I.e., since they are at a renormalization group fixed point, 
increasing/decreasing the number of cites does not change the essential properties of
the state.
For this reason, we always identify MPS chains with different lengths as 
\begin{align}
{\rm Tr} ( A_{m_1} \cdots A_{m_L} ) \ket{m_1 \cdots m_L} \sim {\rm Tr} ( A_{m_1} \cdots A_{m_L} A_{m_{L+1}}) \ket{m_1 \cdots m_L m_{L+1}}. 
\label{eq:mps_equivalence}
\end{align}

\subsection{Topological invariants}
\label{Topological invariants}

In this section, we will construct and discuss topological invariants 
of bosonic SPT phases in (1+1)d using the MPS. 

We start by listing topological invariants.
Detailed descriptions of topological invariants will follow shortly.
There are three types of topological invariants,
which are defined in terms of the data of 2-cocycle $\{b(g,h)\}$:
\begin{itemize}
\item The discrete torsion phase (partition function on $T^2$ with twist) 
\begin{align}
\label{eq:T^2inv}
\epsilon(g,h) = \frac{b(g,h)}{b(h,g)}, 
\quad
V_g V_h = \epsilon(g,h) V_h V_g, 
\quad 
g,h\in G_0, 
\quad 
gh=hg.
\end{align}
\item The crosscap invariant (partition function on $\R P^2$)
\begin{align}
\label{eq:Corsscapinv}
\theta(g) := b(g,g),
\quad V_g V_g^* = \theta(g), 
\quad g \notin G_0, 
\quad g^2 = 1. 
\end{align}
\item The Klein bottle invariant (partition function on the Klein bottle with twist)
\begin{align}
\label{eq:KBinv}
\kappa(g;h) = \frac{b(g,h^{-1}) b(h,h^{-1})}{b(h,g)}, 
\quad 
V_g V_h^T = \kappa(g;h) V_h V_g, 
\quad 
g \notin G_0, h \in G_0, 
\quad g h^{-1} = h g. 
\end{align}
\end{itemize}

Several comments are in order.

-- 
First, in \eqref{eq:T^2inv}, \eqref{eq:Corsscapinv} and \eqref{eq:KBinv}, 
and throughout this subsection, 
we omit the 1-dimensional representation 
$\{ e^{i \theta_g}\}$ for simplicity.  

--
One can check easily that 
these quantities are left unchanged under the 1-coboundary $b(g,h) \mapsto b(g,h) a_g a_g^{\phi(g)} a_{gh}^{-1}$, $a_g \in U(1)$.

--
One can give interpretations to  
these topological invariants 
in terms of spacetime path integrals.
We will mention these interpretations 
later in this subsection, 
and also in Sec.\ 
\ref{$G$-equivariant Topological field theory}
-
\ref{Fukuma-Hosono-Kawai state sum construction}
from the TFT point of view.
In short, 
these three topological invariants are interpreted as
the partition function on the torus,
the projective plane, 
and 
the Klein bottle respectively. 
For this reason, we will often refer the topological invariants
as the partition functions 
(on the torus, the projective plane, and the Klein bottle).

--
 Finally, the above three SPT invariants are not independent. 
One can show 
\begin{align}
&\epsilon(g,1) = \epsilon(1,g) = 1 && {\rm for\ \ } g \in G_0, \\
&\epsilon(h,g) = \epsilon(g,h)^{-1} && {\rm for\ \ } g,h \in G_0, gh=hg, \\
&\epsilon(g,hk) = \epsilon(g,h)\epsilon(g,k) && {\rm for\ \ } g,h,k \in G_0, gh=hg, gk=kg,  \\
&\kappa(g;1) = 1 && {\rm for\ \ } g \notin G_0, \\
&\kappa(g;hk) = \kappa(g;h) \kappa(g;k) && {\rm for\ \ } g \notin G_0, h,k \in G_0, g h^{-1} = h g, g k^{-1}=k g, \\
&\kappa(kg;h) = \kappa(gk^{-1};h) = \epsilon(k,h) \kappa(g;h) && {\rm for\ \ } g \notin G_0, h,k \in G_0, g h^{-1} = h g, h k=k h, \\
&\theta(h g) = \theta(g h^{-1}) = \kappa(g;h) \theta(g) && {\rm for\ \ } g \notin G_0, h \in G_0, g^2=1, g h^{-1} = h g. 
\end{align}
In many cases, the Klein bottle SPT invariant $\kappa(g;h)$ 
can be written in terms of $\epsilon(g,h)$ and $\theta(g)$. 
This is however not always the case. 
A simple example in which the Klein bottle invariant does not reduce to the other invariants 
is an SPT phase protected by   
$G = \Z_4 = \{1,\sigma,\sigma^2, \sigma^3\}$.
Here, the generator $\sigma$ is inversion/time-reversal. 
For a nontrivial projective representation generated by 
$V_{\sigma} = e^{-i s_y \frac{\pi}{4}}$,
where
$s_y = \begin{pmatrix}
0 & -i \\
i & 0 \\
\end{pmatrix}$ 
is the $y$-component of the Pauli matrix, 
from $V_{\sigma} V_{\sigma^2}^T = - V_{\sigma^2} V_{\sigma}$,  
the Klein bottle SPT invariant reads $\kappa(\sigma;\sigma^2) = -1$.

\subsubsection{Topological invariants in terms of ground state 
wave functions}
As mentioned, 
the topological invariants can be interpreted 
by using the path integral formalism.
In the rest of this subsection,
we will instead use the operator formalism,
and in particular 
aim to extract the topological invariant  
solely by using ground state wave functions.
(Apart from our goal of bridging MPSs and TFTs, 
expressing topological invariants solely in terms 
of ground state wave functions may have practical (numerical)
merits. \textcolor{blue}{)}

In order to discuss and define these topological invariants,
one important ingredient 
is {\it gauging} symmetry.
Here, by gauging, we mean coupling the system to the background flat $G$-bundle. 
We will describe how this can be done within MPSs 
for on-site unitary symmetry 
in Sec.\ \ref{Discrete torsion phase: symmetry action on twisted ground state}.
The same gauging procedure can be introduced by using 
the path-integral. 
(In addition, one could promote the background gauge field into a dynamical one.
This procedure is often called {\it orbifolding} 
to distinguish it from gauging.
In this paper, for the purpose of describing SPT phses, 
we will consider gauging but not orbifolding.  
Orbifolding leads to the so-called Dijkgraaf-Wittten
theories, which we briefly discuss in Appendix \ref{Orbifolding: Dijkgraaf-Witten theory in (1+1)d}.)

As for the crosscap invariant,
we need to introduce a ``trick''  
within the operator formalism 
in order to mimic the effect of 
putting the theory on $\mathbb{R}P^2$
in the path integral formalism.
This can be done in two different ways, 
depending on whether the symmetry 
group includes spatial inversion or time-reversal. 
Following Pollmann and Turner
\cite{PollmannTurner2012}
we will introduce two operations, 
``partial inversion'' and ``adjacent partial transposition'',
for spatial inversion and time-reversal, respectively.  
When interpreted in the path integral formalism,
these operations effectively create
$\R P^2$ as the spacetime manifold. 

Introducing such a {\it partial space-time twist operator} 
is a useful way to detect SPT topological invariant 
which cannot be represented by the partition function 
on a mapping torus.~\footnote{
A mapping torus is space-time manifold which takes 
the form 
$M \times_f S^1 := M \times [0,1] / \big\{ (x,0) \sim (f(x), 1) \big\}$, 
where $f : M \to M$ is a diffeomorphism. 
In (invertible) TFTs, the partition function $Z(M \times_f S^1)$ 
on $M \times_f S^1$ is given by the expectation value of 
operator $\hat f$ representing diffeomorphism $f$ on 
the ground state wave function $\ket{\Psi_M}$ on $M$ 
as $Z(M \times_f S^1) = \braket{\Psi_M | \hat f | \Psi_M}$. 
}
This prescription can also be applied to 
fermionic SPT invariants 
\cite{Shapourian-Shiozaki-Ryu}
and SPT phases in more general space dimensions. 
\cite{Shiozaki-Shapourian-Ryu}

For all topological invariants, the fact that they can be extracted from 
ground state wave functions
can be easily proven if we use the fixed point MPS. 

%
%
%

\paragraph{Discrete torsion phase (1): symmetry action on twisted ground state}
\label{Discrete torsion phase: symmetry action on twisted ground state}

Let us first express the torus topological invariant 
\eqref{eq:T^2inv} by using ground state wave functions. 

\begin{figure}[t]
 \begin{center}
  \includegraphics[width=\linewidth, trim=0cm 0cm 0cm 0cm]{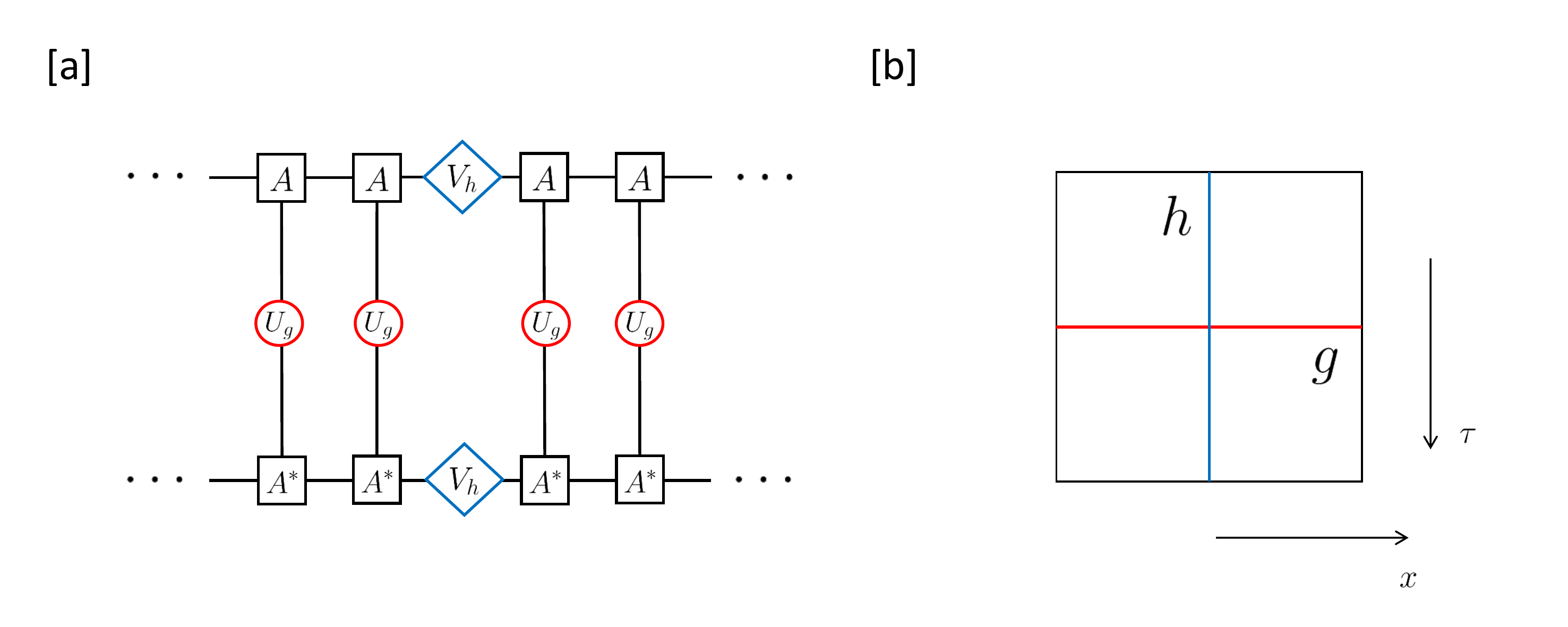}
 \end{center}
 \caption{
[a] MPS expression of the symmetry action on the twisted ground state. 
[b] The equivalent path integral on the 2-torus $T^2$. 
The blue and red lines express the symmetry defect lines.}
 \label{Fig:torus_discrete_torsion}
\end{figure}

From the MPS and the projective representation $\{V_g\}_{g \in G}$, 
we can construct the MPS $\ket{\Psi_h}$ with  
boundary condition twisted by an on-site unitary symmetry $h \in G_0$ as 
\begin{align}
\ket{\Psi_h}
&= \sum_{\{m_i\}} {\rm Tr}\, (A_{m_1} \cdots A_{m_L} V_h) \ket{m_1\cdots  m_L}. 
\end{align}
Then, for
a global unitary symmetry, 
$g \in G_0$, 
\begin{align}
\hat g \ket{\Psi_h}
= \sum_{\{m_i\}} 
{\rm Tr}\, (A_{m_1} \cdots A_{m_L} V_g V_{h} V_g^{-1}) \ket{m_1\cdots  m_L} 
= \frac{b(g,h)}{b(g h g^{-1}, g)} \ket{\Psi_{ghg^{-1}}}. 
\label{Eq:G-action_MPS}
\end{align}
The MPS diagram is shown in Fig.~\ref{Fig:torus_discrete_torsion}[a]. 
If $g h g^{-1} = h$, the $U(1)$ factor is well-defined which is invariant under the 1-coboundary. 
From the TFT point of view, 
this invariant is nothing but the partition function $Z_{T^2}(g,h)$ on the torus $T^2$ with the background $g$ and $h$ twist 
\begin{align}
\epsilon(g,h) = Z_{T^2}(g,h) = \braket{\Psi_h | \hat g | \Psi_h},
\quad (gh=hg). 
\end{align}
  
In the spacetime path integral,
it would be useful to introduce symmetry defect lines to express the background $G$ field. 
The matter field is transformed by $U_g$ when it passes through the symmetry defect line of $g$. 
Fig.~\ref{Fig:torus_discrete_torsion}[b] shows the symmetry defect lines corresponding to the partition function $Z_{T^2}(g,h)$ twisted by $g$ and $h$. 
The discrete torsion phase $\epsilon(g,h)$ arises from the intersection of two symmetry defect lines of $g$ and $h$ with $[g,h]=0$.

\paragraph{Discrete torsion phase (2): partial symmetry action and swapping}
\label{Discrete torsion phase: partial symmetry action and swapping}

\begin{figure}[t]
 \begin{center}
  \includegraphics[width=0.7\linewidth, trim=0cm 0cm 0cm 0cm]{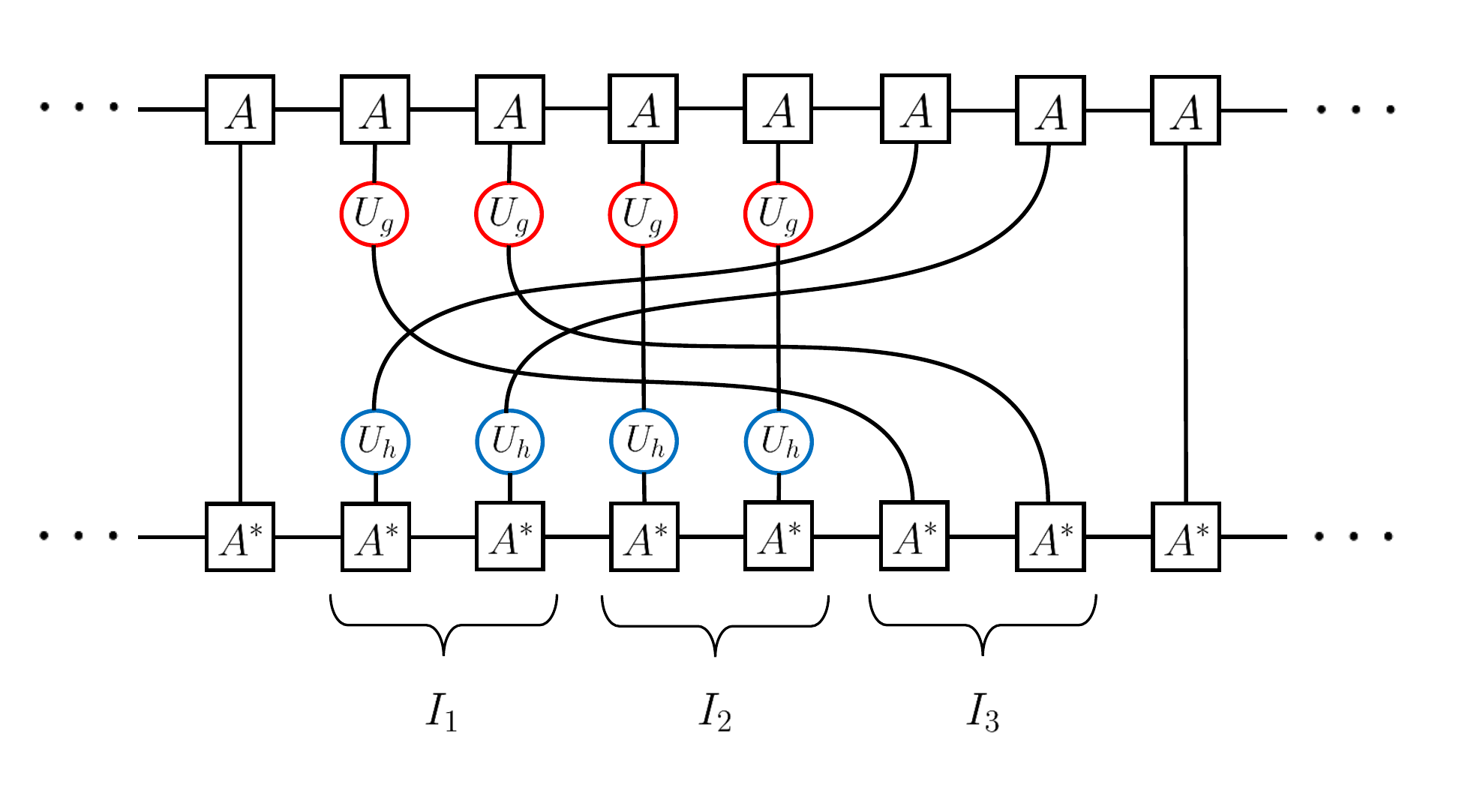}
 \end{center}
 \caption{
MPS expression of the partial symmetry action with swapping defined in (\ref{eq:symmetry_action_and_swap}). 
}
 \label{Fig:symmetry_action_and_swap}
\end{figure}

\begin{figure}[t]
 \begin{center}
  \includegraphics[width=\linewidth, trim=0cm 0cm 0cm 0cm]{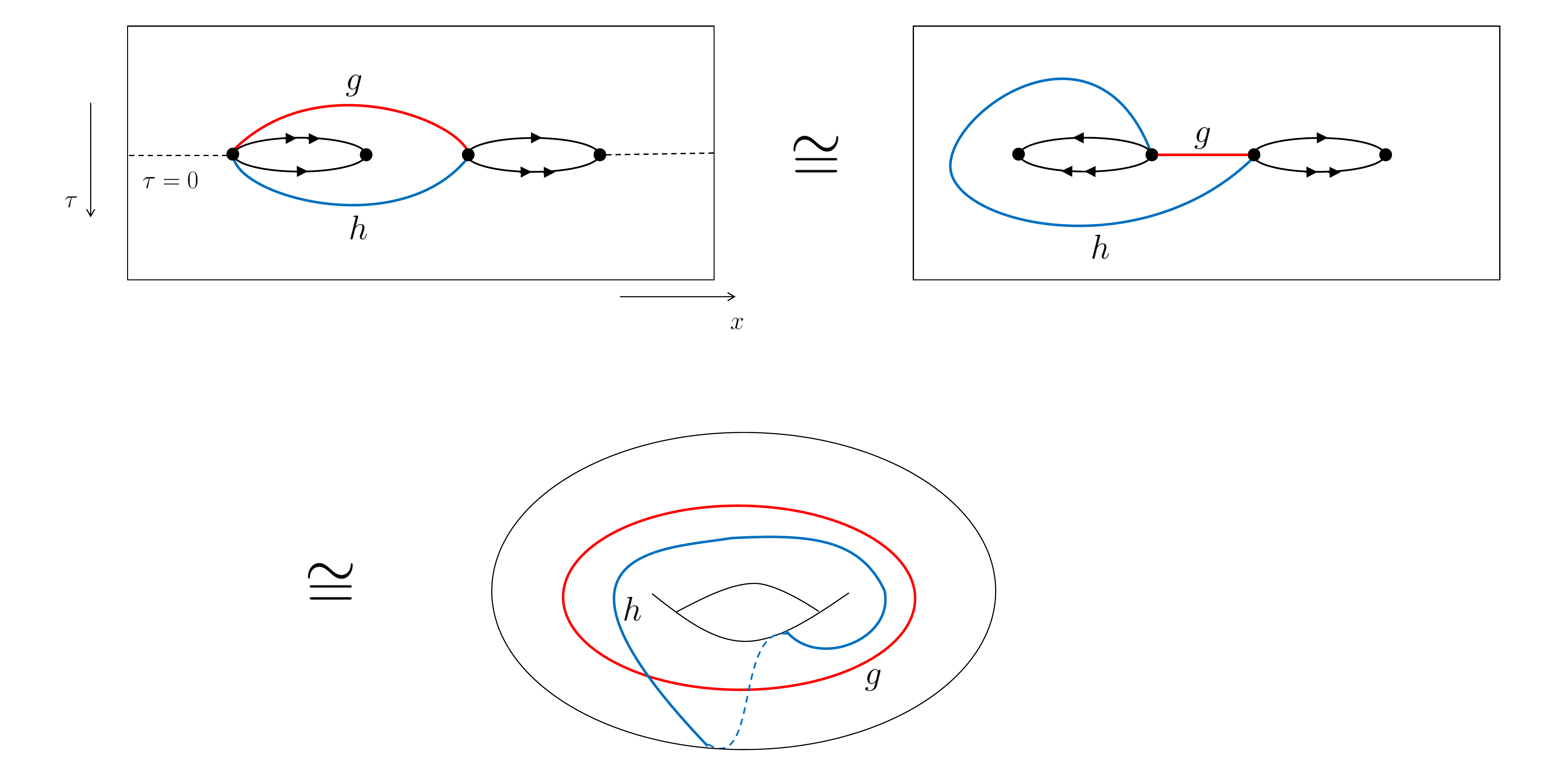}
 \end{center}
 \caption{
The geometry of path integral for the partial symmetry action with swapping operator defined in (\ref{eq:symmetry_action_and_swap}). 
The red and blue lines express the symmetry defect lines. 
The intervals with arrows are identified with ones having the same number of arrows, which results in the 2-torus. 
}
 \label{Fig:symmetry_action_and_swap_path_integral}
\end{figure}

There is an alternative way to detect the discrete torsion phase invariant. 
It is given by the combination of the partial symmetry action and the swapping operator, described as follows.
Let $\ket{\Psi}$ be the ground state on $S^1$ with no flux.
(In the TFT path integral, this state is obtained/defined by the path-integral over the disc.)
We introduce three adjacent intervals $I_1 \cup I_2 \cup I_3$ with $I_1$ and $I_3$ having the same number of sites. 
The discrete torsion phase $\epsilon(g,h)$ is then extracted as the complex $U(1)$ phase of the quantity~\cite{PhysRevLett.109.050402}
\begin{align}
Z = \Braket{\Psi | \prod_{j \in I_1 \cup I_2} (U_h)_j \cdot {\rm Swap}(I_1,I_3) \cdot \prod_{j \in I_1 \cup I_2} (U_g)_j | \Psi}, \qquad 
gh=hg, 
\label{eq:symmetry_action_and_swap}
\end{align}
in the limit $|I_1|,|I_2|, |I_3| \gg \xi$, where $\xi$ is the correlation length of the bulk. 
Here, ${\rm Swap}(I_1,I_3)$ is the operator swapping the two intervals $I_1$ and $I_3$, which is defined by 
\begin{align}
{\rm Swap}(I_1,I_3) \ket{m_j} 
= \left\{\begin{array}{ll}
\ket{m_{j+|I_1|+|I_2|}} & (j \in I_1), \\
\ket{m_{j-|I_1|-|I_2|}} & (j \in I_3), \\
\ket{m_j} & ({\rm otherwise}).
\end{array}\right.
\end{align}
For the MPS representation of the ground state $\ket{\Psi}$, the MPS diagram of $Z$ is written as Fig.~\ref{Fig:symmetry_action_and_swap}. 
For the fixed point MPS (\ref{eq:fixed_point_MPS_def}), it is easy to show that $Z = |Z| \epsilon(g,h)$. 
The path-integral picture also verifies that $Z$ gives the discrete torsion phase. 
See Fig.~\ref{Fig:symmetry_action_and_swap_path_integral}. 
Topologically, the swapping operator ${\rm Swap}(I_1,I_3)$ with the intermediate region $I_2$ is equivalent to adding a genus. 
The background $G$ field obtained by the partial symmetry actions $U_g$ and
$U_h$ on the adjacent intervals $I_1 \cup I_2$
has an intersection between two symmetry defect lines of $g$ and $h$, which leads to the discrete torsion phase $\epsilon(g,h)$.

\paragraph{Crosscap from inversion symmetry: ``partial inversion''}
\begin{figure}[t]
 \begin{center}
  \includegraphics[width=\linewidth, trim=0cm 0cm 0cm 0cm]{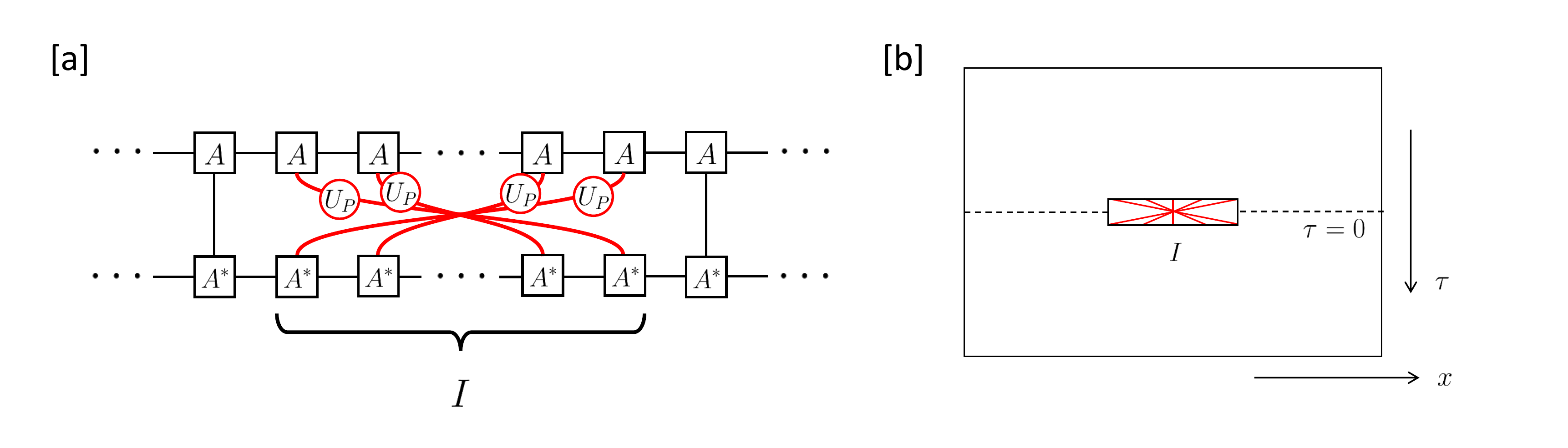}
 \end{center}
 \caption{[a] MPS representation of crosscap. [b] The geometry of path integral for the partial inversion.}
 \label{Fig:Partial_Inversion}
\end{figure}

The crosscap topological invariant
can be defined when the symmetry group $G$
includes spatial inversion or time-reversal. 
The procedures to extract the invariant 
from ground state wave functions
are different for spatial inversion and time-reversal.  
Let us first discuss the crosscap topological
invariant when $G$ includes
spatial inversion. 

Let $\ket{\Psi}$ be a MPS on the closed chain $L$. 
To create the real projective plane $\R P^2$, we take the {\it partial inversion} on the interval $I = \{1, \dots, N\}$,
as shown in Fig.\ \ref{Fig:Partial_Inversion} [a], 
as 
\begin{align}
\hat P_I \ket{\Psi}
= \sum_{\{m_i\}} {\rm Tr}\, (A_{m_1} \cdots A_{m_L}) \ket{\widetilde{m}_1 \cdots \widetilde{m}_N m_{N+1} \cdots m_L}, 
\end{align}
where $\widetilde{\ket{m}}_j$ is the partially inverted physical degrees of freedom, 
\begin{align}
\widetilde{\ket{m}}_j = \ket{n}_{N-j} [U_P]_{nm}. 
\end{align}
We assume the length of the interval $I$ is sufficiently larger than the correlation length. 

The fact that this operation
creates $\mathbb{R}P^2$ as the spacetime manifold 
can be easily understood from Fig.\ \ref{Fig:Partial_Inversion}-[b];
In the path integral representation, the partial inversion is equivalent to inserting 
a one crosscap on the time slice at $\tau=0$. 

One can show that for $P^2 = 1$ the amplitude $\hat P_I \ket{\Psi}$ gives the crosscap invariant~\cite{PollmannTurner2012}
\begin{align}
\theta(P) = Z_{\R P^2}(P) = \frac{\braket{\Psi| \hat P_I |\Psi}}{|\braket{\Psi| \hat P_I |\Psi}|} = b(P,P), 
\quad P^2 = 1, 
\end{align}
in the limit $L \to \infty$ and $N \to \infty$. 

This formula can easily be proven 
for fixed point MPSs introduced in the previous section.
Here we give a proof by using the cut and glue construction.~\cite{Qi2012}
To illustrate the proof, we use the Haldane chain protected by the inversion symmetry with $H^2(\Z_2;U(1)_{\phi}) = \Z_2$. 
First, we cut the chain $L$ by the interval $I$. 
There are four effective low energy degrees of freedom localized at the boundary of $I$, 
\begin{align}
\ket{e^R_0} \otimes \ket{e^L_1} \otimes \ket{e^R_N} \otimes \ket{e^L_{N+1}}, 
\quad 
e^L_j, e^R_j = \uparrow, \downarrow. 
\end{align}
Next, we glue these degrees of freedom to get the original ground state by forming the singlet bond as 
sites $0$-$1$ and $N$-$(N+1)$ as 
\begin{align}
\ket{\Psi} = \frac{1}{\sqrt{2}} (\ket{\uparrow^R_0} \otimes \ket{\downarrow^L_1} - \ket{\downarrow^R_0} \otimes \ket{\uparrow^L_1} ) 
\otimes \frac{1}{\sqrt{2}} (\ket{\uparrow^R_N} \otimes \ket{\downarrow^L_{N+1}} - \ket{\downarrow^R_N} \otimes \ket{\uparrow^L_{N+1}} ). 
\end{align}
The reduced density matrix $\rho_I$ is 
\begin{align}
\rho_I 
&= {\rm tr}_{0,{N+1}} (\ket{\Psi} \bra{\Psi}) 
= \frac{1}{4} \left[ 
(\ket{\uparrow^L_1}\bra{\uparrow^L_1} +\ket{\downarrow^L_1}\bra{\downarrow^L_1}) \otimes 
(\ket{\uparrow^R_N}\bra{\uparrow^R_N} +\ket{\downarrow^R_N}\bra{\downarrow^R_N}) 
\right]
= \frac{1}{4} {\rm Id}^L_{1} \otimes {\rm Id}^R_N. 
\end{align}
The partial inversion $P_I$ acts as 
\begin{align}
\hat P_I \ket{\uparrow^L_1} = \ket{\downarrow^R_N}, 
\quad
\hat P_I \ket{\downarrow^L_1} = -\ket{\uparrow^R_N}, 
\quad 
\hat P_I \ket{\uparrow^R_N} = \ket{\downarrow^L_1}, 
\quad 
\hat P_I \ket{\downarrow^R_N} = -\ket{\uparrow^L_1}, 
\end{align}
from which we read off  
\begin{align}
\braket{\Psi|\hat P_I|\Psi} = {\rm tr}_I( \hat P_I \rho_I) = - \frac{1}{2}. 
\end{align}
We thus obtained the topological invariant $b(P,P) = -1$,
which, as expected, is non-trivial (differs from $b(P,P)=1$), 
and is the $\mathbb{Z}_2$ invariant.

\paragraph{Crosscap from time-reversal symmetry: ``adjacent partial transposition''}

\begin{figure}[!]
 \begin{center}
  \includegraphics[width=\linewidth, trim=0cm 0cm 0cm 0cm]{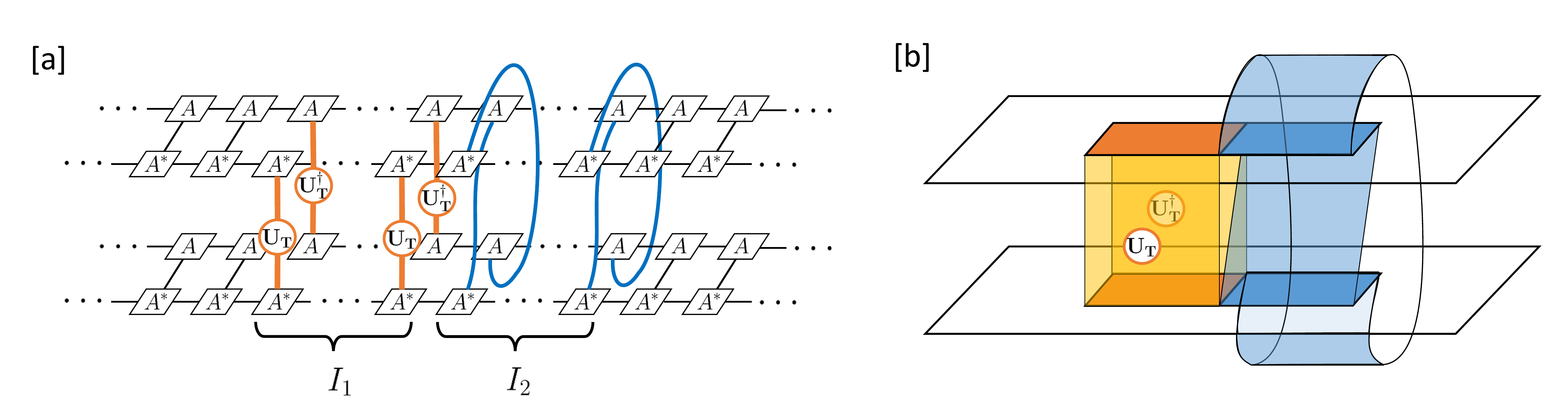}
 \end{center}
 \caption{[a] MPS representation of the crosscap. [b] The path integral representation of partial transposition. }
 \label{Fig:Partial_Transpose}
\end{figure}

\begin{figure}[!]
 \begin{center}
  \includegraphics[width=\linewidth, trim=0cm 0cm 0cm 0cm]{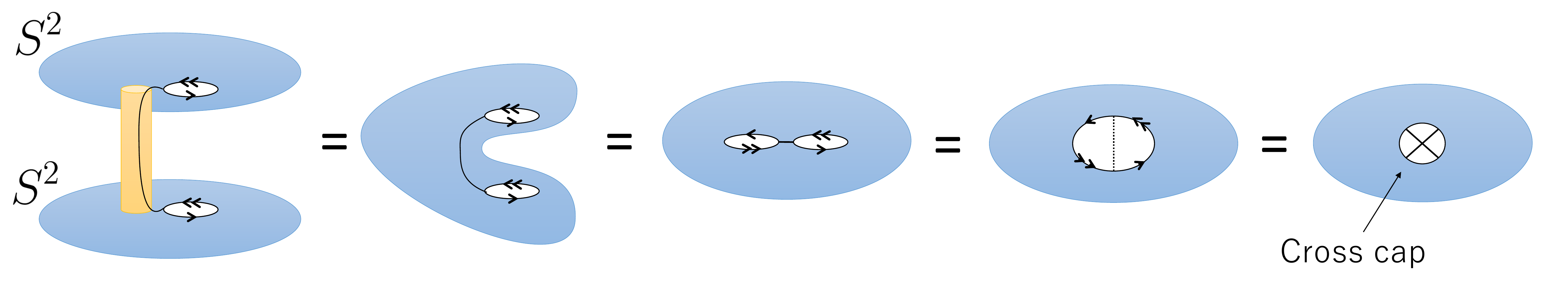}
 \end{center}
 \caption{Equivalence between 
adjacent partial transpose and inserting a cross cap.
 The black line is identified. }
 \label{Fig:TRS_to_RP2}
\end{figure}

Next, we describe the extraction of the crosscap invariant 
when $G$ includes time-reversal.
To this end, we will consider the so-called partial transposition.
The partial transposition has been used, for example, to define 
the entanglement negativity.~\cite{VidalWerner}
Pollmann and Turner~\cite{PollmannTurner2012} 
showed the MPS network 
(Fig.\ \ref{Fig:Partial_Transpose} [a]) of 
the partial transposition on adjacent intervals $I= I_1 \cup I_2$ 
is nothing but the $\Z_2$ topological invariant $b(T,T)$ 
associated with the time-reversal symmetry $T \notin G_0$. 
(Here, to be concrete, let 
$I_1 = \{1, \dots, N\}$ and $I_2 = \{N+1, \dots, M\}$
be two adjacent intervals in the chain of length $L$.) 
In the following, we will review the topological invariant of  
Pollmann and Turner.
We will also note, with an eye toward the TFT descriptions of 
SPT phases and their topological invariants, that 
the Pollmann-Turner can be interpreted as a space-time path integral 
on the real projective plane.

Below, we will express the Pollmann-Turner invariant in terms of a density matrix. 
We introduce the reduced density matrix for the interval $I$ by taking the 
partial trace of the degrees of freedom on living on the compliment of $I$, $L\backslash I$,
as 
\begin{align}
\rho_I = {\rm tr}_{L \backslash I} (\ket{\Psi} \bra{\Psi}). 
\end{align}
The reduced density matrix
$\rho_I$ is expanded in the basis of $I_1$ and $I_2$ as 
\begin{align}
\rho_I = \sum_{ijkl} \ket{e^1_i, e^2_j} \braket{e^1_i, e^2_j | \rho_I | e^1_k, e^2_l} \bra{e^1_k, e^2_l}, 
\end{align}
where $\ket{e^1_i}$ and $\ket{e^2_j}$ are the basis on the intervals $I_1$ and $I_2$, respectively. 
We introduce the partial transposition $\rho_I^{T_1}$ for the interval $I_1$ which is defined by~\cite{VidalWerner}
\begin{align}
\rho_I^{T_1} := \sum_{ijkl} \ket{e^1_i, e^2_j} \braket{e^1_k, e^2_j | \rho_I | e^1_i, e^2_l} \bra{e^1_k, e^2_l}. 
\end{align}
In addition to the partial transposition, 
we also consider the ``unitary part'' of time-reversal,
and consider 
partial time-reversal transformation $\hat T'_{I_1} = \prod_{j \in I_1} (U_T)_j$
action only on $I_1$. 
Note that $\hat T'_{I_1}$ is unitary, i.e.,\ it consists of the only unitary part of the time-reversal transformation $\hat T = (\prod_j (U_T)_j) K$,
where $K$ is complex conjugation.
Putting everything together, 
we consider 
\begin{align}
{\rm tr}_I \big( \rho_I \hat T'_{I_1} \rho_I^{T_1} [\hat T'_{I_1}]^{\dag} \big), \quad \hat T'_{I_1} = \prod_{j \in I_1} (U_T)_j. 
\label{path int PR2} 
\end{align}
Finally, the Pollmann-Turner topological invariant, i.e., 
the crosscap topological invariant 
is given by the phase 
of \eqref{path int PR2}, 
\begin{align}
\theta(T) = \frac{{\rm tr}_I \big( \rho_I \hat T'_{I_1} \rho_I^{T_1} [\hat T'_{I_1}]^{\dag} \big)}{|{\rm tr}_I \big( \rho_I \hat T'_{I_1} \rho_I^{T_1} [\hat T'_{I_1}]^{\dag} \big)|} 
\to b(T,T) , \quad (N,M \to \infty). 
\label{P-T invariant}
\end{align}
In the limit $N,M \gg \xi$, where $\xi$ the correlation length, 
the phase $\theta(T)$ 
is a quantized topological invariant.

The path integral representation of the quantity 
\eqref{path int PR2}
is shown in Fig.\ \ref{Fig:Partial_Transpose} [b], 
which is topologically equivalent to a sphere with one crosscap as shown in Fig.\ \ref{Fig:TRS_to_RP2}.

That \eqref{P-T invariant} is indeed a quantized and topological invariant 
can be proven within the MPS framework. 
Here, we demonstrate this by again using the cut and glue construction~\cite{Qi2012},
and by taking the Haldane chain with time-reversal symmetry as an example.  
Within the cut and glue construction,
there are six active degrees of freedom at low energies
in the reduced density matrix, 
\begin{align}
\ket{e^R_0} \otimes \ket{e^L_1} \otimes \ket{e^R_N} \otimes \ket{e^L_{N+1}}\otimes \ket{e^R_M} \otimes \ket{e^L_{M+1}}, 
\quad 
e^L_j, e^R_j = \uparrow, \downarrow. 
\end{align}
The ground state is a singlet formed from $(\ket{e^R_0} ,\ket{e^L_1})$, $(\ket{e^R_N} ,\ket{e^L_{N+1}})$ and $(\ket{e^R_M} ,\ket{e^L_{M+1}})$ as 
\begin{equation}\begin{split}
\ket{\Psi} 
&= \frac{1}{\sqrt{2}} (\ket{\uparrow^R_0} \otimes \ket{\downarrow^L_1} - \ket{\downarrow^R_0} \otimes \ket{\uparrow^L_1} ) 
\otimes \frac{1}{\sqrt{2}} (\ket{\uparrow^R_N} \otimes \ket{\downarrow^L_{N+1}} - \ket{\downarrow^R_N} \otimes \ket{\uparrow^L_{N+1}} ) \\
&\qquad \otimes \frac{1}{\sqrt{2}} (\ket{\uparrow^R_M} \otimes \ket{\downarrow^L_{M+1}} - \ket{\downarrow^R_M} \otimes \ket{\uparrow^L_{M+1}} ). 
\end{split}\end{equation}
The reduced density matrix $\rho_I$ reads
\begin{align}
\rho_I 
&= {\rm tr}_{0,{M+1}} (\ket{\Psi} \bra{\Psi}) 
\nonumber \\
&= \frac{1}{2} {\rm Id}^L_1 
\otimes \frac{1}{2} \Big[
\ket{\uparrow^R_N} \bra{\uparrow^R_N} \otimes \ket{\downarrow^L_{N+1}} \bra{\downarrow^L_{N+1}} 
+\ket{\downarrow^R_N} \bra{\downarrow^R_N} \otimes \ket{\uparrow^L_{N+1}} \bra{\uparrow^L_{N+1}} 
\nonumber \\
&\ \ \ \ \ \ \ \ \ \ \ -\ket{\uparrow^R_N} \bra{\downarrow^R_N} \otimes \ket{\downarrow^L_{N+1}} \bra{\uparrow^L_{N+1}} 
-\ket{\downarrow^R_N} \bra{\uparrow^R_N} \otimes \ket{\uparrow^L_{N+1}} \bra{\downarrow^L_{N+1}}  
\Big] 
\otimes \frac{1}{2} {\rm Id}^R_M. 
\end{align}
By taking the partial transposition on $I_1 = \{1,N\}$ and 
noting that the unitary part of the time-reversal transformation is given by 
\begin{align}
U_T \ket{\uparrow} = \ket{\downarrow}, 
\quad 
U_T \ket{\downarrow} = -\ket{\uparrow}, 
\end{align}
we have 
\begin{align}
\hat T'_{I_1} \rho^{T_1}_I [\hat T'_{I_1}]^{\dag}
&= \frac{1}{2} {\rm Id}^L_1
\otimes \frac{1}{2} \Big[
\ket{\downarrow^R_N} \bra{\downarrow^R_N} \otimes \ket{\downarrow^L_{N+1}} \bra{\downarrow^L_{N+1}} 
+\ket{\uparrow^R_N} \bra{\uparrow^R_N} \otimes \ket{\uparrow^L_{N+1}} \bra{\uparrow^L_{N+1}}
\nonumber \\
&\ \ \ +\ket{\uparrow^R_N} \bra{\downarrow^R_N} \otimes \ket{\downarrow^L_{N+1}} \bra{\uparrow^L_{N+1}} 
+\ket{\downarrow^R_N} \bra{\uparrow^R_N} \otimes \ket{\uparrow^L_{N+1}} \bra{\downarrow^L_{N+1}}  
\Big] 
\otimes \frac{1}{2} {\rm Id}^R_M , 
\end{align}
which leads to 
\begin{align}
{\rm tr}_I(\rho_I \hat T'_{I_1} \rho^{T_1}_I [\hat T'_{I_1}]^{\dag})
= - \frac{1}{8}. 
\end{align}
The minus sign $(-1)$ is the proper $\Z_2$ invariant for the Haldane chain with time-reversal symmetry.

\paragraph{Klein bottle partition function from inversion symmetry}

Similar to the crosscap topological invariant,
the Klein bottle topological invariant can be
defined both for spatial inversion and 
time-reversal.
Let us start with the case of spatial inversion.  

We act with an inversion transformation $P \notin G_0$ on the twisted MPS $\ket{\Psi_g}$, 
\begin{align}
\hat P \ket{\Psi_g}
= \sum_{ \{m_i\} } {\rm Tr}\, (A_{m_1} \cdots A_{m_L} V_P V^T_{g} V_P^{-1}) \ket{m_1\cdots  m_L} 
= \frac{b(P,g^{-1}) b(g,g^{-1})}{b(P g^{-1} P^{-1},P)} \ket{\Psi_{P g^{-1} P^{-1}}}. 
\end{align}
If $P g^{-1} P^{-1} = g$, the $U(1)$ prefactor is well-defined which is invariant under the 1-coboundary. 
This invariant is nothing but the partition function $Z_{KB}(P;g)$ over the Klein bottle ($KB$) with the background $P$ and $g$ twists 
\begin{align}
\kappa(P;g) = Z_{KB}(P;g) = \braket{\Psi_g | \hat P | \Psi_g}, 
\quad (P \notin G_0, \ g \in G_0, \ P g^{-1}= g P).  
\end{align}

An example is 
a $\Z_2 (= \{1,\sigma\})$ paramagnet with inversion symmetry $\Z_2^P (= \{1,P\})$
where $\Z_2$ charge is preserved under the inversion. 
The topological classification is given by $H^2(\Z_2 \times \Z_2^P,U(1)_{\phi}) = \Z_2 \times \Z_2$. 
The two topological invariants can be seen in $V_P V_P^* = \theta(P)$ and $V_P V^T_{\sigma} = \kappa(P;\sigma) V_{\sigma} V_P$.

\paragraph{Klein bottle partition function from time-reversal symmetry: ``disjoint partial transposition with intermediate twist''}
\begin{figure}[!]
 \begin{center}
  \includegraphics[width=0.9\linewidth, trim=0cm 0cm 0cm 0cm]{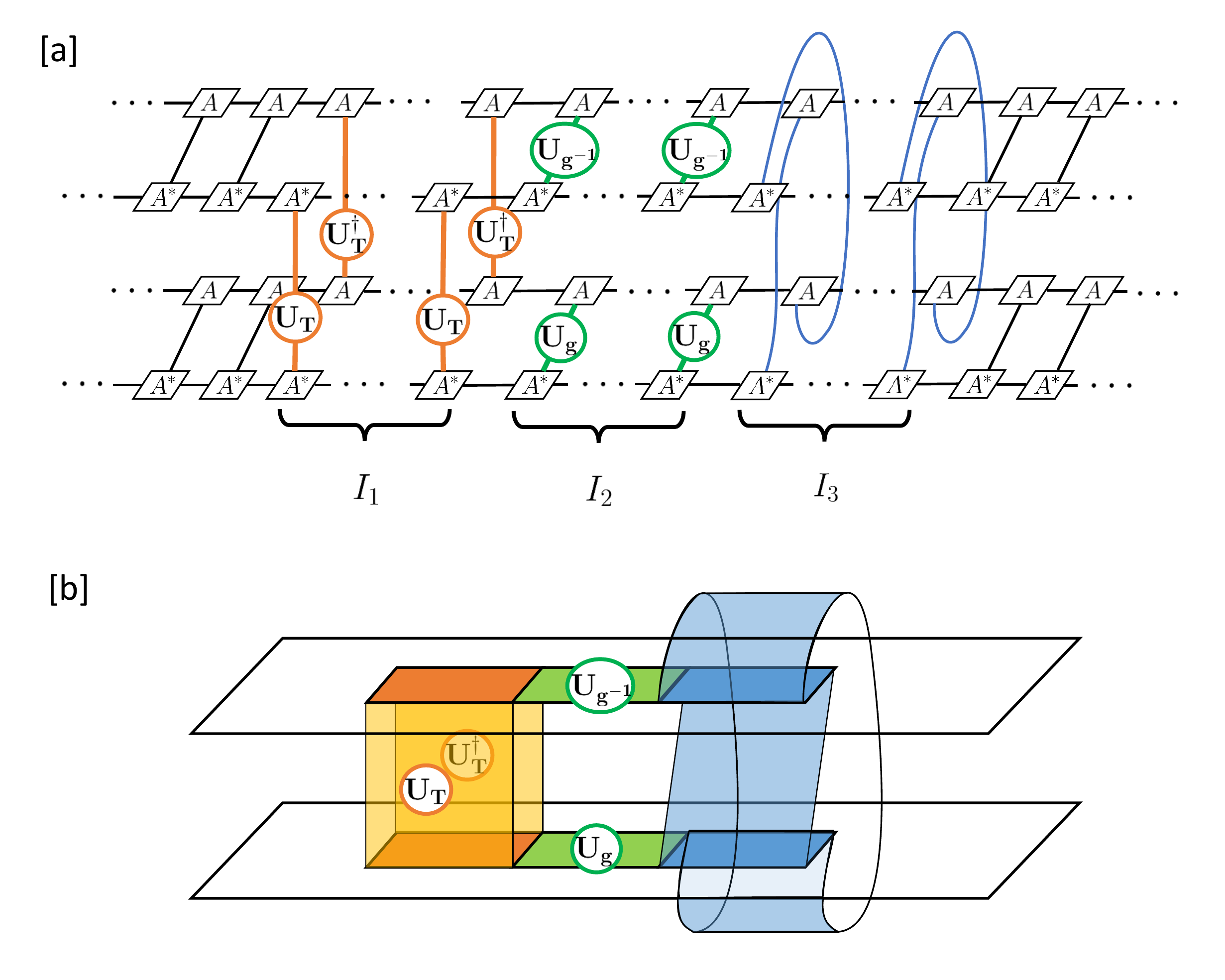}
 \end{center}
 \caption{[a] MPS representation of the Klein bottle partition function with twist. [b] The path integral representation. }
 \label{Fig:Partial_Transpose_Klein}
\end{figure}

Next, as for 
time-reversal symmetry, 
the Klein bottle partition function 
obtained from from time-reversal 
can be represented in terms of MPSs in a way similar to the crosscap partition function. 
First, we divide the closed chain $L$ into three adjacent intervals $I_1 \cup I_2 \cup I_3$, 
$I_1 = \{1, \dots, N_1\}$, $I_2 = \{N_1+1, \dots, N_2\}$, $I_3 = \{N_2+1, \dots, N_3\}$. 
In addition, we introduce one replica. 
We trace out the region except for $I_1 \cup I_3$ {\it with symmetry twist in the interval $I_2$} as 
\begin{align}
\rho_{I_1 \cup I_3}(g):= {\rm tr}_{L \backslash (I_1 \cup I_3)} \Big( \hat g_{I_2} \ket{\Psi} \bra{\Psi} \Big), 
\quad 
\hat g_{I_2} = \prod_{j \in I_2} (U_g)_{j}. 
\end{align}
Then, we consider the following quantity 
\begin{align}
{\rm tr}_{I_1 \cup I_3} \Big( \rho_{I_1 \cup I_3}(g) \hat T'_{I_1} \rho_{I_1 \cup I_3}(g^{-1}) [\hat T'_{I_1}]^{\dag} \Big), 
\quad (T g^{-1} T^{-1} = g), 
\quad
T'_{I_1} = \prod_{j \in I_1} (U_T)_j.
\end{align}
The corresponding MPS network and path integral are shown in Fig.\ \ref{Fig:Partial_Transpose_Klein} [a] and [b], respectively. 
One can show this quantity approaches the Klein bottle partition function 
\begin{align}
\frac{{\rm tr}_{I_1 \cup I_3} \Big( \rho_{I_1 \cup I_3}(g) \hat T'_{I_1} \rho_{I_1 \cup I_3}(g^{-1}) [\hat T'_{I_1}]^{\dag} \Big)}
{\Big| {\rm tr}_{I_1 \cup I_3} \Big( \rho_{I_1 \cup I_3}(g) \hat T'_{I_1} \rho_{I_1 \cup I_3}(g^{-1}) [\hat T'_{I_1}]^{\dag} \Big) \Big|}
\to Z_{KB}(T;g) = \kappa(T;g) 
\end{align}
in the limit $N_1, N_2-N_1, N_3-N_2, L-N_3 \gg \xi$, where $\xi$ is the correlation length. 
It is easy to show the above formula 
for fixed point MPSs
by using the symmetry properties of $A$ matrix.

\section{$G$-equivariant topological field theories and MPSs}
\label{$G$-equivariant Topological field theory}

Having discussed the MPS description
of (1+1)d bosonic SPT phases, 
we now move on (1+1)d $G$-equivaliant TFTs.
In the following sections, 
Sec.\ \ref{Some basics of TFTs} to Sec.\ \ref{G-equivariant open and closed TFTs}, 
we briefly summarize necessary ingredients of  
open and closed $G$-equivariant oriented $(1+1)$d TFTs
following Moore-Segal\cite{Moore-Segal}. 
There are some overlaps with Ref.\ \cite{Kapustin-Turzillo}, 
where they also discuss closed $G$-equivariant $(1+1)$d unoriented TFTs. 
Ref.\ \cite{Moore-Segal} also discusses 
$(1+1)$d open and closed TFTs with spin structure, 
which can describe fermionic SPT phases such as 
class D topological superconductors. 
Here we restrict ourselves to $(1+1)$d bosonic SPT phase protected by on-site unitary $G$-symmetry where $G$ is a finite group. 
In short, 
a $G$-equivariant TFT is a TFT couped with the background $G$-gauge field. 
(Integrating out the background $G$-gauge field, i.e. orbifolding the $G$-symmetry, 
leads to an orbifolded theory which is a TFT without $G$-symmetry.) 



In the following, first, we introduce some general properties of TFTs. 
Next, we summarize 
$(1+1)$d $G$-equivariant closed TFTs 
with an eye toward $(1+1)$d SPT phases. 
Our notations closely follow Moore-Segal\cite{Moore-Segal}. 
Next, we will summarize $(1+1)$d $G$-equivariant open and closed TFTs.

\subsection{Some basics of TFTs}
\label{Some basics of TFTs}

\begin{figure}[!]
 \begin{center}
  \includegraphics[width=0.7\linewidth, trim=0cm 0cm 0cm 0cm]{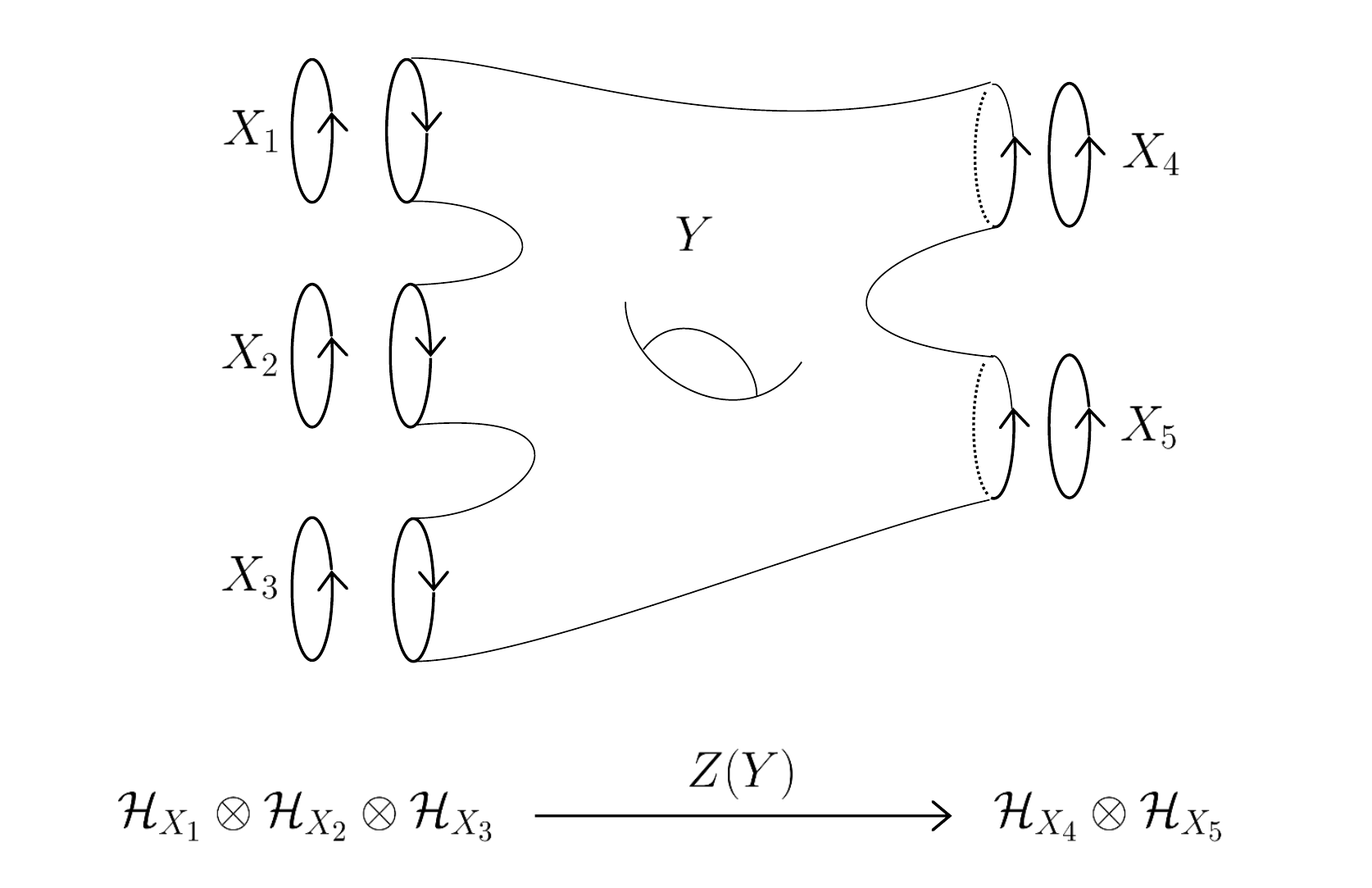}
 \end{center}
 \caption{An example of cobordism.}
 \label{fig:tft_ex}
\end{figure}

In the axiomatic definition,
a TFT in $(d+1)$ dimensions 
is a functor $Z$ from a cobordism category 
$\mathbf{Bord}_{<d,d+1>}$
to the category of finite dimensional complex vector spaces $\mathbf{Vect}$ 
equipped with tensor product.~\cite{Atiyah:1989vu, Quinn}
In $\mathbf{Bord}_{<d,d+1>}$, 
objects are $d$-dimensional manifolds $X_1,X_2,\dots $, 
and morphism is a cobordism $Y: X_1 \to X_2$ 
which is a manifold of dimension $d+1$ 
and has $X_1$ and $X_2$ as its boundary components, 
$\partial Y = (-X_1) \sqcup X_2$, 
where $(-X)$ is $X$ with opposite orientation. 
In general, we can associate a structure 
(e.g. spin structure for spin TFTs, background gauge field for equivariant TFTs, etc.) 
with manifolds. 
For each $d$-dimensional manifold $X$, we associate a 
Hilbert space ${\cal H}_X$ by a functor $Z$. 
A direct sum of manifolds $X_1 \sqcup X_2 \sqcup \cdots $ is 
mapped into a tensor product ${\cal H}_{X_1} \otimes {\cal H}_{X_2} \otimes \cdots$. 
A cobordism $Y$ between $X$ and $X'$ leads to a linear map 
$Z(Y) : {\cal H}_{X} \to {\cal H}_{X'}$. 
See Fig.~ \ref{fig:tft_ex}, for an example.

In any TFT, the cylinder cobordism $X \times I$ leads 
to the identity map $Z(X \times I) = {\rm id}$
\begin{align}
&\vcenter{\hbox{\includegraphics[width=0.2\linewidth]{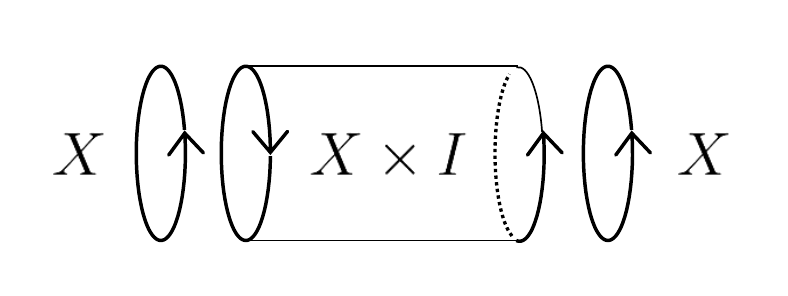}}} 
\ \Longrightarrow \ \ 
{\cal H}_X \overset{\rm id}{\longrightarrow} {\cal H}_X, 
\end{align}
which is equivalent to the fact that the Hamiltonian of TFTs is zero. 

In addition, we have a bilinear form $Q$ and a coform $\Delta$: 
\begin{align}
&\vcenter{\hbox{\includegraphics[width=0.15\linewidth]{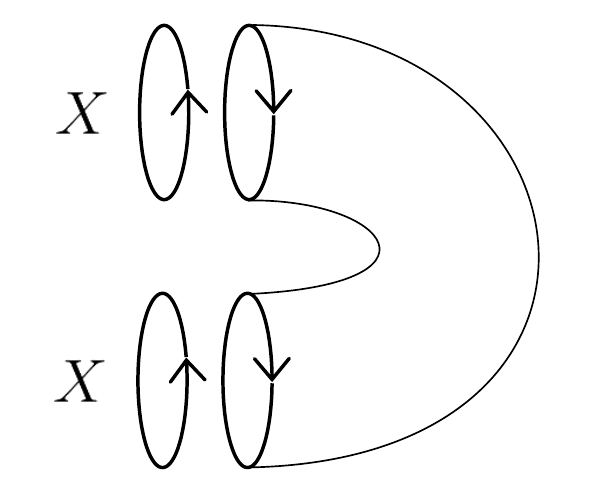}}} && 
\vcenter{\hbox{\includegraphics[width=0.15\linewidth]{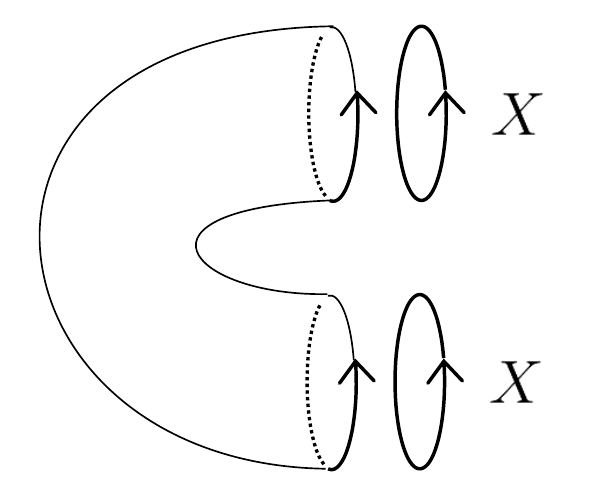}}} \notag \\
&{\cal H}_X \otimes {\cal H}_X \overset{Q}{\longrightarrow} \C, &&
\C \overset{\Delta}{\longrightarrow} {\cal H}_X \otimes {\cal H}_X. 
\end{align}
Let $\{\phi_i \}$ be a basis of ${\cal H}_X$ and write 
$Q(\phi_i, \phi_j) = Q_{ij}, \Delta(1) = \sum_{ij} \Delta_{ij} \phi_i \otimes \phi_j$. 
The equivalence between the ``S-tube'' and the cylinder, 
$$
\includegraphics[width=0.7\linewidth]{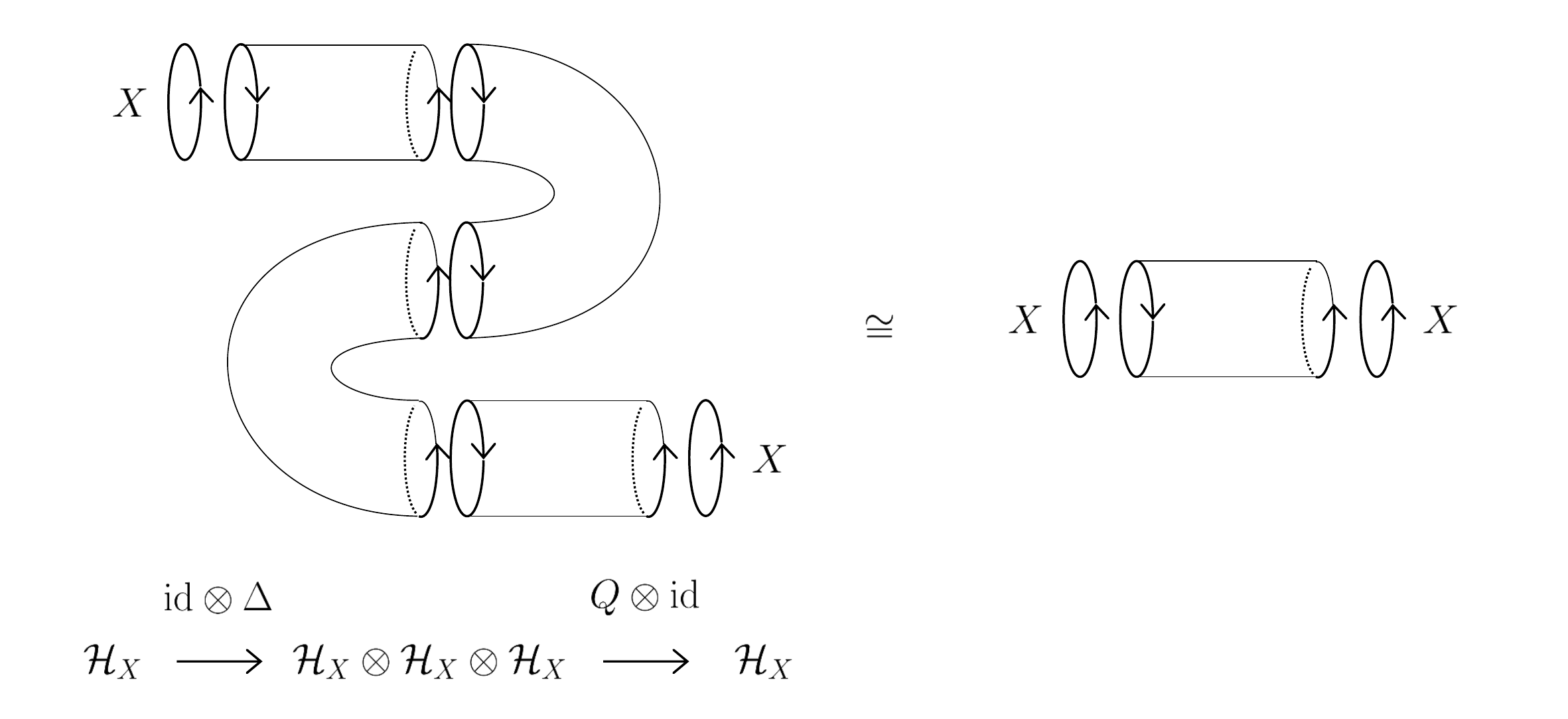}
$$
implies 
\begin{align}
\phi \mapsto \phi \otimes \sum_{jk} \Delta_{jk} \phi_j \otimes \phi_k \mapsto \sum_i Q(\phi, \phi_j) \Delta_{jk} \phi_k = \phi , && \phi \in {\cal H}_X. 
\label{eq:S-condition}
\end{align}
By setting $\phi = \phi_k$, we have $\sum_{j} Q_{ij} \Delta_{jk} = \delta_{ik}$, 
which means $Q$ is nondegenerate. 
Choosing the basis so that $Q_{ij} = \delta_{ij}$, 
in this basis the coform $\Delta$ is simply $\Delta = \sum_{i} \phi_i \otimes \phi_i$.

\subsection{$G$-equivariant oriented closed TFTs}

A $G_0$-equivariant oriented $(1+1)$d TFT 
is a functor $Z$ from 
a cobordism category with a background $G_0$ gauge field
to the category of complex vector spaces. 
(To distinguish on-site unitary symmetries from orientation-reversing symmetries, 
here we use a notation $G_0$ to denote on-site unitary symmetries.) 
For $(1+1)$d TFTs, the minimum object is an oriented circle 
$(S^1,pt,g)$ with background $g \in G_0$ flux 
together with a trivialization at a base point $pt \in S^1$, 
which is specified by a twisted boundary condition by an element $g \in G_0$ at $pt$. 
We denote the Hilbert space associated with $(S^1, pt, g \in G_0)$ by ${\cal C}_g$:
\begin{align}
&\vcenter{\hbox{\includegraphics[width=0.3\linewidth]{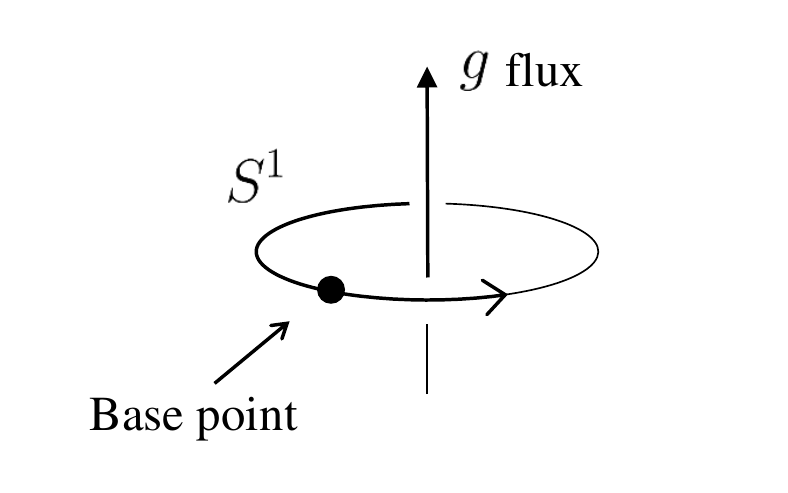}}} 
\ \Longrightarrow \ \ \ \ 
{\cal C}_g = {\cal H}_{(S^1,pt,g)}, \ (g \in G_0).
\end{align}
For an unpointed circle $S^1$, 
a background flux inserted in the circle $S^1$ is characterized by 
a conjugacy class $[g] = \{ h g h^{-1} | h \in G\}$ 
rather than an element $g \in G_0$. 
For point circles $(S^1,pt)$ with trivialization of background $G_0$ gauge field at $pt$, 
Hilbert spaces are labeled by elements $g \in G_0$. 
We have thus a $G_0$-graded Hilbert space 
\begin{align}
{\cal C} = \bigoplus_{g \in G_0} {\cal C}_g.
\end{align}

\begin{figure}[!]
 \begin{center}
  \includegraphics[width=0.8\linewidth, trim=0cm 0cm 0cm 0cm]{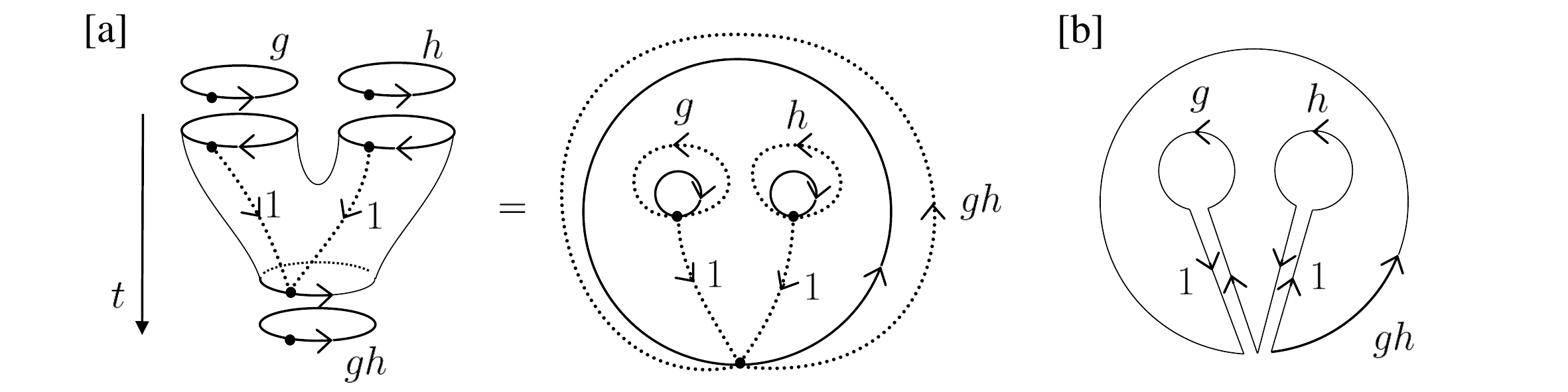}
 \end{center}
 \caption{
 [a] A fusion process ${\cal C}_g \otimes {\cal C}_h \to {\cal C}_{gh}$. 
 Dashed lines with group elements represent holonomies. 
 The figure [b] shows a holonomy around the boundary of simply connected space 
 that is obtained by cutting the surface in [a] at the lines connecting base points. 
 }
 \label{fig:tft_fusion_detail}
\end{figure}

\begin{table}[!]
\caption{Building blocks of $G$-equivariant closed $(1+1)$d TFTs. 
Building blocks (a)-(f) define $G$-equivariant oriented TFTs in $(1+1)$d.
$G$-equivariant unoriented TFTs in $(1+1)$d
are defined by including (g) and (h), in addition to (a)-(f). 
$G_0 \subset G$ represents orientation preserving symmetries. 
The fourth column shows corresponding simple and fixed point MPS representations (see Sec.~\ref{sec:tft-mps}). }
\begin{center}
\scalebox{0.8}{
\hspace*{-2cm}
\begin{tabular}{| >{\centering\arraybackslash}m{1cm} | >{\centering\arraybackslash}m{3cm} | >{\centering\arraybackslash}m{3.5cm} | >{\centering\arraybackslash}m{5cm} | >{\centering\arraybackslash}m{5cm} | }
\hline
& Manifolds & Hilbert spaces & Simple and fixed point MPS & Comment \\
\hline
(a) & \includegraphics[width=0.5\linewidth, trim=0cm 0cm 0cm 0cm]{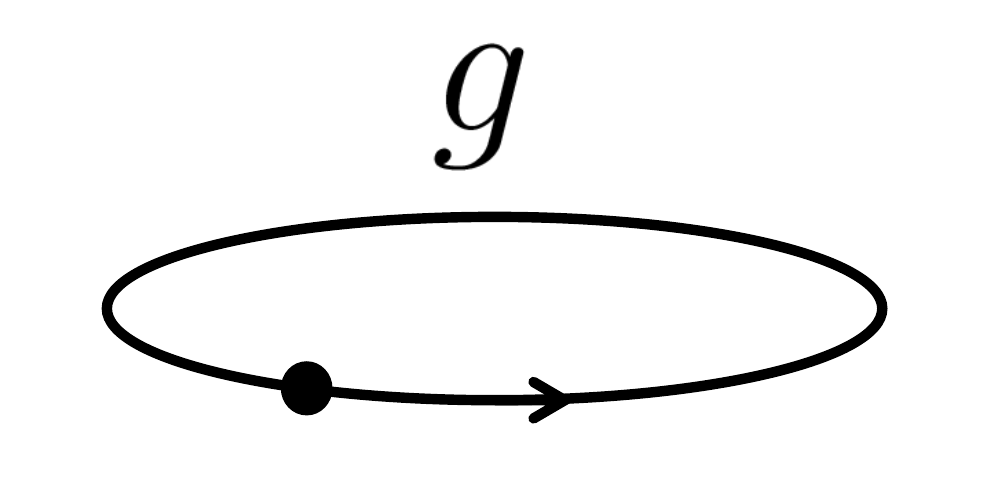} & 
${\cal C}_g \ (g \in G_0)$ & Hilbert space generated by $\ket{\Psi_g} = {\rm tr}(A_m V_g) \ket{m}, g \in G_0$ & Hilbert space over a space circle with $g$-flux \\
\hline 
(b) & \includegraphics[width=0.5\linewidth, trim=0cm 0cm 0cm 0cm]{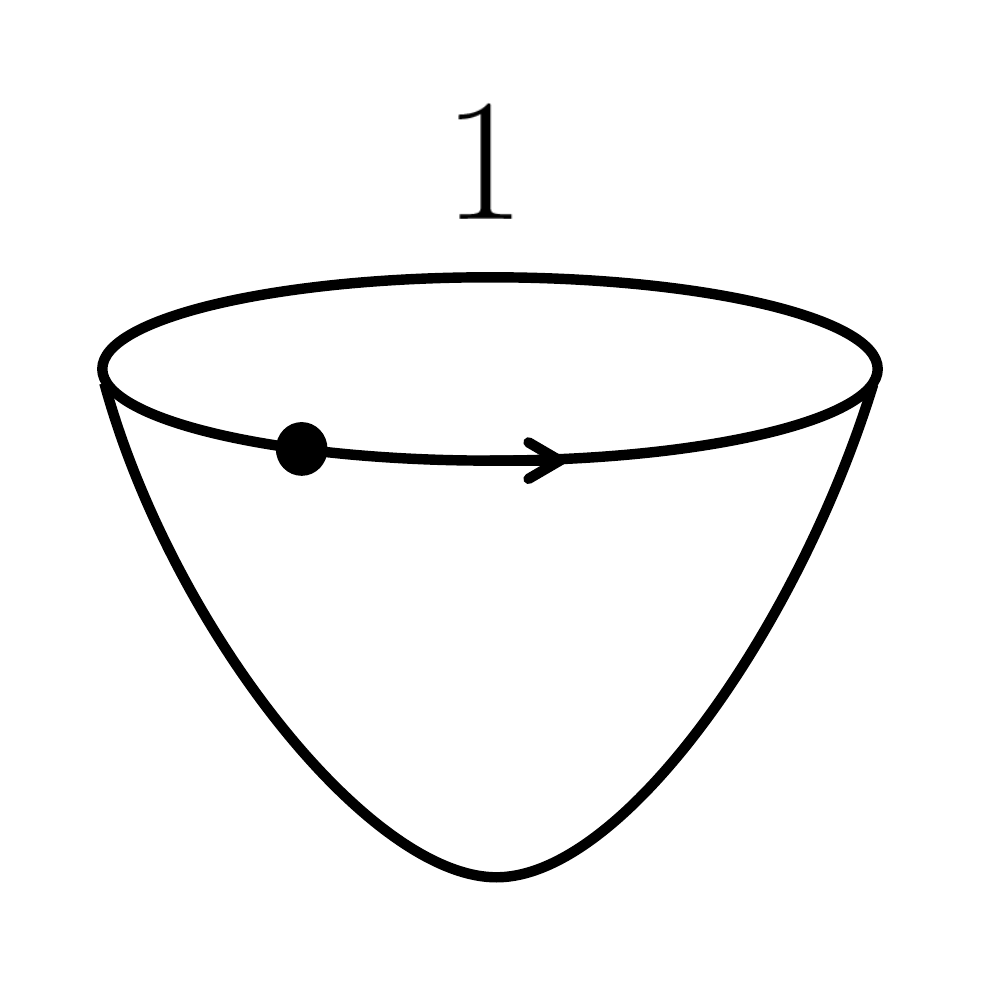} & 
$$
\begin{array}{l}
\theta_{\cal C}: {\cal C}_1 \to \C, \\
\phi \mapsto \theta_{\cal C}(\phi) \\
\end{array}
$$
& $\theta_{\cal C} \big( {\rm tr} (A_m) \ket{m} \big) = 1$ & 
$Q(\phi_1, \phi_2) := \theta_{\cal C}(\phi_1 \phi_2), (\phi_1 \in {\cal C}_g , \phi_2 \in {\cal C}_{g^{-1}})$ is a bilinear nondegenerate form. \\
\hline
(c) & \includegraphics[width=0.5\linewidth, trim=0cm 0cm 0cm 0cm]{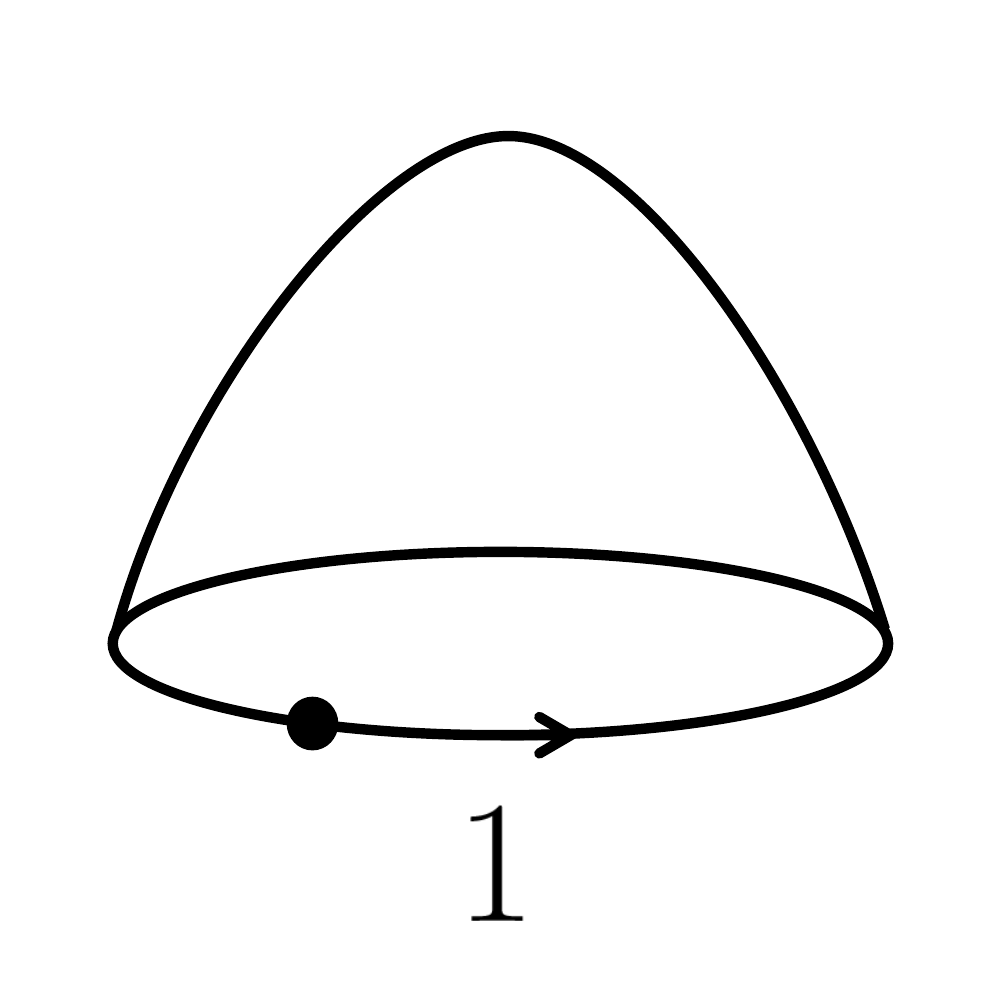} & 
$$
\begin{array}{l}
\C \to {\cal C}_1, \\
1 \mapsto 1_{\cal C} \\
\end{array}
$$
& $1_{\cal C} = {\rm tr} (A_m) \ket{m}$ & State on the boundary of disc
$1_{\cal C} \phi = \phi 1_{\cal C} = \phi$. \\
\hline
(d) & \includegraphics[width=0.7\linewidth, trim=0cm 0cm 0cm 0cm]{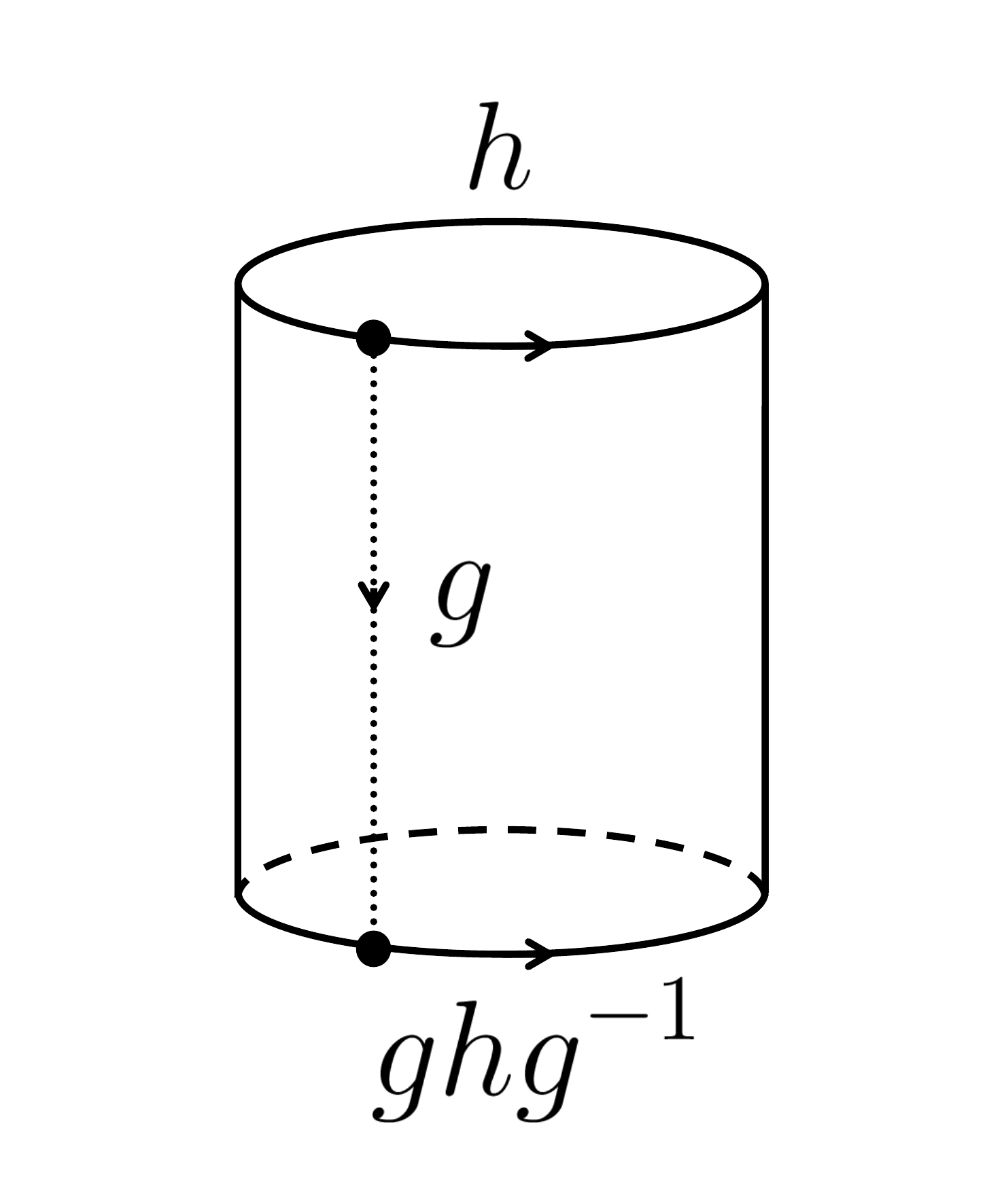} & 
$$
\begin{array}{ll}
\alpha_{g\in G_0} : & {\cal C}_h \to{\cal C}_{ghg^{-1}}, \\
& \phi \mapsto \alpha_g(\phi) \\
\end{array}
$$
& 
$$
\begin{array}{l}
{\rm tr} (A_m V_h) \ket{m} \\
\mapsto {\rm tr} (A_m V_h) \ \hat g (\ket{m}) \\
= {\rm tr} (A_m V_g V_h V_g^{\dag}) \ket{m} \\
\end{array}
$$
& 
On site unitary $g \in G_0$ symmetry action \\
\hline
(e) & \includegraphics[width=\linewidth, trim=0cm 0cm 0cm 0cm]{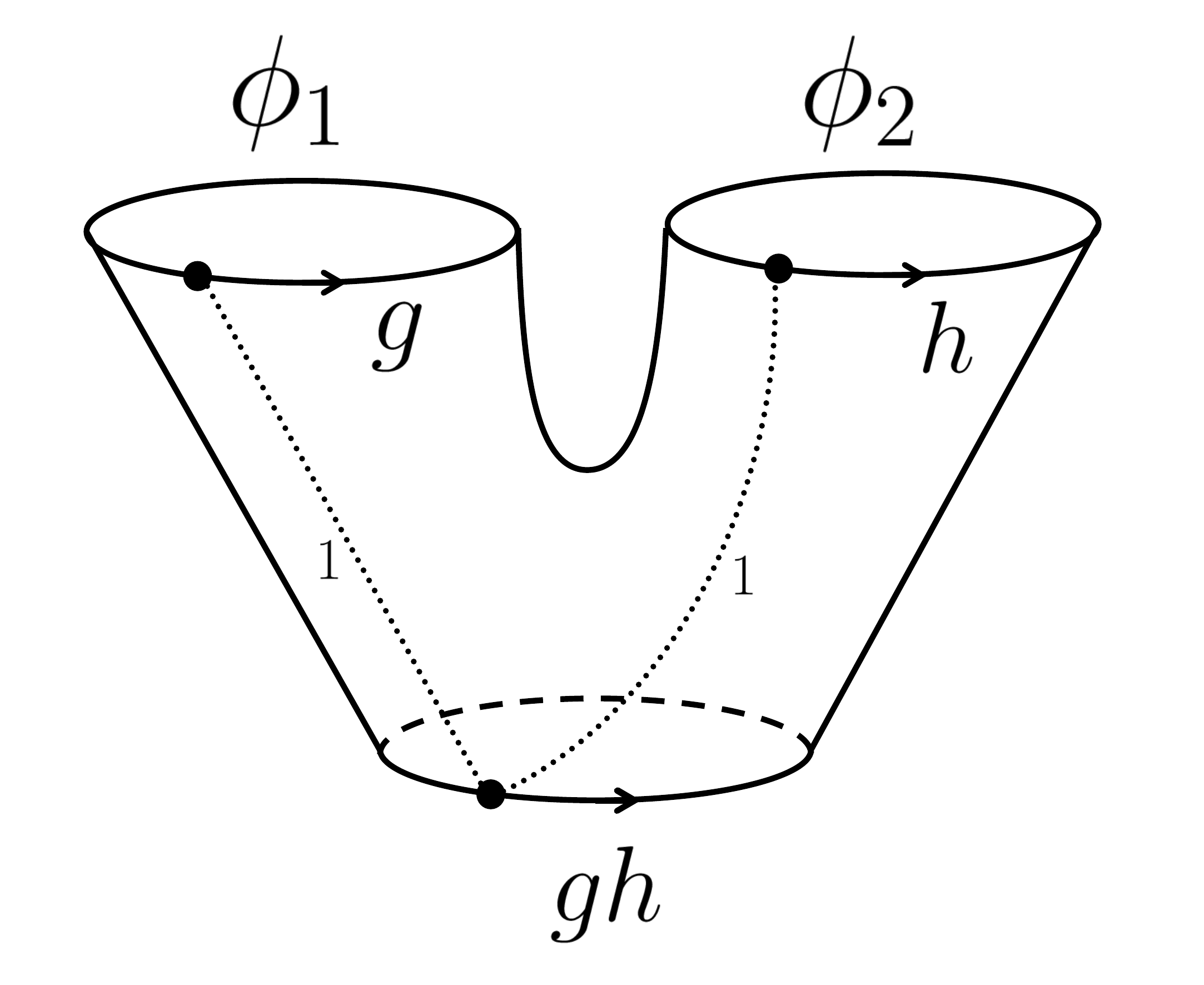} & 
$$
\begin{array}{l}
{\cal C}_g \otimes {\cal C}_h \to {\cal C}_{gh}, \\
\phi_1 \otimes \phi_2 \mapsto \phi_1 \phi_2 \\
\end{array}
$$
& 
$$
\begin{array}{l}
{\rm tr} (A_{m_1} V_g) \ket{m_1} \\
\otimes {\rm tr} (A_{m_2} V_h) \ket{m_2} \\
\mapsto {\rm tr} (A_m V_g V_h) \ket{m} \\
\end{array}
$$
& ``Fusion'' of two closed chains \\
\hline
(f) & \includegraphics[width=\linewidth, trim=0cm 0cm 0cm 0cm]{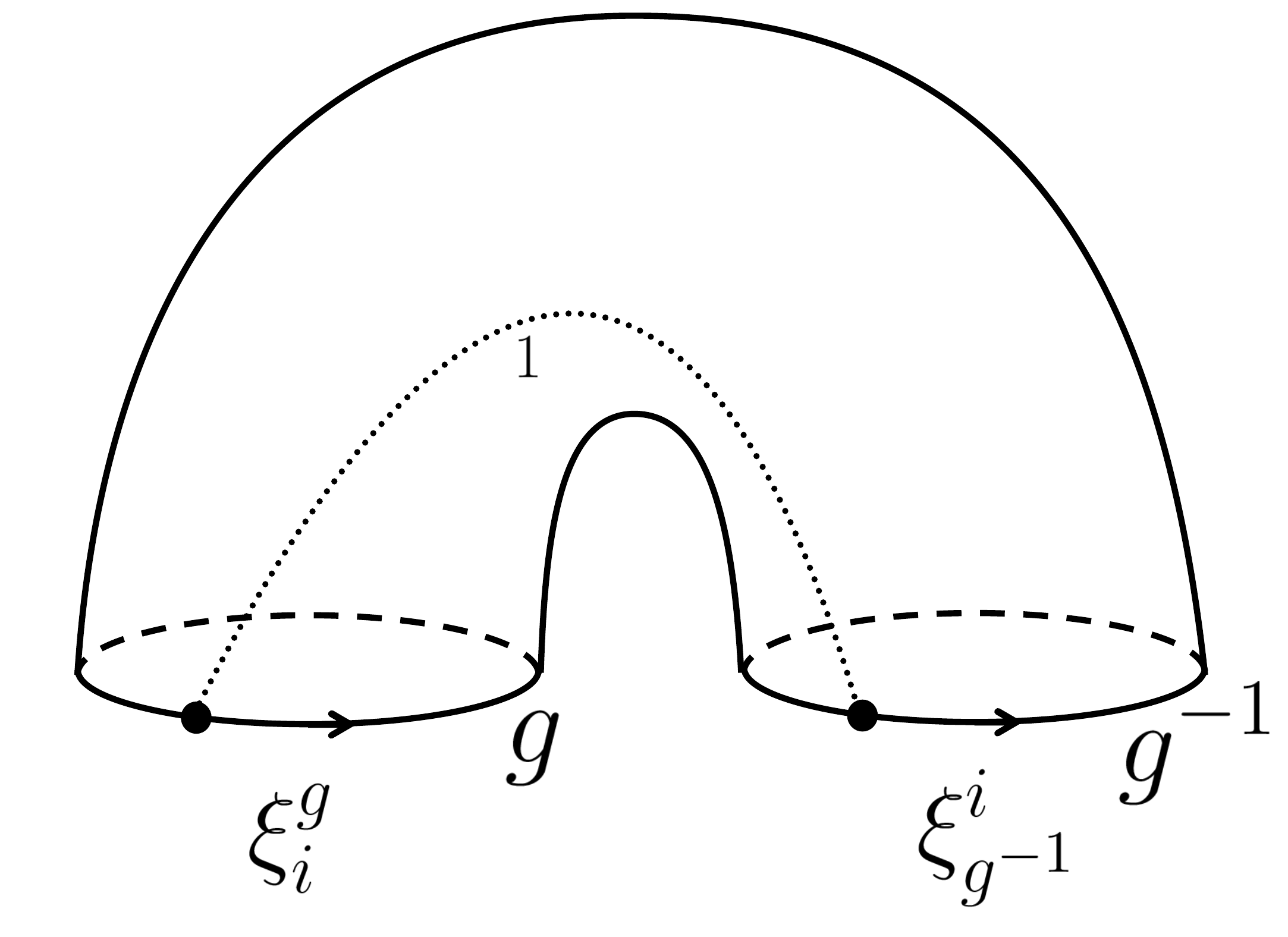} & 
$$
\begin{array}{l}
\Delta_g : \C \to {\cal C}_g \otimes {\cal C}_{g^{-1}}, \\
\Delta_g(1) = \sum_i \xi^g_i \otimes \xi_i^{g^{-1}} \\
\end{array}
$$
& ${\rm tr} (A_{m_1} V_g) \ket{m_1} \otimes {\rm tr} (A_{m_2} V_{g^{-1}}) \ket{m_2}$ & 
$\xi^g_{i} \in {\cal C}_{g}$ are basis of ${\cal C}_{g}$ and 
$\xi_{i}^{g^{-1}} \in {\cal C}_{g^{-1}}$ 
are their dual basis of ${\cal C}_{g^{-1}}$ that satisfy 
$\theta_{\cal C} (\xi^g_{i} \xi_{j}^{g^{-1}}) = \theta_{\cal C} (\xi_{j}^{g^{-1}} \xi^g_{i}) = \delta_{i j}$.  \\
\hline
\hline
(g) & \includegraphics[width=0.6\linewidth, trim=0cm 0cm 0cm 0cm]{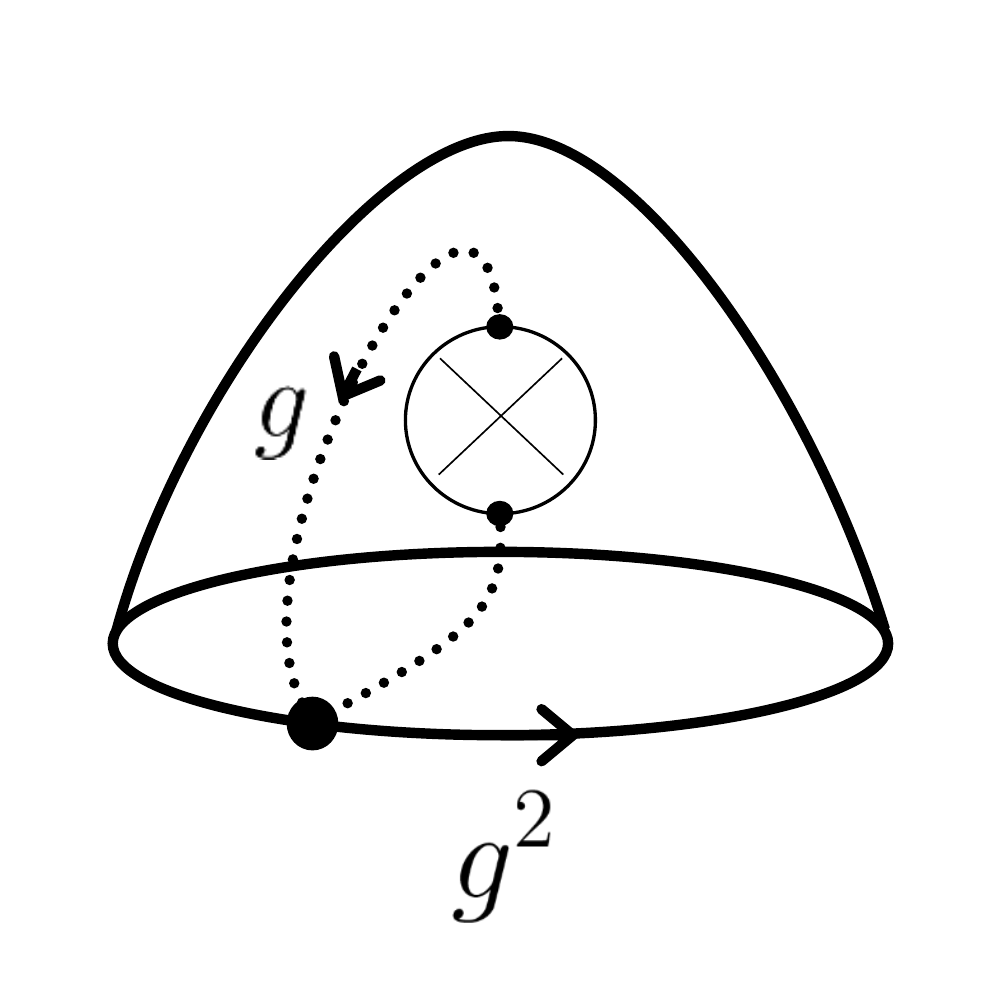} & 
$$
\begin{array}{l}
\C \to {\cal C}_{g^2}, g \notin G_0, \\
1 \mapsto \theta_{g} \\
\end{array}
$$
& $b(g,g) {\rm tr} (A_m V_{g^2}) \ket{m}$ & 
State on the boundary state of M\"{o}bius strip, ``cross cap state''. \\
\hline
(h) & \includegraphics[width=0.7\linewidth, trim=0cm 0cm 0cm 0cm]{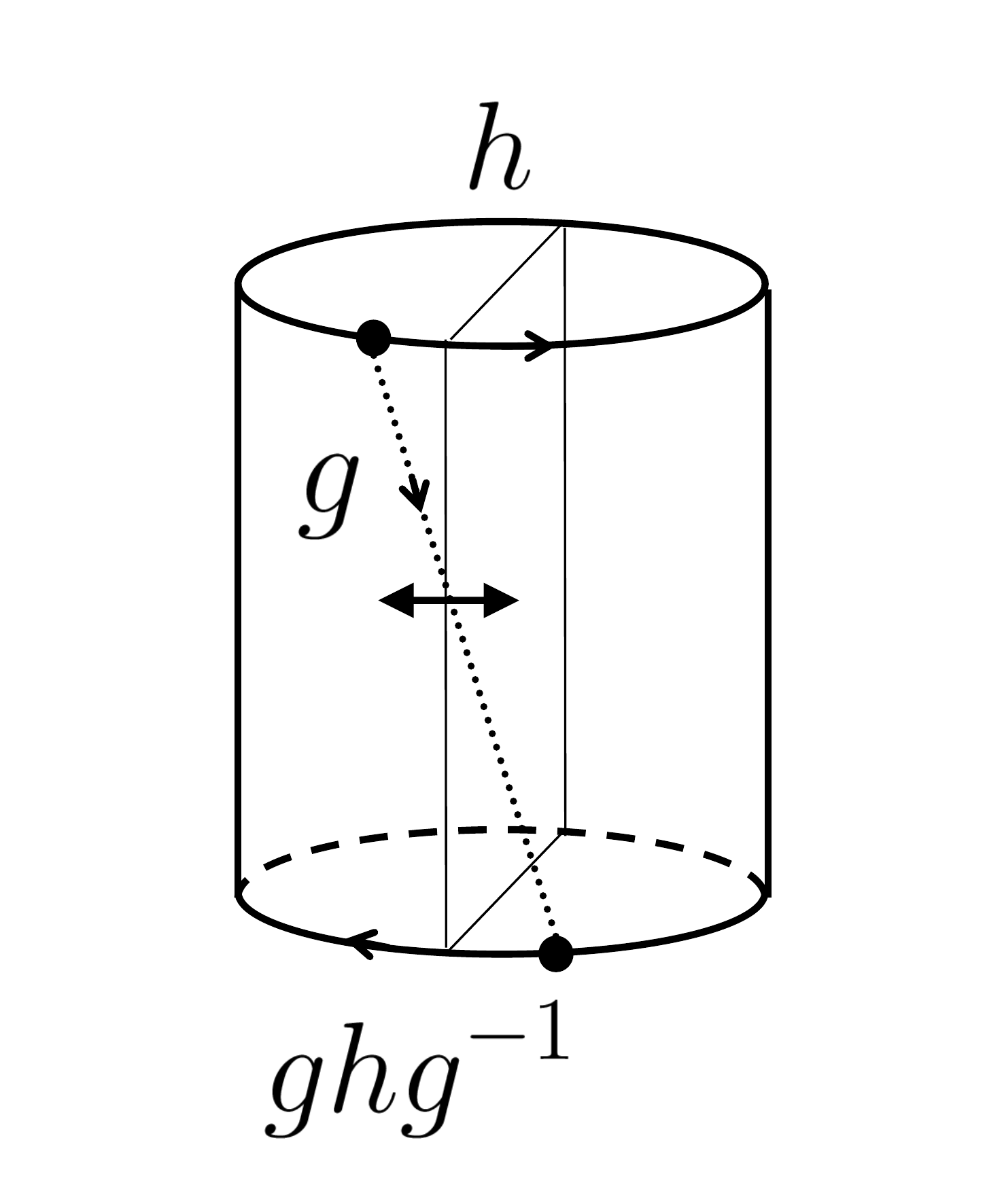} & 
$\alpha_{g\notin G_0} : {\cal C}_h \to{\cal C}_{gh^{-1}g^{-1}}$
& 
$$
\begin{array}{l}
{\rm tr} (A_m V_h) \ket{m} \\
\mapsto {\rm tr} (A_m V_h) \ \hat g (\ket{m}) \\
= {\rm tr} (A^T_m V_g V_h V_g^{\dag}) \ket{m} \\
\end{array}
$$
& 
$g$ reflection \\
\hline
\end{tabular}
\hspace*{-2cm}
}
\end{center}
\label{tab:functor_ori_tft}
\end{table}

In graphical representations of morphisms, 
we specify the background gauge field by 
holonomies connecting base points on initial and mapped circles. 
For example, the fusion process of two circles with $g$ and $h$ fluxes 
is represented in Fig.\ \ref{fig:tft_fusion_detail} [a]. 
The $G_0$ flux of mapped pointed circle $(S^1,pt)$ 
is determined by holonomies along base points as 
shown in Fig.~\ref{fig:tft_fusion_detail} [b]. 
Recall that a holonomy around a boundary of simply connected spaces is trivial. 
In short, we simply write the bordsim of the fusion process as 
\begin{align}
&\vcenter{\hbox{\includegraphics[width=0.4\linewidth]{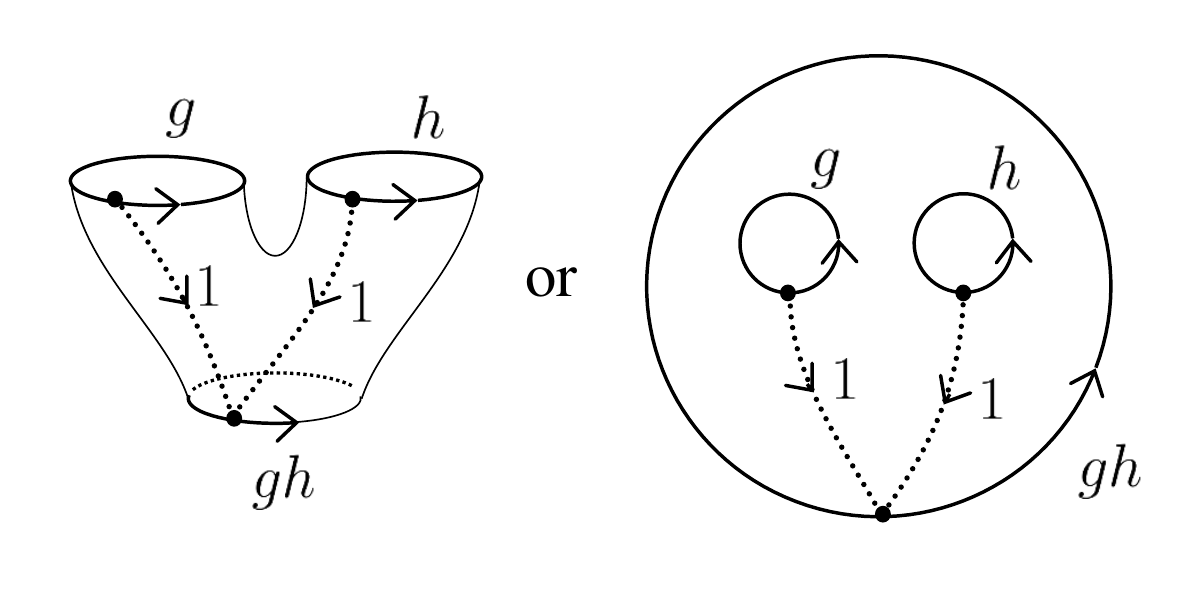}}} 
\ \Longrightarrow \ \ \ \ 
{\cal C}_g \otimes {\cal C}_h \to {\cal C}_{gh}, \quad (g,h \in G_0). 
\end{align}

In Table \ref{tab:functor_ori_tft},
we show building blocks of $G_0$-equivariant oriented (1+1)d TFTs. 
All other cobordisms and partition functions can be constructed by processes in Table \ref{tab:functor_ori_tft}. 
For example, 
the 
``branching'' process 
is given by
\begin{align}
&\vcenter{\hbox{\includegraphics[width=0.8\linewidth]{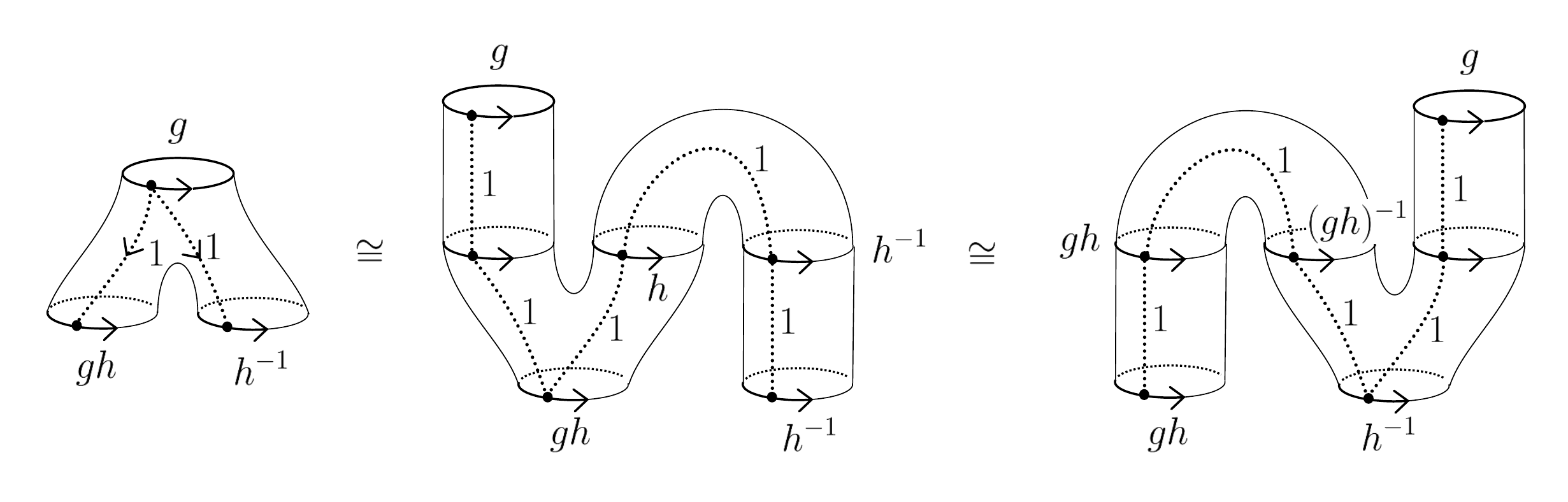}}}
\label{fig:tft/separate}
\end{align}
\begin{align}
\Longrightarrow Z \Big( \vcenter{\hbox{\includegraphics[width=0.15\linewidth]{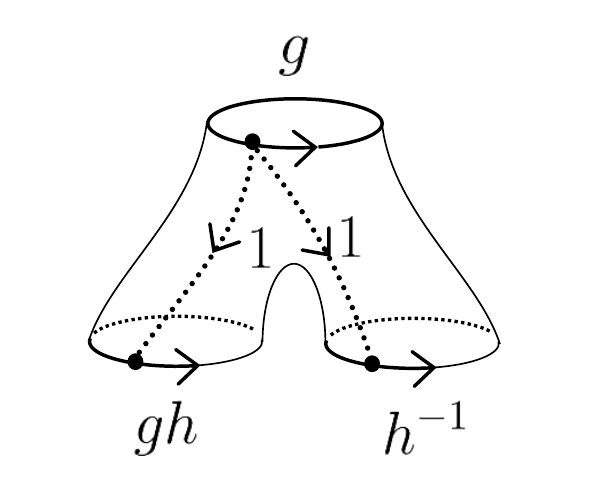}}} \Big): 
{\cal C}_g \to {\cal C}_{gh} \otimes {\cal C}_{h^{-1}}, && 
\phi \mapsto \sum_i \phi \xi_i^h \otimes \xi_i^{h^{-1}}
= \sum_i \xi_i^{gh} \otimes \xi_i^{(gh)^{-1}} \phi, && 
(g,h \in G_0). 
\end{align}
Here, 
we made use of Items (e) and (f) in Table\ref{tab:functor_ori_tft},
and 
$\Delta_g(1) = \sum_i \xi_i^g \otimes \xi_i^{g^{-1}} \in {\cal C}_g \otimes {\cal C}_{g^{-1}}$ 
is the coform defined in Item (f). 

The fusion process (e), 
which, by an axiom of TFTs, 
is associated to a map $\mathcal{C}_g \otimes \mathcal{C}_h \to \mathcal{C}_{gh}$,
makes the Hilbert space $\mathcal{C}$ into an algebra.  
There are several constraints on the algebra,
which are obtained, e.g., 
by considering different factorizations
of surfaces into building blocks in Table \ref{tab:functor_ori_tft}.
Due to Turaev,~\cite{Turaev} 
we have the minimum defining algebraic relations~\cite{Moore-Segal, Kapustin-Turzillo}: 
To give a $G_0$-equivariant oriented TFT is
equivalent to give a $G_0$-graded algebra ${\cal C} = \bigoplus_{g \in G_0} {\cal C}_g$ 
together with a group homomorphism $\alpha : G_0 \to {\rm Aut}({\cal C})$ 
such that ${\rm Aut}({\cal C}) \ni \alpha_g : {\cal C}_h \to {\cal C}_{g h g^{-1}}$, and 
\begin{itemize}
\item[(1)] There is a $G_0$-invariant trace $\theta_{\cal C} : {\cal C}_1 \to \C, \ \theta_{\cal C} \circ \alpha_g = \theta_{\cal C}$,  
such that the induced paring ${\cal C}_g \otimes {\cal C}_{g^{-1}} \to \C$ is nondegenerate. 
\item[(2)] For $\phi \in {\cal C}_g$, \ $\alpha_g (\phi) = \phi$. 
\item[(3)] For $\phi_1 \in {\cal C}_{g_1}, \phi_2 \in {\cal C}_{g_2}$, $\alpha_{g_2}(\phi_1) \phi_2 = \phi_2 \phi_1$. 
\item[(4)] (Punctured Torus) $\sum_i \alpha_{h}(\xi^g_i) \xi_i^{g^{-1}} = \sum_i \xi^h_i \alpha_{g}(\xi_i^{h^{-1}}) \in {\cal C}_{hg h^{-1} g^{-1}}$. 
\end{itemize}
The non-degenerate property in (1) 
is followed by the same way as (2). 
Derivations of (1-4) 
are summarized in Appendix \ref{app:Closed TFT}. 

The state in (4) is 
the handle adding operator 
\begin{align}
\vcenter{\hbox{\includegraphics[width=0.2\linewidth]{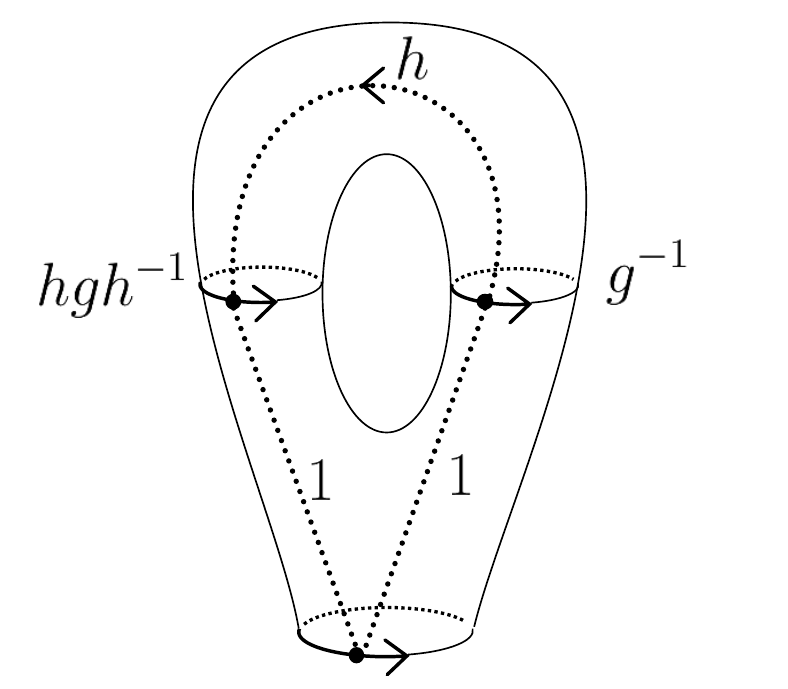}}}
= \sum_i \alpha_{h}(\xi^g_i) \xi_i^{g^{-1}}, 
\label{eq:tft_handle}
\end{align}
which enables us to 
compute all possible partition functions on surfaces of genus $g$ with twist. 
For example, 
the partition function on torus $T^2$ with twist is given by
\begin{align}
\vcenter{\hbox{\includegraphics[width=0.2\linewidth]{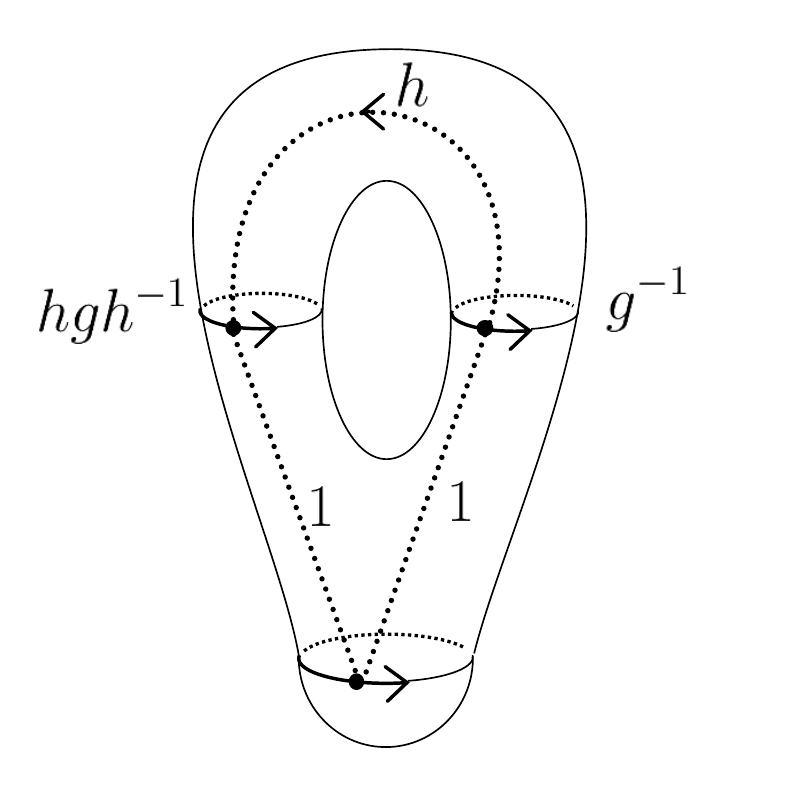}}}
= Z_{T^2}(h,g) 
= \sum_i \theta_{\cal C} (\alpha_h(\xi^g_i) \xi_i^{g^{-1}}), \ \ (hg=gh). 
\end{align}

\subsubsection{General solution for semi simple cases}

If ${\cal C}_1$, the untwisted sector Hilbert space, 
is semisimple, 
${\cal C}_1 \cong \bigoplus_{x \in X} \C \epsilon_x$, 
$\epsilon_x \epsilon_y = \delta_{x,y} \epsilon_x$, 
we have general solutions for the 
algebraic constraints 
(1-4) as follows.~\cite{Turaev, Moore-Segal}
Here, $X$ is a finite set equipped with $G_0$-action 
$g \cdot (h \cdot x) = (gh) \cdot x$. 

For a given $G_0$-set $X$, 
the twisted sector Hilbert space ${\cal C}_g$ 
consists of little group at $x$ as 
${\cal C}_g = \bigoplus_{x \in X, g \cdot x = x} L_{g,x}$, 
where $L_{g,x} \cong \C$ are lines. 
The multiplication of ${\cal C} = \bigoplus_{g \in G_0} {\cal C}_g$ is determined by 
a given group cocycle $b_x(g,h) \in Z^2(G_0,C(X,U(1))) \big( \cong Z^2_{G_0}(X,U(1)) \big)$ as 
\footnote{$C(X,U(1))$ is the $G_0$-module consisting of 
$U(1)$-valued functions on $X$. 
The $G_0$-structure is defined by $(g \cdot f)(x) := f(g \cdot x)$, $g \in G_0, x \in X$. 
The group cohomology $H^2(G_0,C(X,U(1)))$ classifies the following extension 
$$
1 \to C(X,U(1)) \to \hat G \to G_0 \to 1.
$$
}
\begin{align}
\ell_{g_2,x_2} \ell_{g_1,x_1} = 
\left\{\begin{array}{ll}
b_{x_1}(g_2, g_1) \ell_{g_2 g_1,x_1} & (x_2 = g_1 \cdot x_1) \\
\\
0 & (\mbox{otherwise})
\end{array}\right. && 
(\ell_{g,x} \in L_{g,x}). 
\end{align}
The associativity condition $\ell_{g_3,x_3} (\ell_{g_2,x_2} \ell_{g_1,x_1}) = (\ell_{g_3,x_3} \ell_{g_2,x_2}) \ell_{g_1,x_1}$ 
corresponds to the 2-cocycle condition 
\begin{align}
b_{g_1 \cdot x}(g_3,g_2) b_x(g_3 g_2, g_1) = b_x(g_2,g_1) b_x(g_3,g_2 g_1). 
\end{align}
In short, $G_0$-equivariant TFTs are classified by 
the group cohomology $H^2(G_0,C(X,U(1))) \big( \cong H^2_{G_0}(X,U(1)) \big)$. 

To make a contact with physics of SPT phases, 
let us specialize to the case where ${\cal C}_1$ is simple 
${\cal C}_1 \cong \C$. 
In this case, the ground state in the untwisted sector is unique, 
and the classification is reduced into group cohomology with 
$U(1)$ coefficient $H^2(G_0,U(1))$. 

On the other hand, in semisimple cases, 
we have a combination of symmetry breaking 
and symmetry fractionalization discussed in 
Refs.~\cite{Schuch2011, chen2011complete}.
Since the group cohomology $H^2(G_0, C(X,U(1)))$ 
splits into $G_0$-orbits, 
we can simply assume that $X$ consists of a single $G_0$-orbit. 
Let subgroup $G' \subset G_0$ be an unbroken symmetries, 
then, we have a bijection $X \cong G_0/G'$ as a set, 
which is a ``Nambu-Goldstone manifold''. 
Each element $x \in G_0/G'$ represents 
a vacuum which partially breaks $G_0$ symmetry and 
retains $G'$ symmetry. 
All the elements $G_0/G'$ are permuted by broken symmetries 
in $G_0$. 
The topological classification is given by 
\begin{align}
H^2_{G_0}(G_0/G', U(1)) \cong H^2_{G'}(pt, U(1)) \cong H^2(G',U(1)), 
\end{align}
says, the group cohomology classification for 
unbroken symmetries.

\begin{figure}[!]
 \begin{center}
  \includegraphics[width=0.3\linewidth, trim=0cm 0cm 0cm 0cm]{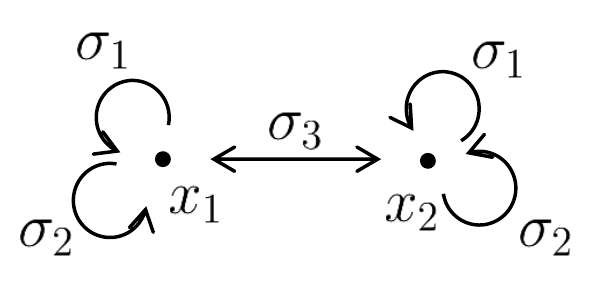}
 \end{center}
 \caption{Example of a $G_0$-set $X$ for a combination of symmetry broken and 
 symmetry fractionalization. 
 In this example, $X$ consists of two points $X = \{x_1, x_2\}$ 
 which are permuted by the broken symmetry $\sigma_3 \in G_0$. 
 }
 \label{fig:tft/spec}
\end{figure}

For example, 
let full symmetry be $G_0 = \Z_2[\sigma_1] \times \Z_2[\sigma_2] \times \Z_2[\sigma_3]$ and 
unbroken symmetry be $G' = \Z_2[\sigma_1] \times \Z_2[\sigma_2]$,  
where $\sigma_i (i=1,2,3)$ are generators of $\Z_2$. 
In this case, $X$ consists of two points $\{x_1, x_2\}$ which 
are exchanged by the broken symmetry as $x_2 = \sigma_3 \cdot x_1$ 
as shown in Fig.~\ref{fig:tft/spec}. 
The topological classification is given by 
that for the unbroken symmetry as 
$H^2(\Z_2[\sigma_1] \times \Z_2[\sigma_2],U(1)) = \Z_2$. 
See Fig.~\ref{fig:tft/spec}. 
%
%

\subsection{$G$-equivariant unoriented closed TFTs}

$(1+1)$d oriented (closed) TFTs were extended to 
unoriented TFTs by Turaev-Turner~\cite{turaev2006unoriented} 
and equivariant unoriented TFTs by Kapsutin-Turzillo~\cite{Kapustin-Turzillo}. 
See also Refs.~\cite{tagami2012unoriented, sweet2013equivariant}. 
Here we review $G$-equivariant unoriented $(1+1)$d TFTs. 

As before,
let $G$ be a full symmetry group including orientation-reversing symmetries 
and $G_0 \subset G$ be the orientation-preserving subgroup. 
There are two new ingredients to define 
 (equivariant) unoriented $(1+1)$d TFTs: 
the crosscap state and reflection transformation 
((g) and (h) in Table \ref{tab:functor_ori_tft}, respectively). 
As for Item (g), 
the boundary state of the M\"obius strip defines the 
crosscap state $\theta_g \in {\cal C}_{g^2} (g \notin G_0)$ 
\begin{align}
\vcenter{\hbox{\includegraphics[width=0.4\linewidth]{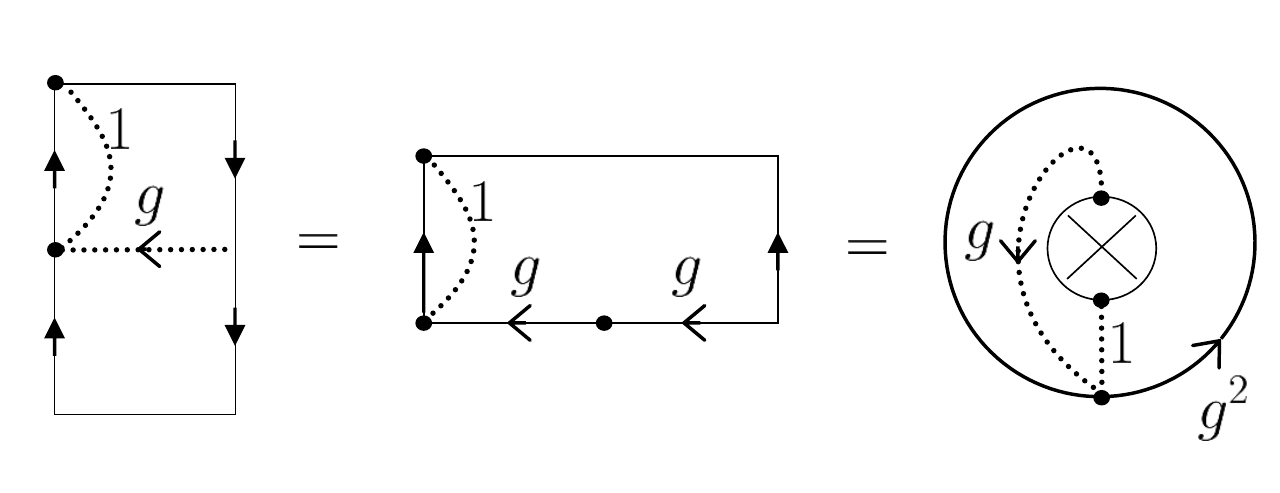}}}
\ = \theta_{g} \in {\cal C}_{g^2}, \quad (g \notin G_0). 
\label{eq:tft_crosscap}
\end{align}
Notice that the crosscap state $\theta_{g}$ belongs to 
the twisted sector of $g^2 \in G_0$. 
As for Item (h), 
the presence of an orientation-reversing symmetry $g \notin G_0$ 
can be used to 
consider reflection of the circle 
\begin{align}
\vcenter{\hbox{\includegraphics[width=0.15\linewidth]{figs/tft/cylinder_P}}}
\ \Rightarrow \alpha_g : {\cal C}_h \to {\cal C}_{g h^{-1} g^{-1}}, 
\quad (g \notin G_0). 
\end{align}

In a way similar to $G_0$-equivariant oriented TFTs, 
we have several constraints on the algebraic category. 
Kapustin-Turzillo\cite{Kapustin-Turzillo} showed that 
to give a $G$-equivariant unoriented $(1+1)$d TFT 
is equivalent to give a $G_0$-graded algebra
${\cal C} = \bigoplus_{g \in G_0} {\cal C}_g$ 
together with a group homomorphism 
$\alpha : G \to {\rm Aut}({\cal C})$ 
such that $\alpha_{g \in G_0} : {\cal C}_h \to {\cal C}_{g h g^{-1}}$, 
$\alpha_{g \notin G_0} : {\cal C}_h \to {\cal C}_{g h^{-1} g^{-1}}$.
They must satisfy 
(1)-(4), and 
\begin{itemize}
\item[(5)] $\alpha_g(\phi_1 \phi_2) = \alpha_g(\phi_2) \alpha_g(\phi_1)$, $g \notin G_0$. 
\item[(6)] $\alpha_{h \in G_0}(\theta_g) = \theta_{h g h^{-1}}$ and $\alpha_{h \notin G_0}(\theta_g) = \theta_{h g^{-1} h^{-1}}$. 
\item[(7)] (Punctured M\"obius strip) $\theta_g \phi = \alpha_g(\phi) \theta_{gh}$, $\phi \in {\cal C}_h$. 
\item[(8)] (Punctured Klein bottle) $\sum_i \alpha_{g}(\xi^{(gh)^{-1}}_i) \xi_i^{gh} = \theta_g \theta_h$, $g,h \notin G_0$. 
\end{itemize}
Derivations of these constraints\cite{Kapustin-Turzillo} are summarized in Appendix \ref{app:Closed TFT}. 

All possible partition functions are constructed from 
the handle adding operator (\ref{eq:tft_handle}) 
and crosscap adding operator (\ref{eq:tft_crosscap}). 
For example, 
the partition function on real projective plane reads 
\begin{align}
Z_{\R P^2}(g) 
= \vcenter{\hbox{\includegraphics[width=0.15\linewidth]{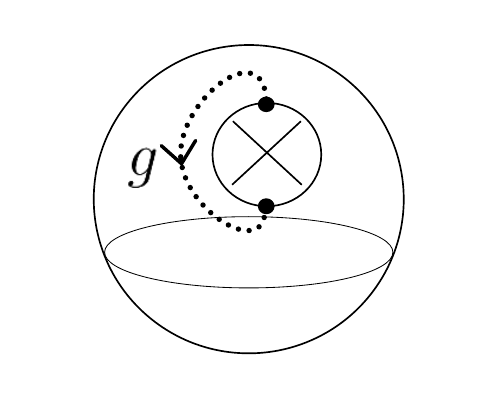}}}
= \theta_{\cal C}(\theta_g), \qquad (g \notin G_0, g^2 = 1). 
\end{align}
The Klein bottle partition function is 
\begin{align}
Z_{KB}(g;h) 
= \vcenter{\hbox{\includegraphics[width=0.2\linewidth]{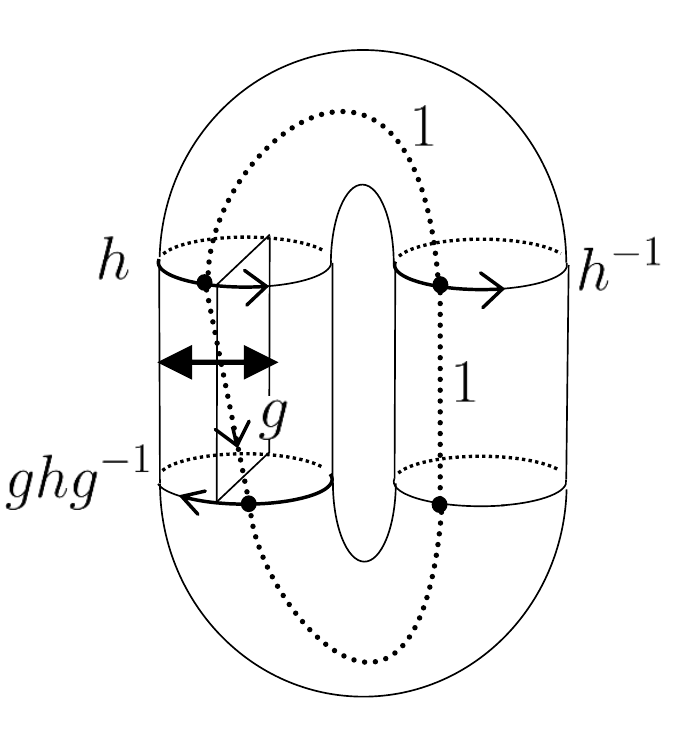}}}
= \theta_{\cal C}\big( \sum_i \alpha_g(\xi_i^h) \xi_i^{h^{-1}} \big) 
= \theta_{\cal C}\big( \theta_g \theta_{g^{-1} h^{-1}} \big), \quad (g \notin G_0, h \in G_0). 
\end{align}

\subsubsection{General solution for simple cases}

In the cases where ${\cal C}_1$ is simple ${\cal C}_1 \cong \C$, 
i.e., there is a unique ground state, 
Kapustin-Turzillo~\cite{Kapustin-Turzillo} showed general solutions 
of the algebraic constraints (1) - (8). 
They showed that to give a $G$-equivariant 
unoriented $(1+1)$d simple TFT is to give a 
2-group cycle $b(g,h) \in Z^2(G,U(1)_{\phi})$.~\footnote{
$U(1)_{\phi}$ is equipped with $G$-action defined by (\ref{eq:app_g_action}). }
This is consistent with the  group cohomology classification of 
bosonic $(1+1)$d SPT phases with reflection or time-reversal symmetry~\cite{chen2011complete}.

\subsubsection{Relation to MPS}
\label{sec:tft-mps}

In the SPT context, the spatial circle $S^1$ with $g$-flux in TFTs is 
identified with a bulk SPT phase with $g$-twisted boundary condition. 
The uniqueness condition of 
the ground state in SPT phases
implies that the corresponding TFTs are invertible, 
i.e.,
we have a simple algebra of untwisted sector ${\cal C}_1 \cong \C$. 
In TFTs, 
there is no excited state and the Hilbert space consists only of ground states. 
The correlation length of the bulk is zero, 
so a TFT is represented by a fixed point MPS 
introduced in Sec.~\ref{Fixed point MPSs} 
\begin{align}
\ket{\Psi_g} 
= \sum_m {\rm Tr}\, (A_{m_1} \cdots A_{m_L} V_g) \ket{m_1 \cdots m_L} 
\sim \sum_m {\rm Tr}\, (A_m V_g) \ket{m}. 
\end{align}
Here we used the equivalence relation of fixed point MPSs (\ref{eq:mps_equivalence}). 
Only one physical site is sufficient to describe the MPS representation of a TFT's ground state. 
The correspondence between MPSs and  equivariant TFTs, 
can be pictorially represented as
\begin{align}
\vcenter{\hbox{\includegraphics[width=0.3\linewidth]{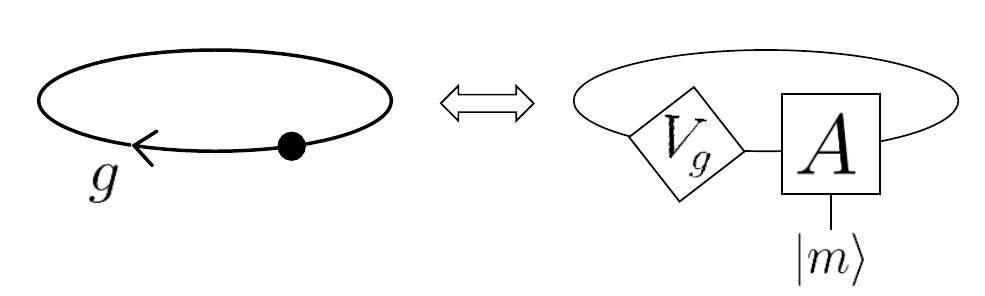}}}
\end{align}

Cobordisms in $G$-equivariant TFTs correspond to 
various ``adiabatic deformations'' of closed chains, e.g., 
``fusion'' and ``separating'', and symmetry operations. 
The fourth column in Table \ \ref{tab:functor_ori_tft} 
summarizes correspondences between cobordisms in $G$-equivariant TFTs 
and MPS representations. 

For example, the fusion process of two 
closed chain is {\it formally} represented in MPS networks as follows. 
For two MPSs
\begin{align}
\ket{\Psi_g} = {\rm Tr} \big[ A_1 V_g \big], 
\quad 
\ket{\Psi_h} = {\rm Tr} \big[ A_2 V_h \big], 
\end{align}
the fusion $\ket{\Psi_g} \cdot \ket{\Psi_h}$ is given by 
\begin{align}
&\vcenter{\hbox{\includegraphics[width=0.6\linewidth]{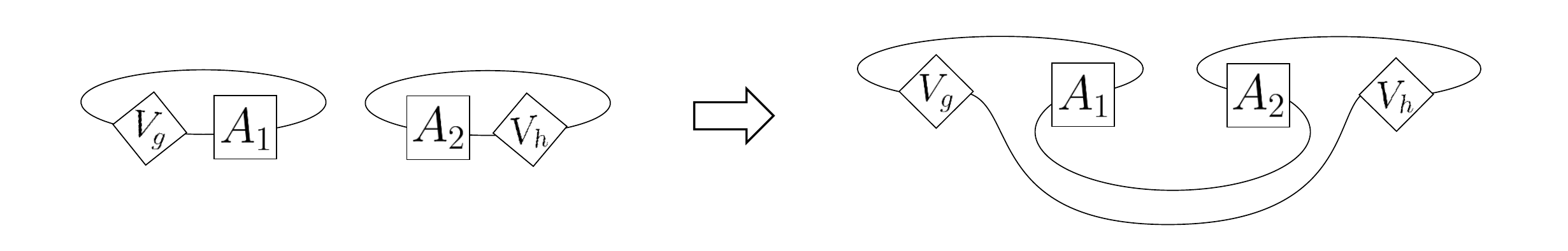}}} \\
&\Rightarrow 
\ket{\Psi_g} \cdot \ket{\Psi_h} = {\rm Tr} \big[ A_2 A_1 V_g V_h \big] \ket{m_2 m_1} \sim b(g,h) {\rm Tr} \big[ A V_{gh} \big] \ket{m} = b(g,h) \ket{\Psi_{gh}}. 
\end{align}
Here we used the equivalence relation of fixed point MPSs (\ref{eq:mps_equivalence}).

\subsection{$G$-equivariant open and closed TFTs}
\label{G-equivariant open and closed TFTs}

\begin{table}[h!]
\caption{Building blocks of $G$-equivariant open and closed $(1+1)$d TFTs. 
In the forth column, MPS representations are shown. 
In figures, dashed lines with group elements represent holonomies. 
(r) is the definition of the boundary state for a boundary condition $a$. 
}
\begin{center}
\scalebox{0.8}{
\hspace*{-2cm}
\begin{tabular}{| >{\centering\arraybackslash}m{1cm} | >{\centering\arraybackslash}m{3cm} | >{\centering\arraybackslash}m{3cm} | >{\centering\arraybackslash}m{6cm} | >{\centering\arraybackslash}m{5cm} | }
\hline
& Manifolds & Hilbert spaces & Simple and fixed point MPS & Comment \\
\hline
(i) & \includegraphics[width=0.2\linewidth, trim=0cm 0cm 0cm 0cm]{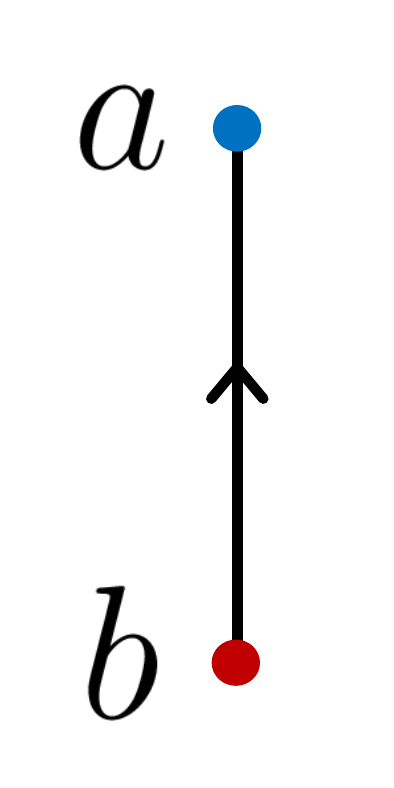} & 
${\cal O}_{ab}$ & 
$$
\begin{array}{c}
\mbox{Hilbert space spanned by} \\
\Big\{ \big( L_i^T A_m R_j \big) \ket{m} \Big\}, \\
L_i \in V_a^*, R_j \in V_b. 
\end{array}
$$ & Open chain. $L_i$ ($R_j$) are basis of 
$V_a^*$ ($V_b$). \\
\hline
(j) & \includegraphics[width=0.3\linewidth, trim=0cm 0cm 0cm 0cm]{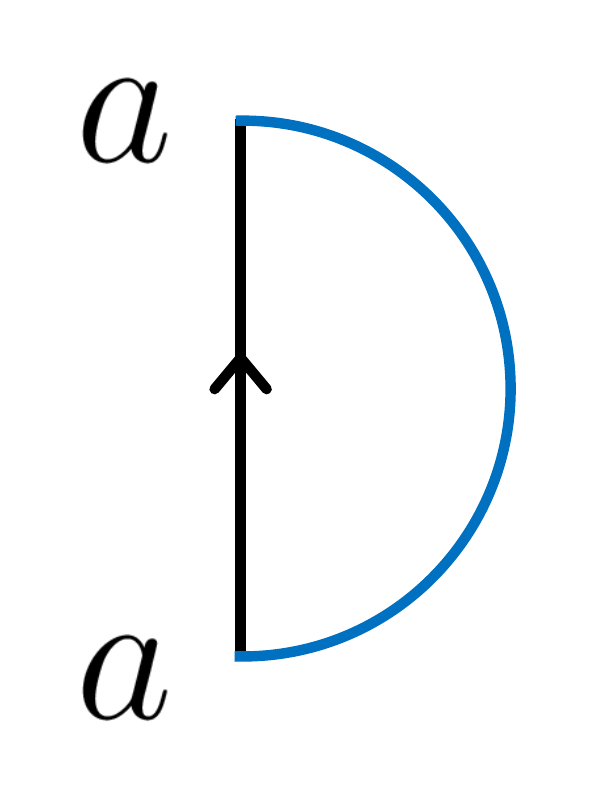} & 
$$
\begin{array}{c}
\theta_{a}: {\cal O}_{aa} \to \C, \\
\psi \mapsto \theta_{a} (\psi) \\
\end{array}
$$
& 
$$
\begin{array}{c}
\theta_a \Big( \big( v_L^T A_m v_R \big) \ket{m} \Big) = (v_L,v_R), \\
v_L \in V_a^* , v_R \in V_a \\
\end{array}
$$ & 
$(v_L,v_R) = \sum_i [v_L]_i [v_R]_i$. 
Notice that $\theta_a (1_a) = {\rm dim} V_a$ for simple and fixed point MPS. 
\\
\hline
(k) & \includegraphics[width=0.3\linewidth, trim=0cm 0cm 0cm 0cm]{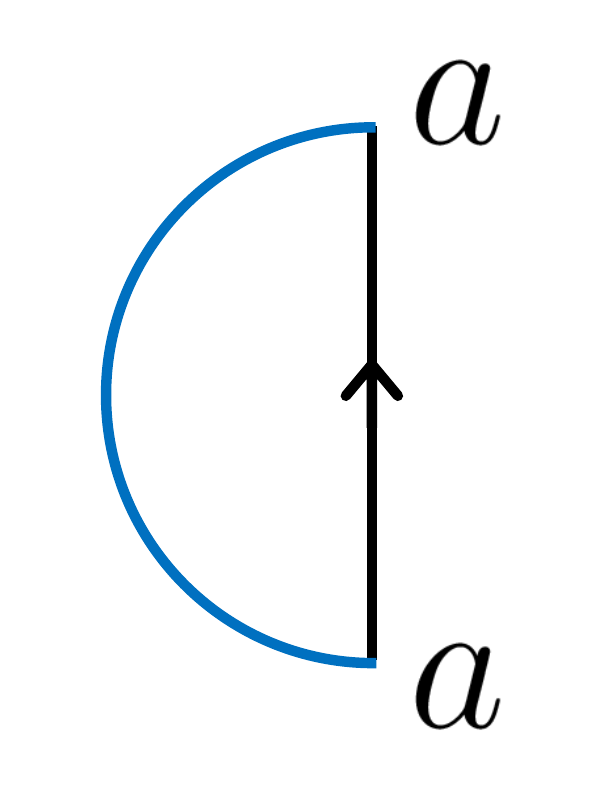} & 
$$
\begin{array}{c}
\C \to {\cal O}_{aa}, \\
1 \mapsto 1_{a} \\
\end{array}
$$
& $\sum_i \big( L_i^T A_m R_i \big) \ket{m}$ & 
$1_{a}$ is the unit satisfying $1_{a} \psi = \psi 1_{b} = \psi, \psi \in {\cal O}_{ab}$. \\
\hline
(l) & \includegraphics[width=0.6\linewidth, trim=0cm 0cm 0cm 0cm]{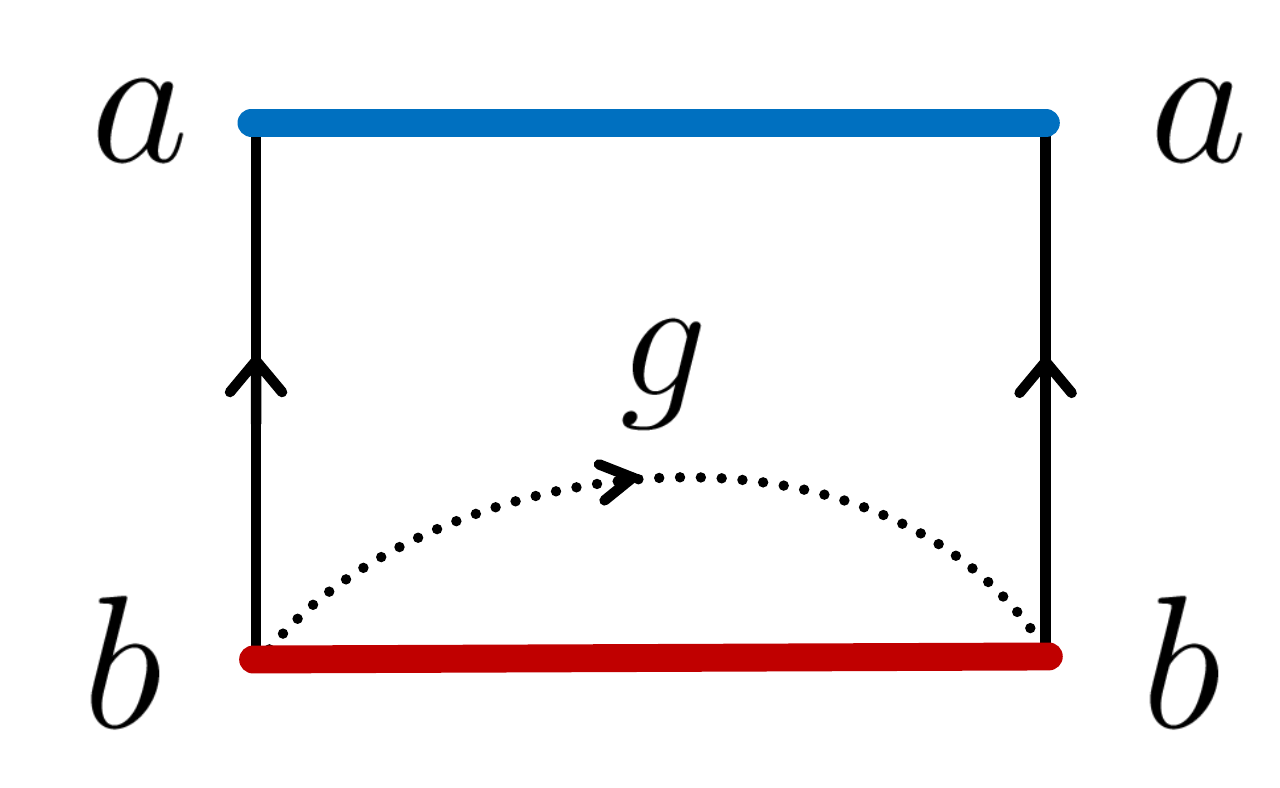} & 
$$
\begin{array}{c}
\rho_{g \in G_0}: {\cal O}_{ab} \to {\cal O}_{ab}, \\
\psi \mapsto \rho_g \psi \\
\end{array}
$$
& 
$$
\begin{array}{l}
\big( v_L^T A_m v_R \big) \ket{m} \\
\mapsto \big( v_L^T V_{g,a}^{\dag} A_m V_{g,b} v_R \big) \ket{m}, \\
v_L \in V_a^*, v_R \in V_b. 
\end{array}
$$
& $g$-action on open chain. $V_{g,a}$ is representation matrix of $V_a$. \\
\hline
(m) & \includegraphics[width=0.6\linewidth, trim=0cm 0cm 0cm 0cm]{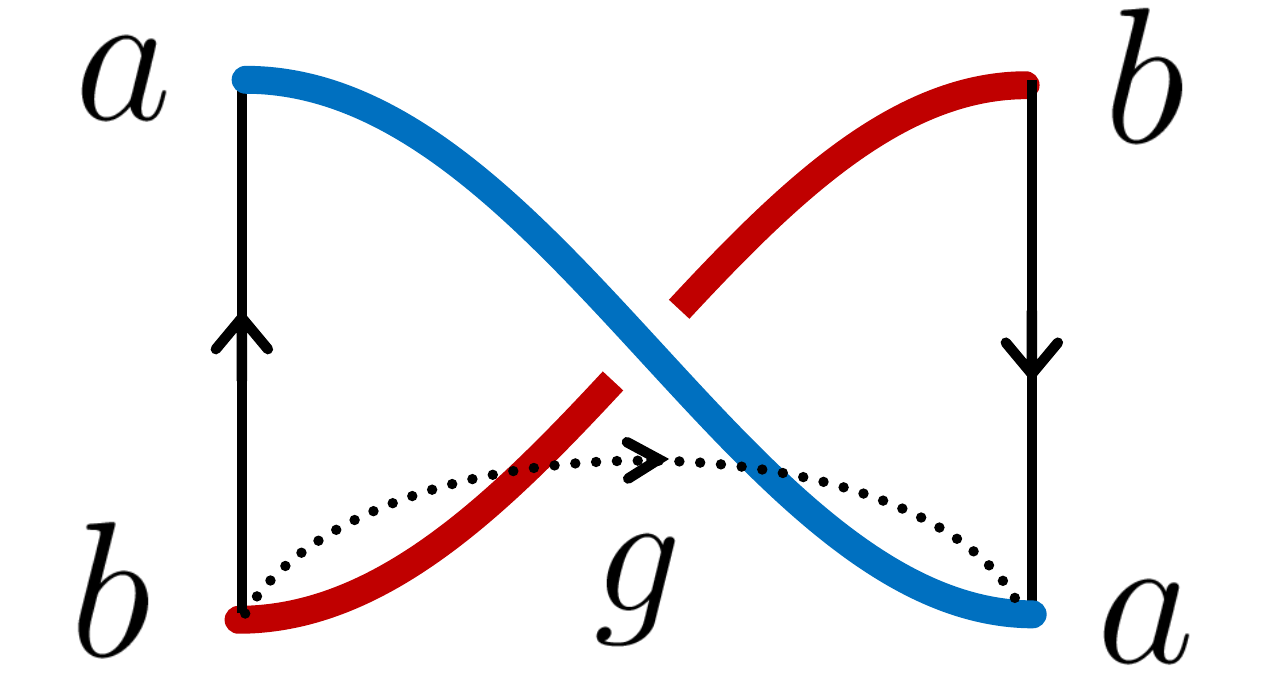} & 
$$
\begin{array}{c}
\rho_{g \notin G_0}: {\cal O}_{ab} \to {\cal O}_{ab}, \\
\psi \mapsto \rho_g \psi \\
\end{array}
$$
& 
$$
\begin{array}{l}
\big( v_L^T A_m v_R \big) \ket{m} \\
\mapsto \big( v_R^T V_{g,b}^{\dag} A_m V_{g,a} v_L \big) \ket{m}, \\
v_L \in V_a^*, v_R \in V_b. 
\end{array}
$$
& $g$-reflection on an open chain \\
\hline
(n) & \includegraphics[width=0.6\linewidth, trim=0cm 0cm 0cm 0cm]{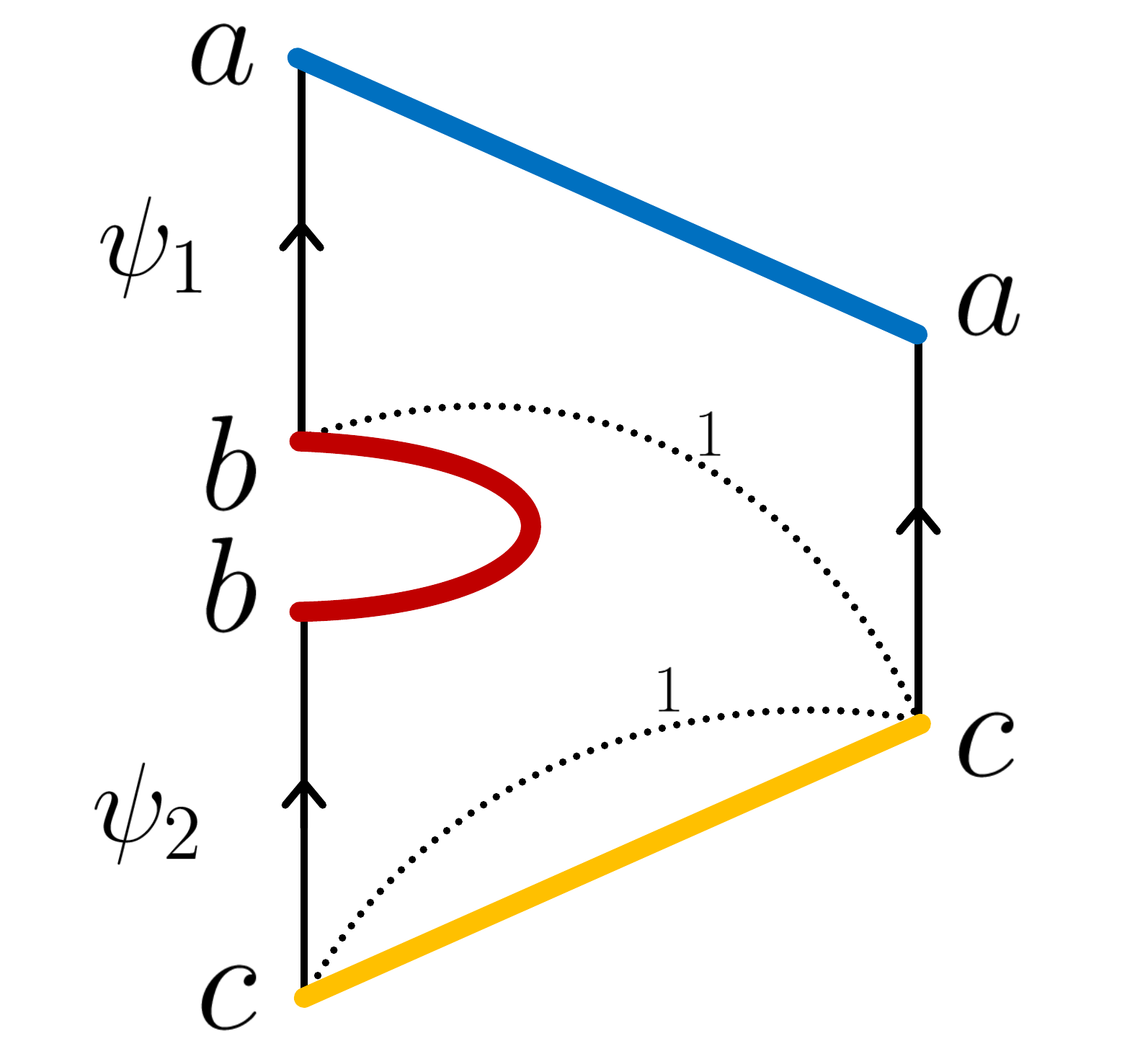} & 
$$
\begin{array}{c}
{\cal O}_{ab} \otimes {\cal O}_{bc} \to {\cal O}_{ac}, \\
\psi_1 \otimes \psi_2 \mapsto \psi_1 \psi_2 \\
\end{array}
$$
& 
$$
\begin{array}{c}
\big( v_L^T A_{m_1} v_R \big) \ket{m_1} \otimes \big( w_L^T A_{m_2} w_R \big) \ket{m_2} \\
\mapsto (w_L, v_R) \big( v_L^T A_m w_R \big) \ket{m}, \\
v_L \in V_a^*, v_R \in V_b, w_L \in V^*_b, w_R \in V_c. 
\end{array}
$$
& Fusion of two open chains  \\
\hline
(o) & \includegraphics[width=0.5\linewidth, trim=0cm 0cm 0cm 0cm]{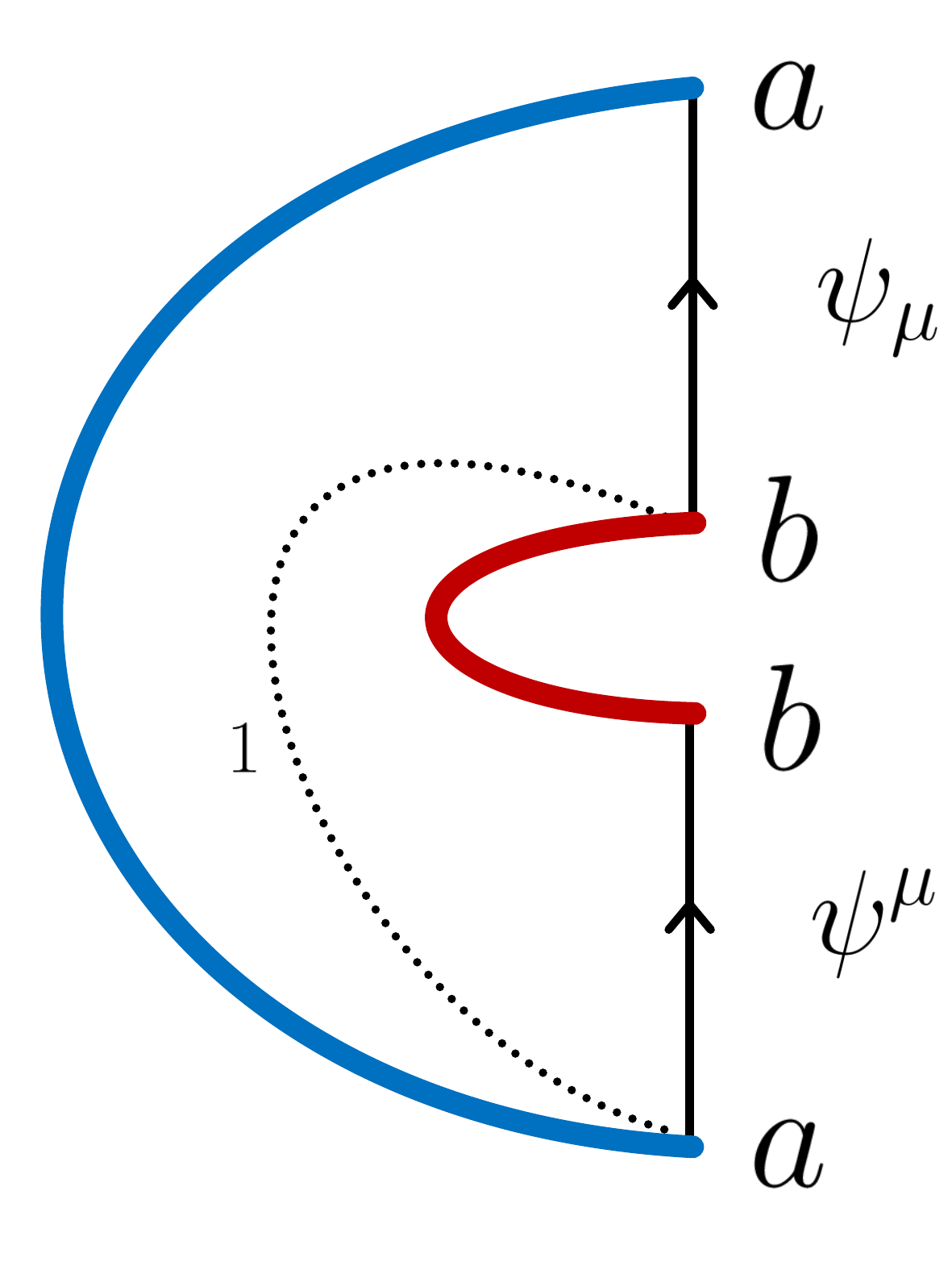} & 
$$
\begin{array}{c}
\C \to {\cal O}_{ab} \otimes {\cal O}_{ba}, \\
1 \mapsto \sum_{\mu} \psi_{\mu} \otimes \psi^{\mu} \\
\end{array}
$$ 
& $\sum_{ij} \big( L_i^T A_{m_1} R_j \big) \ket{m_1} \otimes \big( L_j^T A_{m_2} R_i \big) \ket{m_2}$ & 
$\psi_{\mu} \in {\cal O}_{ab}$ are basis of ${\cal O}_{ab}$ and $\psi^{\mu} \in {\cal O}_{ba}$ are their dual of ${\cal O}_{ba}$ that satisfy $\theta_{a} (\psi_{\mu} \psi^{\nu}) = \delta_{\mu}^{\nu}$.  \\
\hline
(p) & \includegraphics[width=0.66\linewidth, trim=0cm 0cm 0cm 0cm]{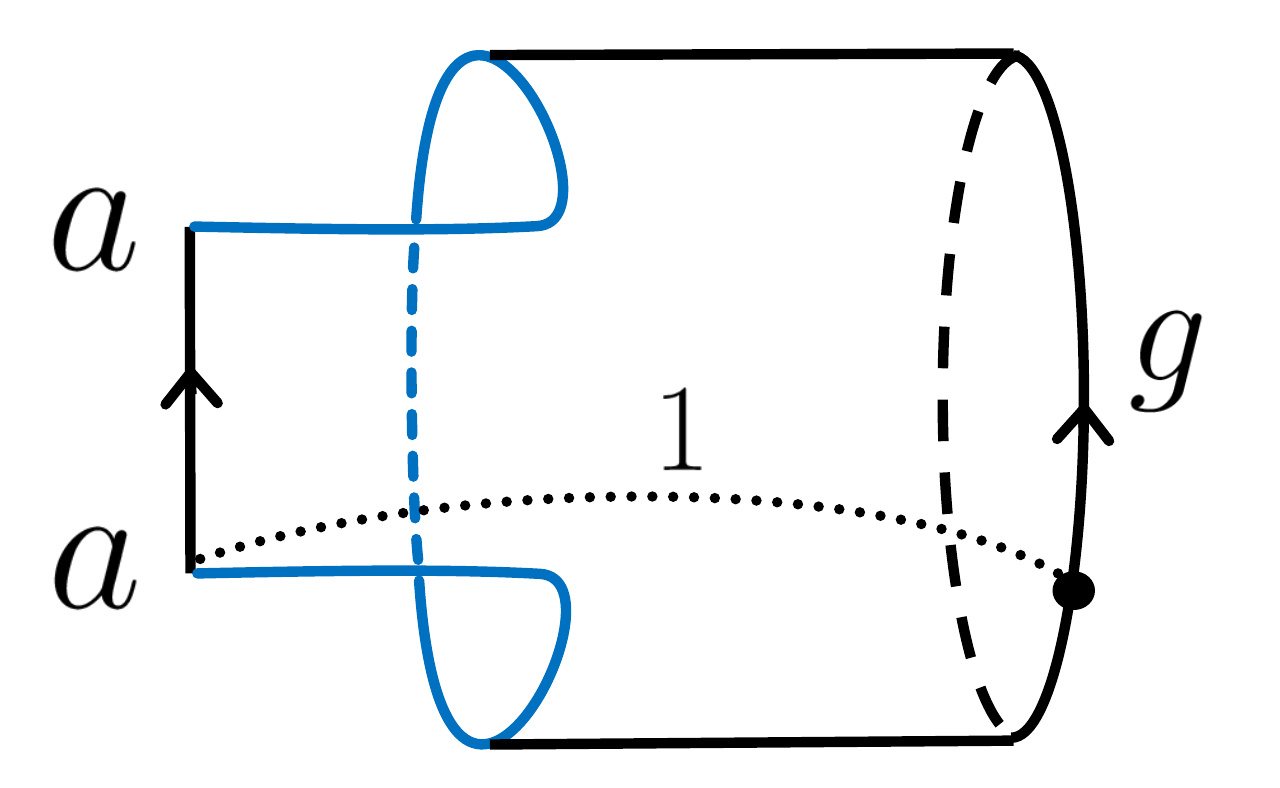} & 
$$
\begin{array}{c}
\imath^{g,a} : {\cal O}_{aa} \to {\cal C}_g, \\
\psi \mapsto \imath^{g,a}(\psi) \\
\end{array}
$$ 
& $\big( v_L^T A_m v_R \big) \ket{m} \mapsto (V_g^* v_L, v_R) {\rm Tr}(A_m V_g) \ket{m}$ & Open to closed map \\
\hline
(q) & \includegraphics[width=0.6\linewidth, trim=0cm 0cm 0cm 0cm]{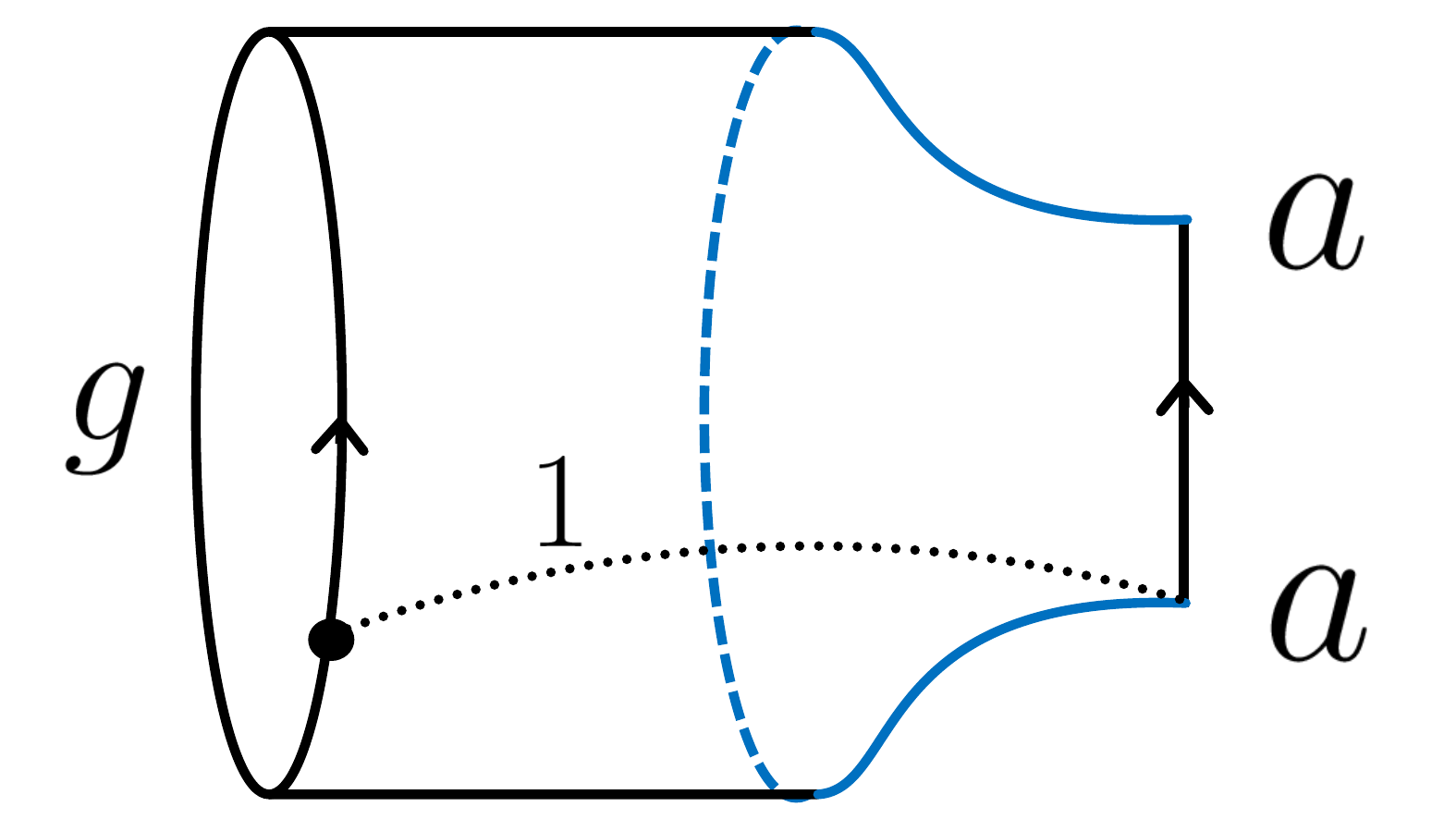} & 
$$
\begin{array}{c}
\imath_{g,a} : {\cal C}_{g} \to {\cal O}_{aa}, \\
\phi \mapsto \imath_{g,a}(\phi) \\
\end{array}
$$ 
& 
${\rm Tr}(A_m V_g) \ket{m} \mapsto \sum_i \big( L_i^T A_m V_g R_i \big) \ket{m}$ & Closed to open map \\
\hline \hline
(r) & \includegraphics[width=0.7\linewidth, trim=0cm 0cm 0cm 0cm]{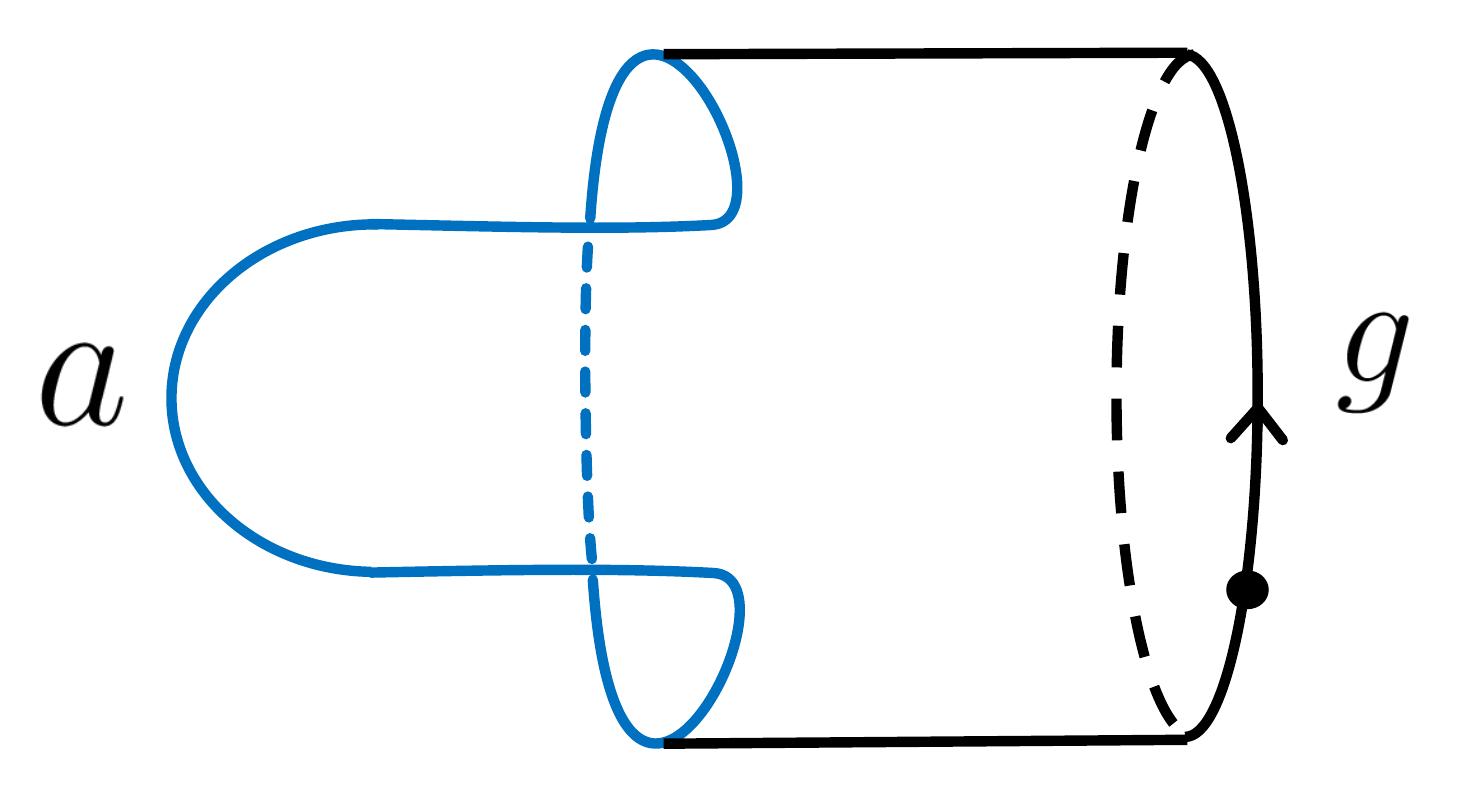} & 
$B_a = \imath^{g,a}(1_a) \in {\cal C}_g$ 
& 
${\rm tr}\big( V_{g,a}^{\dag} \big) {\rm Tr}(A_m V_{g,a}) \ket{m}$ & 
Boundary state for boundary condition $a$. 
\\
\hline
\end{tabular}
\hspace*{-2cm}
}
\end{center}
\label{Tab:Functor_Open}
\end{table}
\clearpage

Next, we extend closed TFTs to include open chains (intervals). 
A new object is an oriented interval $I_{ab} = [0,1]$ with boundary conditions 
$a,b$ as shown in Table.\ \ref{Tab:Functor_Open} (i).~\footnote{
In string theory, 
the boundary conditions $a, b, \dots$ are 
Chan-Paton factors associated with the 
endpoints of open strings. }
We denote the Hilbert space associated with the interval $I_{ab}$ by ${\cal O}_{ab}$. 
An element $\psi \in {\cal O}_{ab}$ represents 
a state living in the open chain with boundary conditions $a$ and $b$. 
(Note that the boundary conditions $a, b$ do not represent 
some states in the open chain. )

Similar to closed TFTs, 
we have several cobordisms in open and closed TFTs. 
Table \ref{Tab:Functor_Open} summarizes the building blocks. 
We have some remarks in order. 
\begin{itemize}
\item 
We use the same notation as Moore-Segal~\cite{Moore-Segal}. 
The fusion process is represented as 
\begin{align}
\vcenter{\hbox{\includegraphics[width=0.2\linewidth]{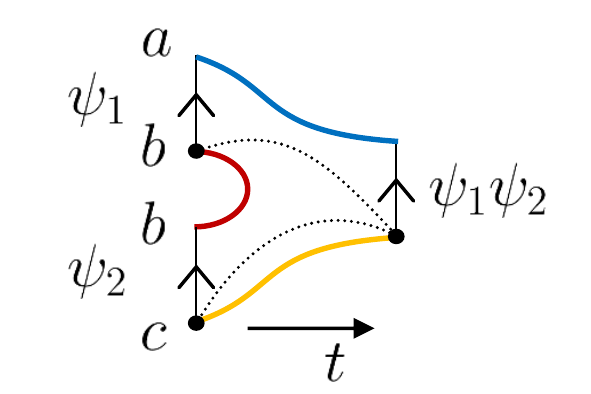}}} 
\Rightarrow \ 
{\cal O}_{ab} \otimes {\cal O}_{bc} \to {\cal O}_{ac}, \ \  \psi_1 \otimes \psi_2 \mapsto \psi_1 \psi_2.  
\end{align}
Note the order of two intervals $I_{ab}$ and $I_{bc}$. 
\item 
In addition to on-site symmetry transformation
$\rho_{g \in G_0} : {\cal O}_{ab} \to {\cal O}_{ab}$, 
we have reflection on an open chain 
$\rho_{g \notin G_0}: {\cal O}_{ab} \to {\cal O}_{ba}$ 
which exchanges the boundary conditions $a,b$. 
$\rho_g$ satisfies $\rho_g \circ \rho_h = \rho_{gh}$ $(g,h \in G)$. 
\item 
Essentially new ingredients are 
the open-to-closed map $\imath^{g,a}$ 
and the closed-to-open map $\imath_{g,a}$
which connect closed chains and open chains as~\cite{Moore-Segal} 
\begin{align}
\imath^{g,a} : {\cal O}_{aa} \to {\cal C}_g, && 
\imath_{g,a} : {\cal C}_g \to {\cal O}_{aa}, && 
(g \in G_0). 
\end{align}
Here, to glue back to a closed chain from a open chain, 
the boundary conditions should agree. 
\end{itemize}

All bordsims can be constructed by using building blocks listed in Table \ref{Tab:Functor_Open}. 
For example,
a ``branching'' process of an open chain 
is given by the same way as (\ref{fig:tft/separate}), 
\begin{align}
&
\vcenter{\hbox{\includegraphics[width=0.6\linewidth]{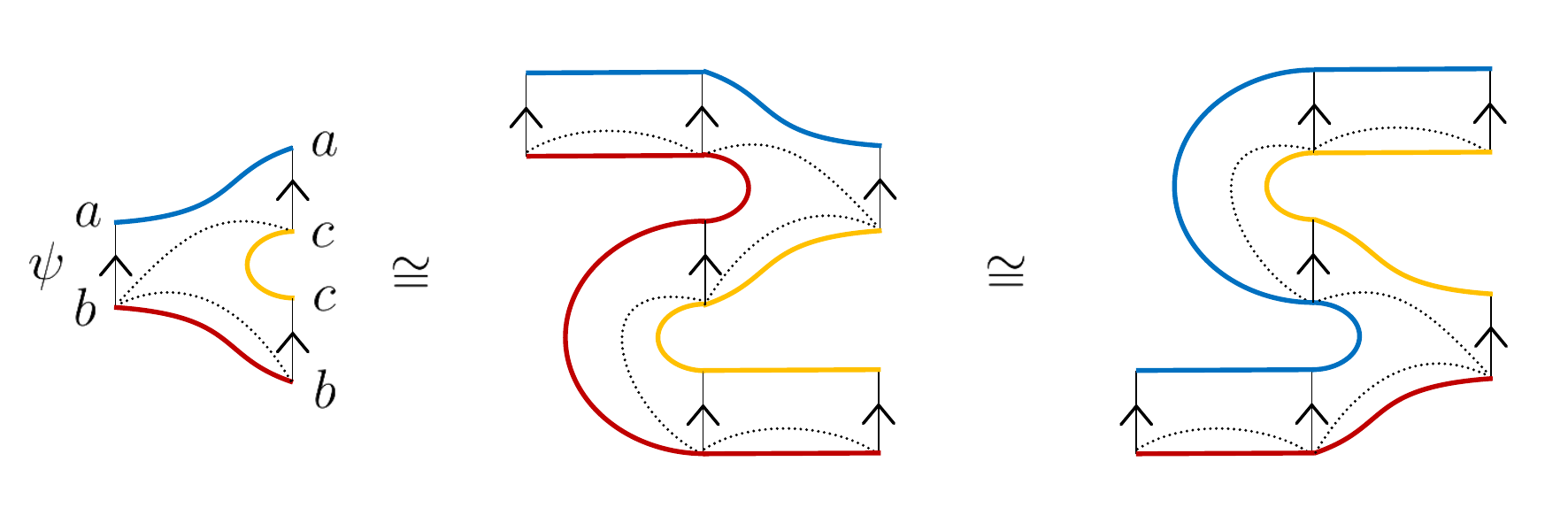}}} 
\nonumber \\
& \Rightarrow \ \ 
{\cal O}_{ab} \to {\cal O}_{ac} \otimes {\cal O}_{cb}, \ \ 
\psi \mapsto 
\sum_{\mu} \psi \psi_{\mu} \otimes \psi^{\mu} 
= \sum_{\mu} \tilde \psi_{\mu} \otimes \tilde \psi^{\mu} \psi. 
\end{align}
Here, $\sum_{\mu} \psi_{\mu} \otimes \psi^{\mu} (\psi_{\mu} \in {\cal O}_{bc}, \psi^{\mu} \in {\cal O}_{cb})$ 
and $\sum_{\mu} \tilde \psi_{\mu} \otimes \tilde \psi^{\mu} (\tilde \psi_{\mu} \in {\cal O}_{ac}, \tilde \psi^{\mu} \in {\cal O}_{ca})$ 
are coform defined in (o) of Table \ref{Tab:Functor_Open}.

In the target algebraic category, 
there are constraints from the open and closed cobordism category. 
We have the following constraints for oriented open and closed TFTs by Moore-Segal~\cite{Moore-Segal}: 
\begin{itemize}
\item[(9)] \ 
$\imath_{1,a}(1_{\cal C}) = 1_{{\cal O}_{aa}}$. 
\item[(10)]\ 
$\rho_g (\psi_1 \psi_2) = (\rho_g \psi_1) (\rho_g\psi_2), \  g \in G_0, \psi_1 \in {\cal O}_{ab}, \psi_2 \in {\cal O}_{bc}$. 
\item[(11)]\ 
$\imath_{g_1,a}(\phi_1) \imath_{g_2,a}(\phi_2) = \imath_{g_2 g_1,a}(\phi_2 \phi_1), \ \phi_1 \in {\cal C}_{g_1}, \phi_2 \in {\cal C}_{g_2}$. 
\item[(12)]\ 
$\imath_{g,a}(\phi) (\rho_g \psi) = \psi \imath_{g,a}(\phi), \ \phi \in {\cal C}_g, \psi \in {\cal O}_{aa}$. 
\item[(13)]\ 
$\theta_{a} ( \psi \imath_{g^{-1},a}(\phi) ) = \theta_{\cal C} ( \imath^{g,a}(\psi) \phi ), \ \phi \in {\cal C}_{g^{-1}}, \psi \in {\cal O}_{aa}$. 
\item[(14)]\ ($G$-equivariant Cardy condition) 
$\pi_{g,b}^a = \imath_{g,b} \circ \imath^{g,a} \mbox{ with } \pi_{g,b}^a(\psi) = \sum_{\mu} \psi^{\mu} \psi (\rho_g \psi_{\mu}),\  g \in G_0$. 
\end{itemize}
For unoriented  open and closed TFTs, 
one can find the following additional constraints:
\begin{itemize}
\item[(15)]\ 
$\rho_g (\psi_1 \psi_2) = (\rho_g \psi_2) (\rho_g \psi_1), \  g \notin G_0, \psi_1 \in {\cal O}_{ab}, \psi_2 \in {\cal O}_{bc}$. 
\item[(16)]\ 
$\sum_{\mu} (\rho_g \psi_{\mu} ) \psi^{\mu} = \imath_{g^2,a} (\theta_g), \ \ g \notin G_0, \ \psi_{\mu}, \psi^{\mu} \in {\cal O}_{aa}$.
\end{itemize}
In Appendix \ref{app:open TFT}, we summarize the derivations 
of these constraints. 

By solving these constraints, we can determine the 
general properties of the target algebraic category for a 
given $G$-equivariant closed TFT $b \in Z^2_G(X,U(1)_{\phi})$ 
with $G$-set $X$. 
In the cases where ${\cal C}_1$ is semisimple, i.e., 
combination of symmetry breaking and symmetry fractionalization, 
and there is no orientation-reversing symmetry, 
Moore-Segal~\cite{Moore-Segal} gives the complete solution: 
$b$-twisted equivariant vector bundles over $X$. 
Here, for simplicity, we assume the ground state of closed chain is unique, 
${\cal C}_1 \cong \C$, 
and there are only on-site symmetries $G_0$.
We have~\cite{Moore-Segal}
\begin{itemize}
\item 
The category of boundary conditions $\{a,b,\dots \}$ 
is equivalent to the category of $b$-projective 
representations $\{V_a, V_b, \dots \}$. 
\item 
${\cal O}_{ab} \cong {\rm Hom}(V_b^*,V_a^*) = V_a^* \otimes V_b$. 
\end{itemize}
This is precisely the boundary degrees of freedom that appear when one 
introduce a boundary in SPT phases. 
In the next section, 
we describe 
how to represent elements of ${\cal O}_{ab}$ and cobordisms 
by using simple and fixed point MPS for open chains. 


\subsubsection{Relation to open MPS}

In the SPT context, an interval $I_{ab}$ is 
identified with an open SPT phase with 
boundary condition $a$ and $b$. 
An element of $\psi \in {\cal O}_{ab} \cong V_a^* \otimes V_b$ 
is identified with a state of the open chain Hilbert space~\footnote{
Note that $a,b$ do not specify a state
in the representation space of the $b(g,h)$-projective representations. 
For example, the dihedral group $D_4 = \{1, C_4, C_2, C_4^{-1}, \sigma_x, \sigma_y, \sigma_d, \sigma_d'\}$ has two 
$b$-irreps $E_{\frac{1}{2}}$ and $E_{\frac{3}{2}}$ for the nontrivial two cocycle $[b] \in H^2(D_4,U(1)) \cong \Z_2$. 
In this case, $a, b$ specify $E_{\frac{1}{2}}$ or $E_{\frac{3}{2}}$. 
}
\begin{align}
\psi 
= \sum_m \big( v_L^T A_m v_R \big) \ket{m} 
= \sum_m [v_L]_i [A_m]_{ij} [v_R]_j \ket{m}, 
\quad 
v_L \in V_a^*, 
\quad 
v_R \in V_b. 
\end{align}
The correspondence between MPSs and  equivariant TFTs, 
can be pictorially represented as
\begin{align}
\vcenter{\hbox{\includegraphics[width=0.2\linewidth]{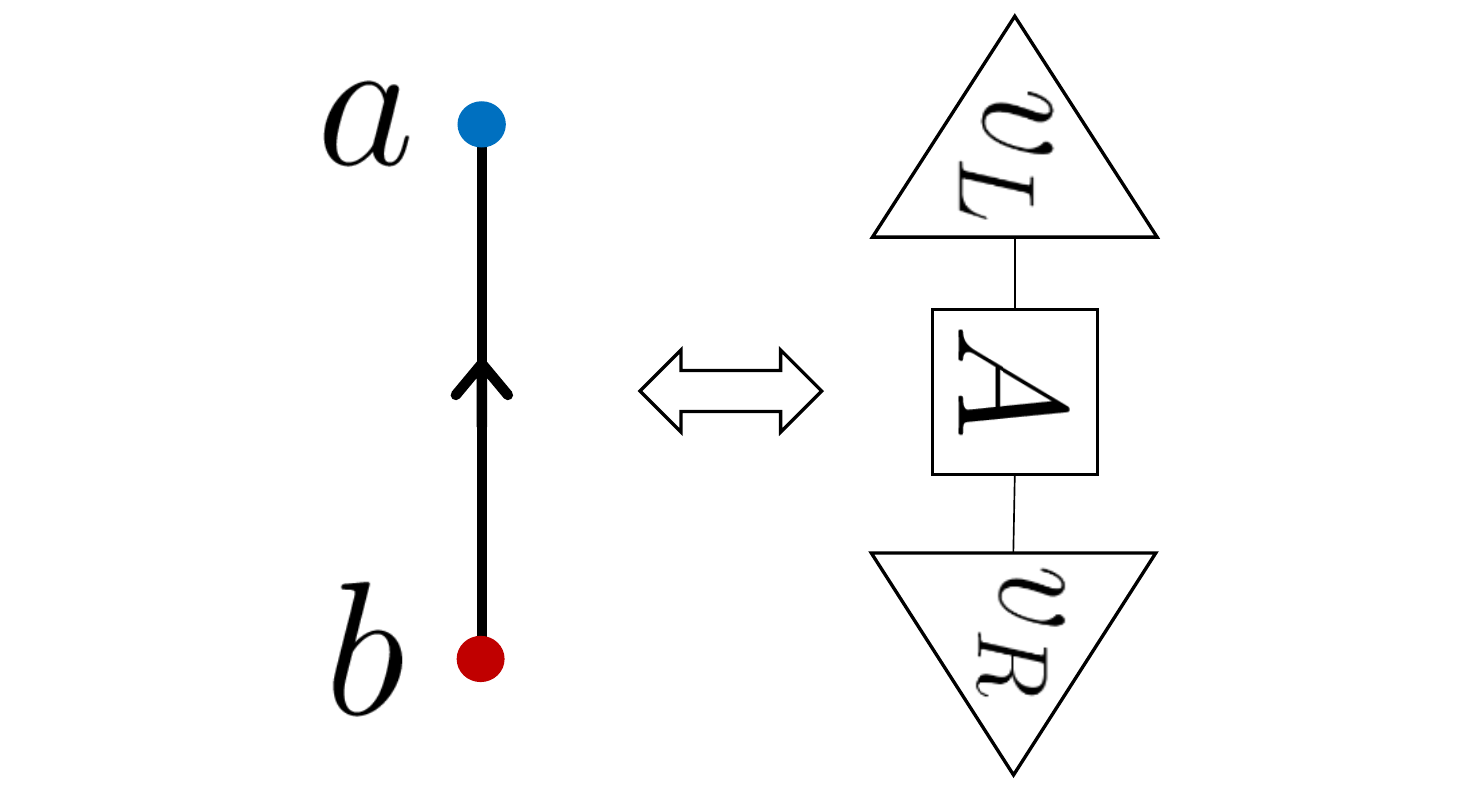}}}
\end{align}
Cobordisms in $G$-equivariant open and closed TFTs correspond to 
various ``adiabatic deformations'' of open chains and closed chains. 
The fourth column in Table \ \ref{Tab:Functor_Open} 
summarizes MPS representations, 
which satisfy algebraic constraints (9) - (16). 

For example, the fusion process of two 
open chains is represented in MPS networks as follows. 
For two open MPSs
\begin{align}
\psi_1 &= \sum_m v_L^T A^{(1)}_m v_R \ket{m}, \quad v_L \in V_a^*, v_R \in V_b, \\
\psi_2 &= \sum_m w_L^T A^{(2)}_m w_R \ket{m}, \quad w_L \in V_b^*, w_R \in V_c, 
\end{align}
the fusion $\psi_1 \psi_2$ is given by 
\begin{align}
\vcenter{\hbox{\includegraphics[width=0.4\linewidth]{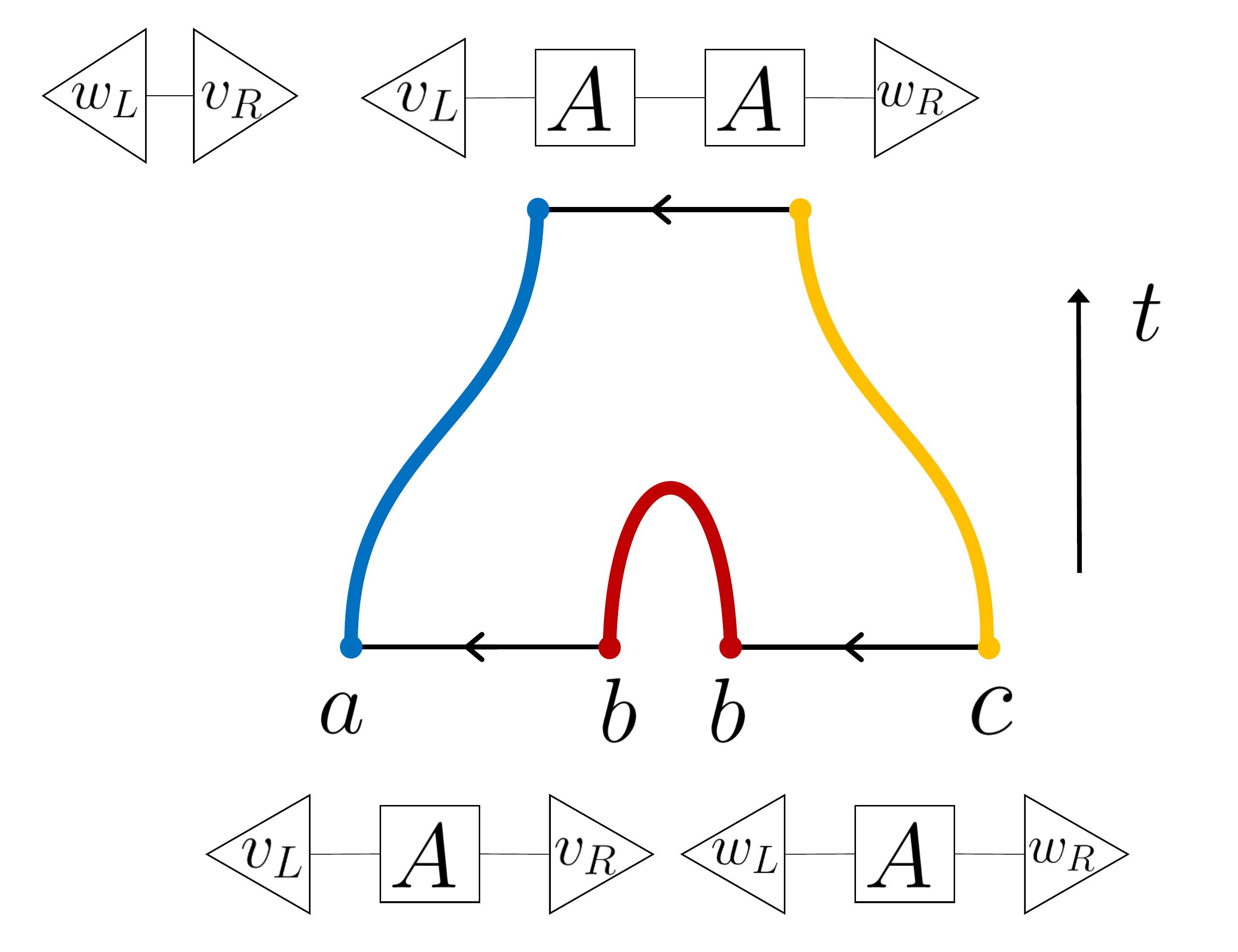}}}  && 
\Longrightarrow && 
\begin{array}{ll}
\psi_1 \psi_2  & = (w_L,v_R) \sum_{m_1 m_2} v_L^T A^{(1)}_{m_1} A^{(2)}_{m_2} w_R \ket{m_1 m_2} \\
& \\
 & \sim (w_L,v_R) \sum_{m} v_L^T A_{m} w_R \ket{m} \\
\end{array}
\end{align}
Here we introduced a notation $(w_L,v_R) = \sum_i [w_L]_i [v_R]_i$ 
and used an equivalence relation of fixed point MPSs (\ref{eq:mps_equivalence}).

\subsubsection{Equivariant Cardy conditions and boundary states}
\begin{figure}[!]
 \begin{center}
  \includegraphics[width=0.7\linewidth, trim=0cm 2cm 0cm 0cm]{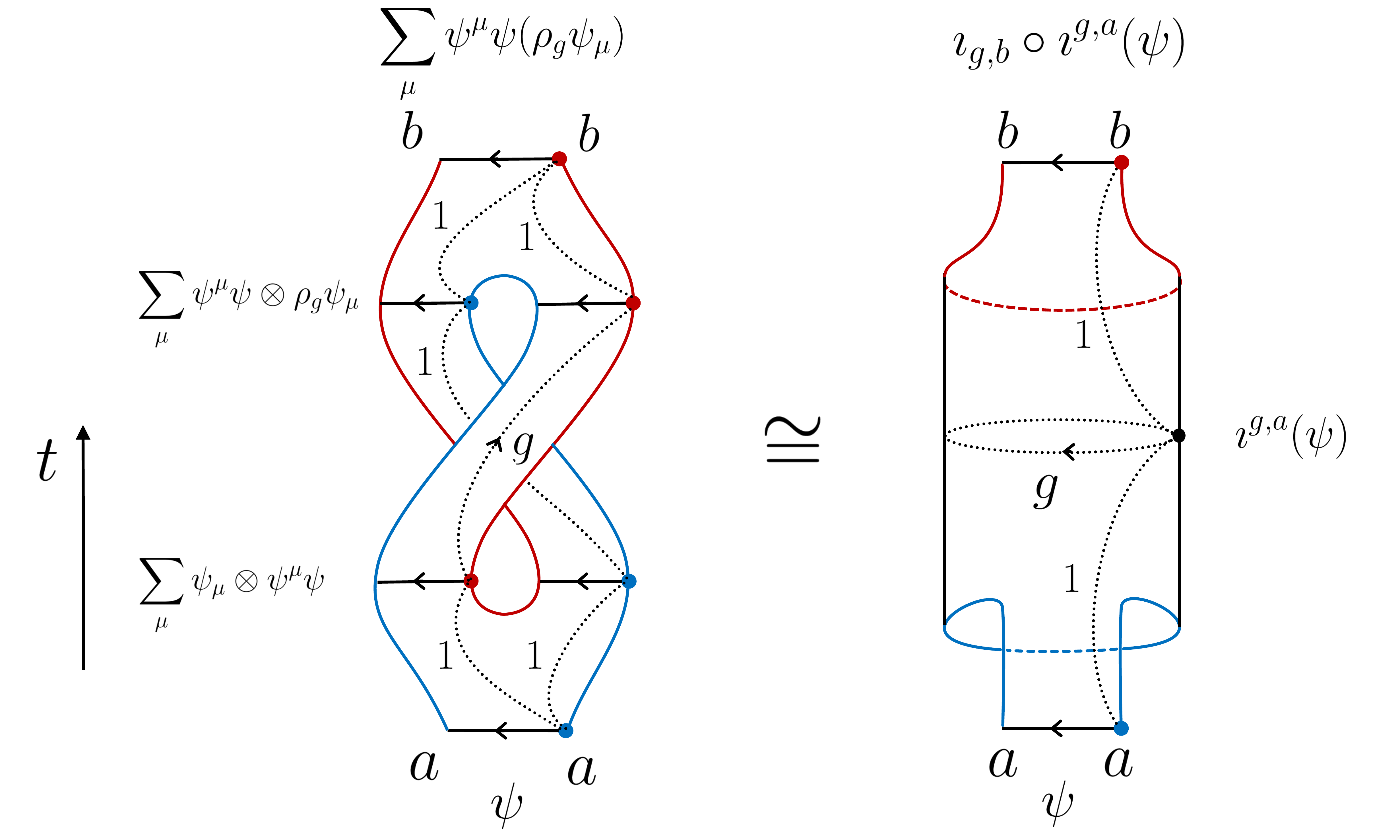}
 \end{center}
 \caption{
 $G$-equivariant Cardy condition.}
 \label{fig:G_Cardy}
\end{figure}

In the derivation of the category of boundary conditions by 
Moore-Segal~\cite{Moore-Segal}, 
the (generalized) $G$-equivariant Cardy condition (14) plays an essential role. 
Here, we show MPS representations listed in the fourth column 
in Table \ref{Tab:Functor_Open} satisfy the $G$-equivariant Cardy condition. 
The Cardy condition comes from the equivalence between 
(i) the double twist diagram shown in the left of Fig. \ref{fig:G_Cardy} 
and 
(ii) closed string channel shown in the right of Fig. \ref{fig:G_Cardy}. 
These diagram can be interpreted in the context of 
opne SPT chains : 
\begin{itemize}
\item[(i)]
Cutting an open SPT chain $I_{aa}$ into two open chains $I_{ab} \sqcup I_{ba}$ 
and taking the $G_0$-action on the left chain ${\cal O}_{ab}$ and 
exchanging two open chains and gluing back at $a$. 
This process is written as 
\begin{align}
\psi 
= \big( v_L^T A_m v_R \big) \ket{m} 
&\mapsto \sum_{j} \big( v_L^T A_{m_1} R_j \big) \ket{m_1} \otimes \big( L_j^T A_{m_2} v_R \big) \ket{m_2} 
\nonumber \\
&\mapsto \sum_{j} \big( v_L^T V_g^{\dag} A_{m_1} V_g R_j \big) \ket{m_1} \otimes \big( L_j^T A_{m_2} v_R \big) \ket{m_2} 
\nonumber \\
&\mapsto \sum_{j} \big( L_j^T A_{m_2} v_R \big) \ket{m_2} \otimes \big( v_L^T V_g^{\dag} A_{m_1} V_g R_j \big) \ket{m_1} 
\nonumber \\
&\mapsto (V_g^* v_L, v_R) \sum_{j} \big( L_j^T A_m R_j \big) \ket{m}.
\end{align}

\item[(ii)] 
Gluing the both ends of open SPT chain $I_{aa}$ to the $g$-twisted closed chain $(S^1,g)$ and cutting into the open chain $I_{bb}$. 
This process is expressed as 
\begin{align}
\psi 
= \big( v_L^T A_m v_R \big) \ket{m} 
&\mapsto (V_g^* v_L, v_R) {\rm tr} \big( A_m V_g \big) \ket{m} 
\mapsto (V_g^* v_L, v_R) \sum_{j} \big( L_j^T A_m R_j \big) \ket{m}. 
\end{align}
\end{itemize}

\begin{figure}[!]
 \begin{center}
  \includegraphics[width=1\linewidth, trim=0cm 2cm 0cm 0cm]{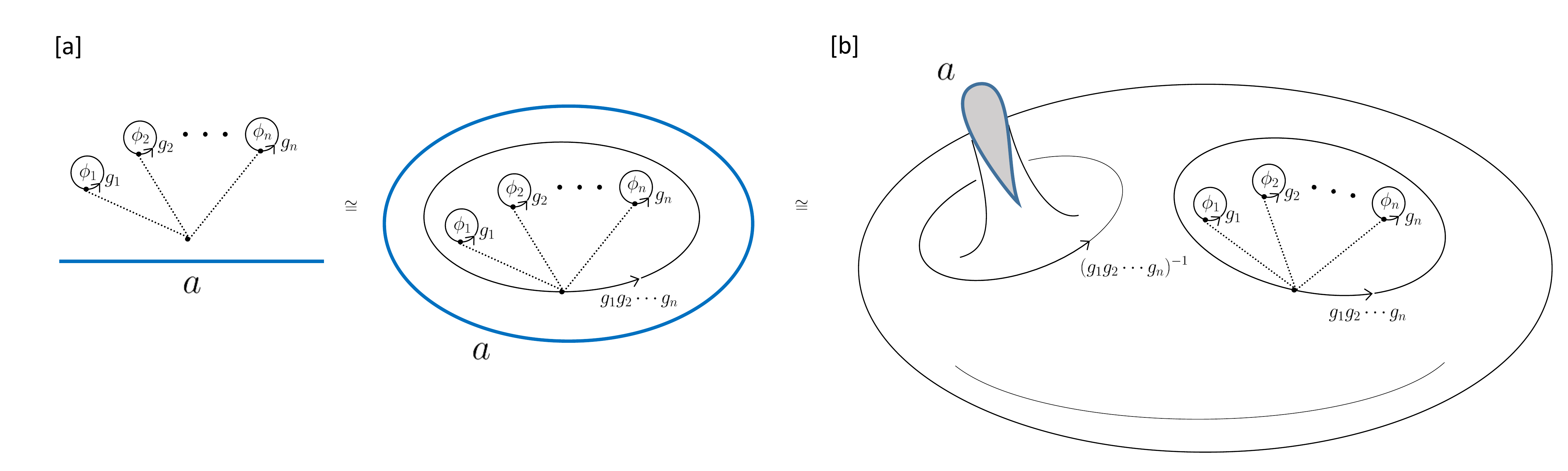}
 \end{center}
 \caption{
The definition of $G$-equivariant boundary state $B_{g,a}$. 
[a] Correlation functions on the upper half plane with boundary condition $a$. 
[b] Amplitude of closed sector with insertion of the boundary state. 
}
 \label{fig:bs_def}
\end{figure}

It is useful to introduce the equivariant boundary state 
$B_{g,a} \in {\cal C}_g$ in a way similar to 
usual boundary state $B_a$ for non-equivariant TFTs. 
Defining property of boundary state is that 
the correlation functions on upper half plane with boundary condition $a$ (Fig.~\ref{fig:bs_def} [a])
is the same as the closed string amplitude with insertion of the boundary state (Fig.~\ref{fig:bs_def} [b]): 
\begin{align}
\theta_a \big( \imath^{h,a}(\phi_1 \phi_2 \cdots \phi_n) \big) 
= \theta_{\cal C} \big( B_{h^{-1}, a} \phi_1 \phi_2 \cdots \phi_n \big), && 
\phi_i \in {\cal C}_{g_i}, h = g_1 g_2 \cdots g_n. 
\end{align}
From the algebraic constraint (13), 
the $G$-equivariant boundary state 
is given by 
the image of open to closed map on the unit element $1_a$ of the open chain ${\cal O}_{aa}$, 
\begin{align}
B_{g,a} 
:= \vcenter{\hbox{\includegraphics[width=0.1\linewidth]{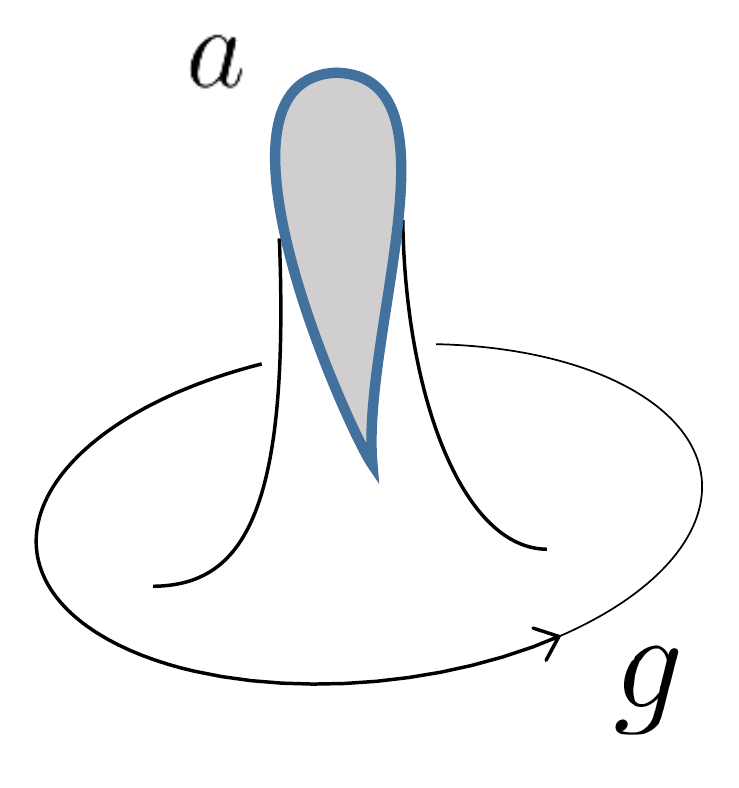}}}
= \imath^{g,a}(1_{a}) \in {\cal C}_g. 
\end{align}
For simple and fixed point MPSs, we have 
\begin{align}
B_{g,a} 
= {\rm tr} \big( V_{g,a}^{\dag} \big) {\rm Tr} \big( A_m V_{g,a} \big) \ket{m}, 
\end{align}
where $V_{g,a}$ is the representation matrix of the $V_a$ representation. 
Notice that $\chi_a(g)^* = {\rm tr} \big( V_{g,a}^{\dag} \big)$ 
is the character of $V_a$ representation, which is 
vacuous if there is a group element element $h \in G_0, [g,h]=0$ 
with nontrivial discrete torsion phase $b(g,h) \neq b(h,g)$.~\cite{cho2016relationship} 

If we insert the boundary states in the Cardy condition (14), 
we get a more familiar form 
\begin{align}
\Braket{B_{g,b} | B_{g,a}}
= \theta_{\cal C} \big( B_{g^{-1},b} B_{g,a} \big)
= \theta_b \big( \imath_{g,b} \circ \imath^{g,a} (1_a) \big) 
= \theta_b \big( \pi_{g,b}^a (1_a) \big) 
= {\rm Tr}_{{\cal O}_{ab}} (\rho_g), 
\end{align}
which is the character of $G$-action on the 
open chain Hilbert space ${\cal O}_{ab}$, 
\begin{align}
\chi_{{\cal O}_{ab}}(g) 
= {\rm Tr}_{{\cal O}_{ab}} (\rho_g)  
= \chi_a(g)^* \chi_b(g) . 
\end{align}

\subsubsection{Crosscap invariant in open chain}

The partition function on $\R P^2$, 
$Z_{\R P^2}(g) = \theta(g), g \notin G_0, g^2 = 1$,  
can be detected in open chains. 
Making use of the algebraic relation (16), 
one can find the M\"obius strip with boundary condition $a$ 
is equivalent to the closed string amplitude from 
crosscap $\theta_g$ to boundary state $B_{g^2,a}$, 
\begin{align}
&\vcenter{\hbox{\includegraphics[width=0.5\linewidth]{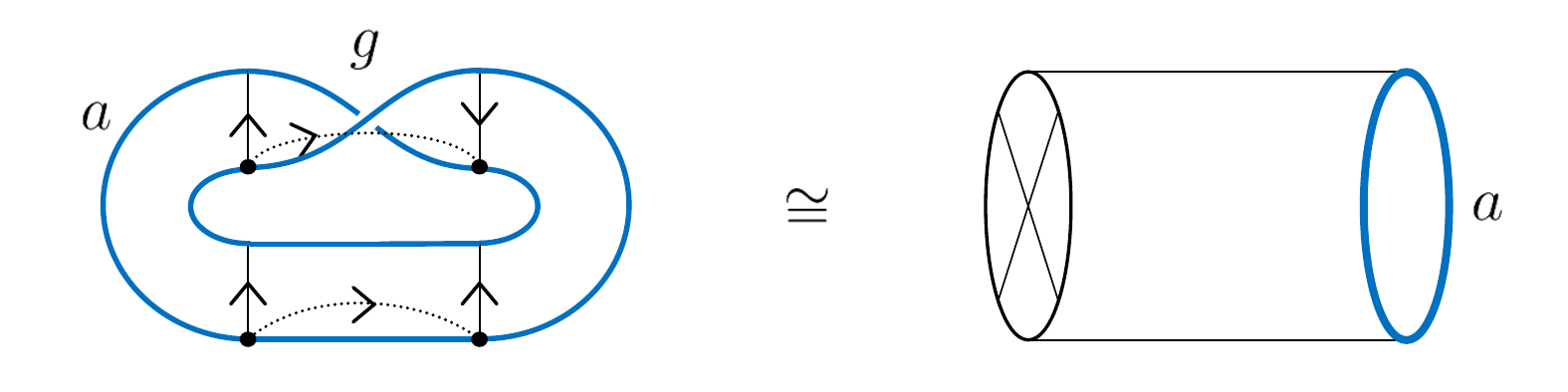}}} 
\nonumber \\
&\Rightarrow \ 
\braket{B_{g^2,a} | \theta_g} 
= \theta_{\cal C}(B_{g^{-2},a} \theta_g) 
= \theta_a(\imath_{g^2,a}(\theta_g)) 
= {\rm Tr}_{{\cal O}_{aa}} (\rho_g) , \ \  g \notin G_0. 
\end{align}
For $g^2 = 1$ and unique ground state ${\cal C}_1 \cong \C$, 
we have the topological invariant on the real projective plane, 
which can be confirmed in simple and fixed point MPS as 
\begin{align}
\theta_{a} \Big( \sum_{ij} \big( R_j^T V_g^{\dag} A_{m_1} V_g L_i \big) \ket{m_1} \cdot \big( L_j^T A_{m_2} R_i \big) \ket{m_2} \Big) 
= [V_g]_{ji} [V_g^{\dag}]_{ij}
= {\rm dim}(V_a) \ \theta(g).  
\end{align}

\subsection{State sum construction}
\label{Fukuma-Hosono-Kawai state sum construction}

In this section, we discuss the so-called state sum construction of TFTs. 
Compared with the axiomatic approaches discussed previously,  
the state sum construction exploits specific discretizations (triangulations)
of spacetime. 
We will first review this construction for standard (non-equivariant) TFTs
following Fukuma-Hosono-Kawai \cite{FHK}.
We will then consider the state sum construction of $G$-equivariant TFTs,
and compute, among others, the partition functions on the torus, Klein bottle, and real
projective plane. 
As promised earlier,  we will confirm that they match precisely with the
topological (SPT) invariants derived from MPSs. 

\subsubsection{Fukuma-Hosono-Kawai state sum construction}

\begin{figure}[t]
 \begin{center}
  \includegraphics[width=0.5 \linewidth, trim=0cm 0cm 0cm 0cm]{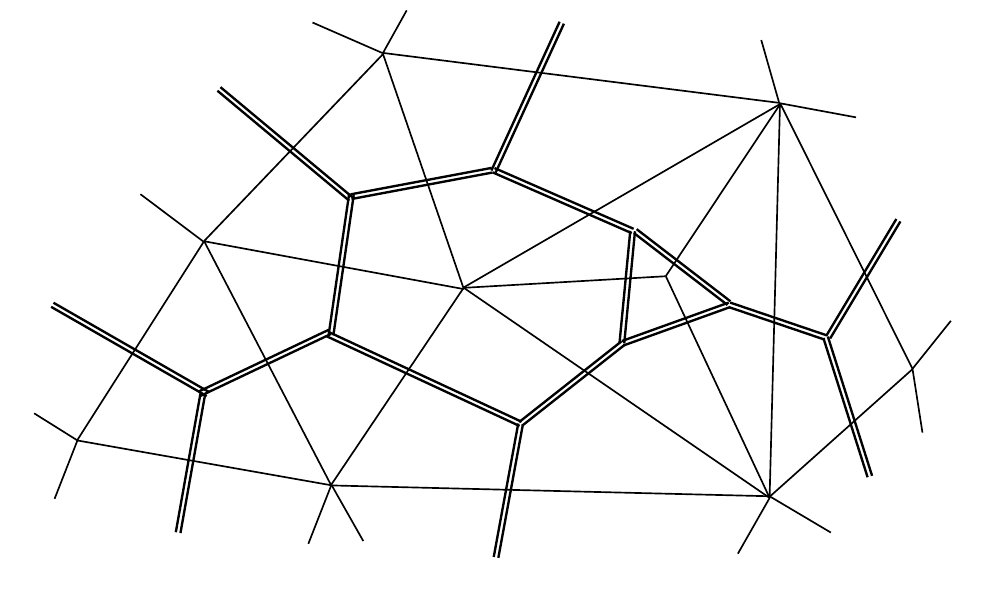}
 \end{center}
 \caption{Triangulation of two-dimensional spacetime 
 and its dual (represented by double lines).}
 \label{fig:state_sum/triangulation}
\end{figure}

Let us start by briefly reviewing the Fukuma-Hosono-Kawai 
state sum construction.
\cite{FHK}
In the state sum construction of oriented 2d TFTs,
we consider a triangulation of 2d spacetime.
For a given triangulation, we can consider its dual, the dual
triangulation -- see Fig.\ \ref{fig:state_sum/triangulation}. 
For faces and edges of the triangulation,
we associate 
$\C$ numbers $C_{\mu \nu \rho}$ and $g^{\mu \nu}$ $(\mu,\nu,\rho = 1, \dots, N)$ as 
\begin{align}
&
\vcenter{\hbox{\includegraphics[width=0.15\linewidth]{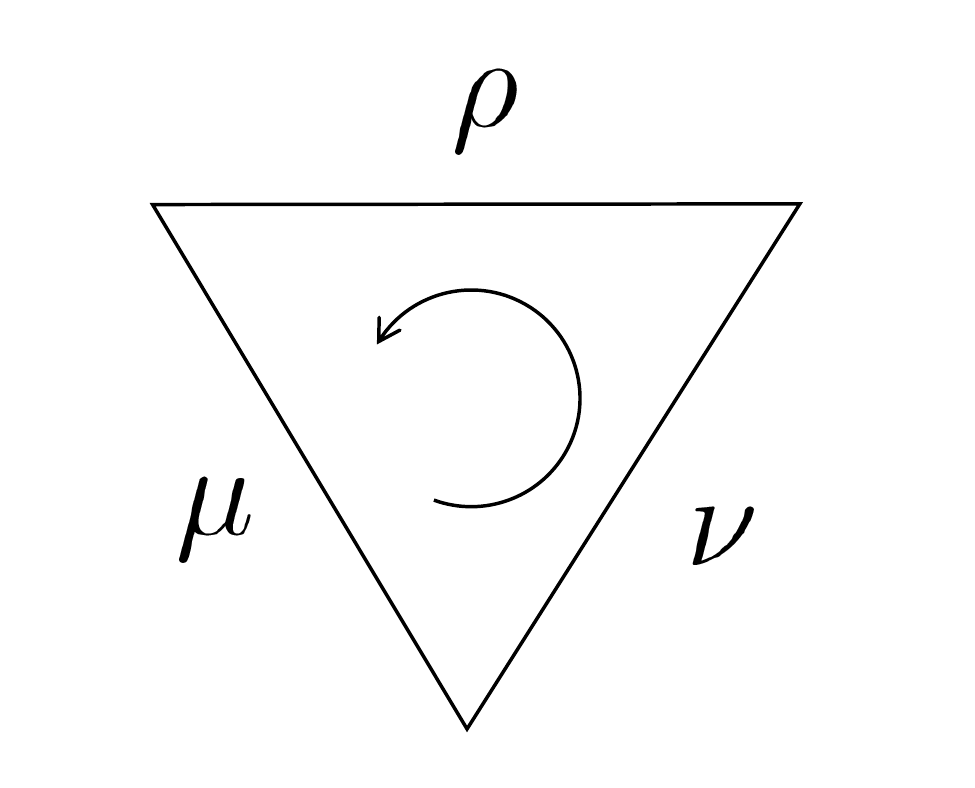}}} \ 
= \vcenter{\hbox{\includegraphics[width=0.15\linewidth]{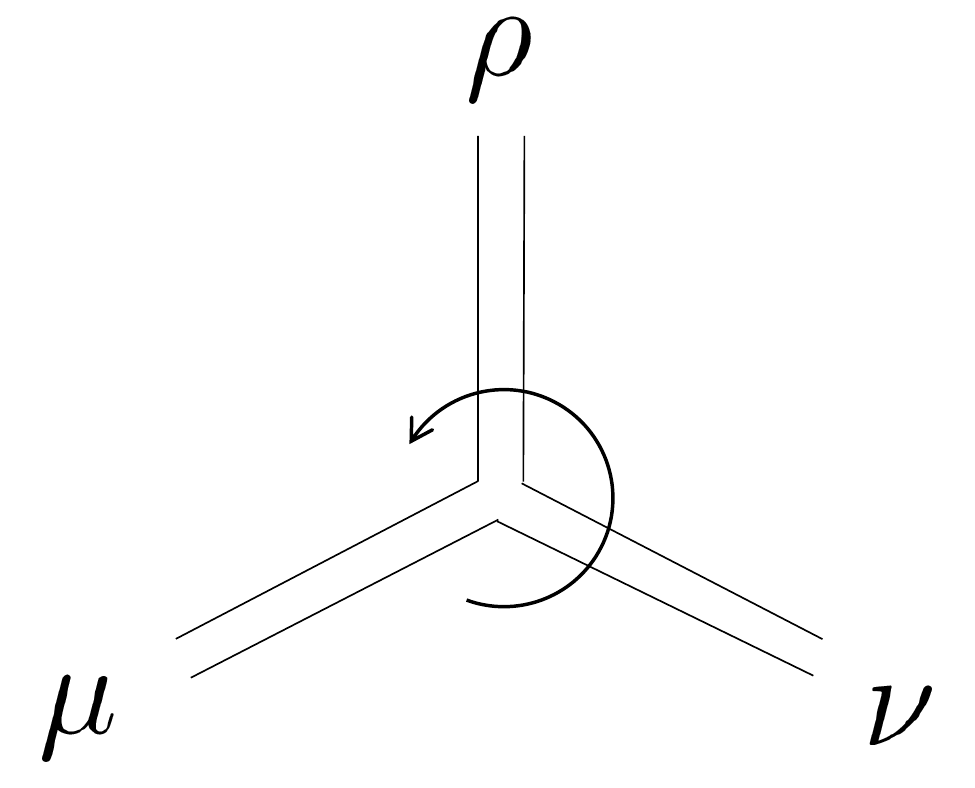}}} \ 
= \ C_{\mu \nu \rho}, 
\\
&
\vcenter{\hbox{\includegraphics[width=0.1\linewidth]{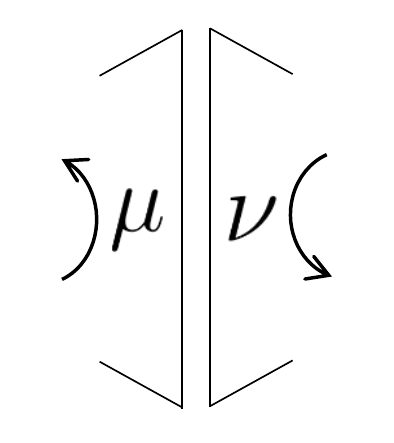}}} \ 
= \vcenter{\hbox{\includegraphics[width=0.15\linewidth]{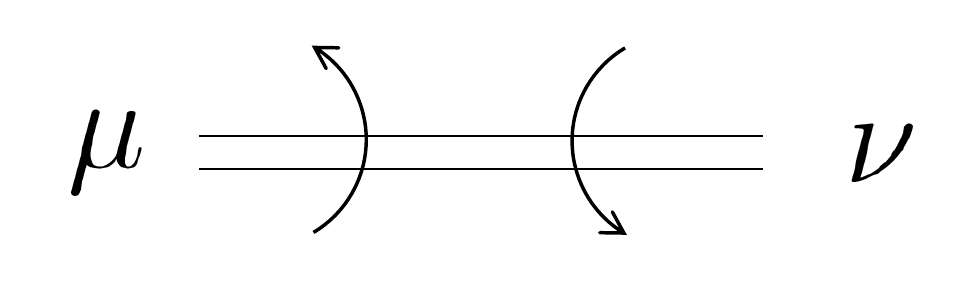}}} \ 
= \ g^{\mu \nu}. 
\end{align}
We demand that 
$C_{\mu \nu \rho}$ is cyclically symmetric 
$C_{\mu \nu \rho} = C_{\nu \rho \mu} = C_{\rho \mu \nu}$, 
and $g^{\mu \nu}$ is symmetric $g^{\mu \nu} = g^{\nu \mu}$. 
$g_{\mu \nu}$ is defined as the inverse of $g^{\mu \nu}$, 
$g^{\mu \nu} g_{\nu \rho} = \delta^{\mu}_{\rho}$. 
$g_{\mu \nu}$ and $g^{\mu \nu}$ are used for raising and lowering indices. 
For example,
we introduce ${C_{\mu \nu}}^{\rho} = C_{\mu \nu \sigma} g^{\sigma \rho}$. 
For a given triangulation $\Sigma_T$ 
of a surface $\Sigma$, 
the partition function on $\Sigma_T$ 
is given by
\begin{align}
Z(\Sigma_T) = \sum_{\rm faces} C_{\mu \nu \rho} \sum_{\rm edges} g^{\eta \epsilon}. 
\end{align}

In order to make $Z(\Sigma_T)$ independent of triangulations, 
$C$ and $g$ have to satisfy the fusion and bubble conditions
\begin{align}
\label{eq:fusion_cond}
&\vcenter{\hbox{\includegraphics[width=0.3\linewidth]{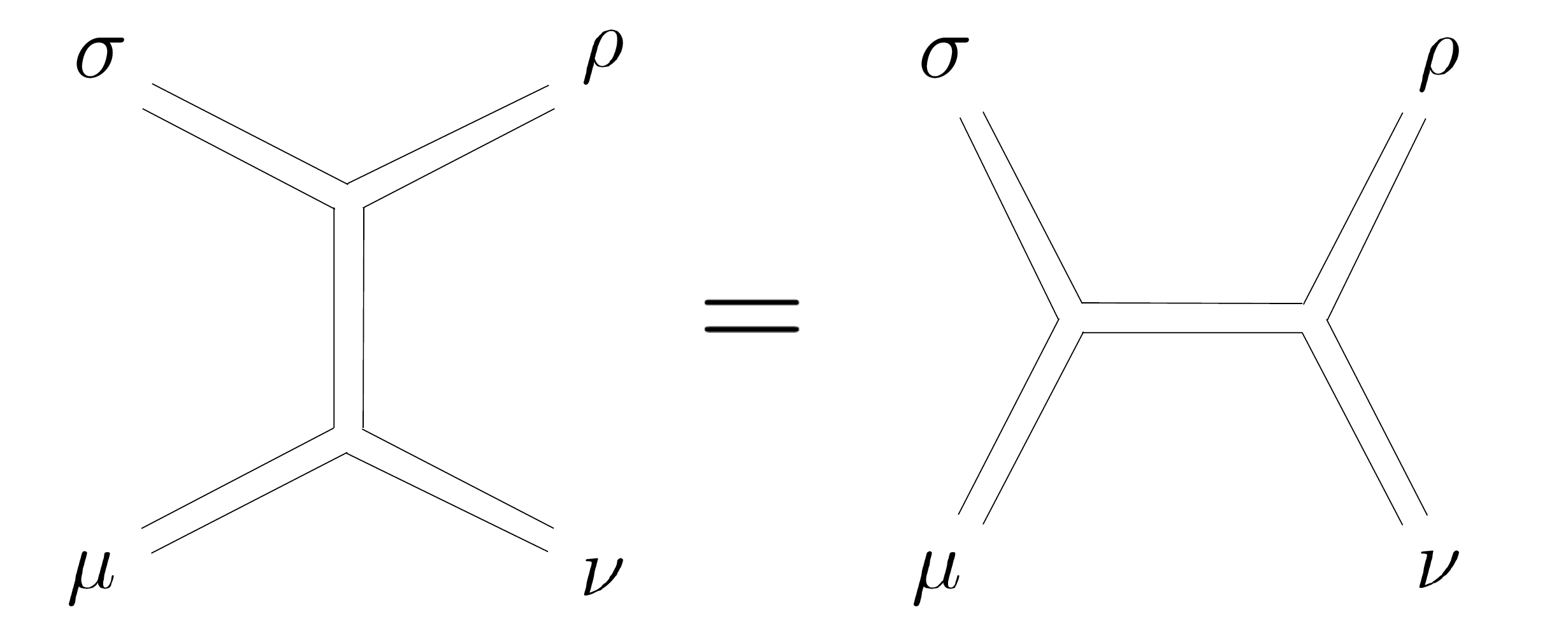}}} : \qquad 
{C_{\mu \nu}}^{\eta} {C_{\eta \rho}}^{\sigma} = {C_{\nu \rho}}^{\eta} {C_{\mu \eta}}^{\sigma},
\\
\label{eq:bubble_cond}
&\vcenter{\hbox{\includegraphics[width=0.3\linewidth]{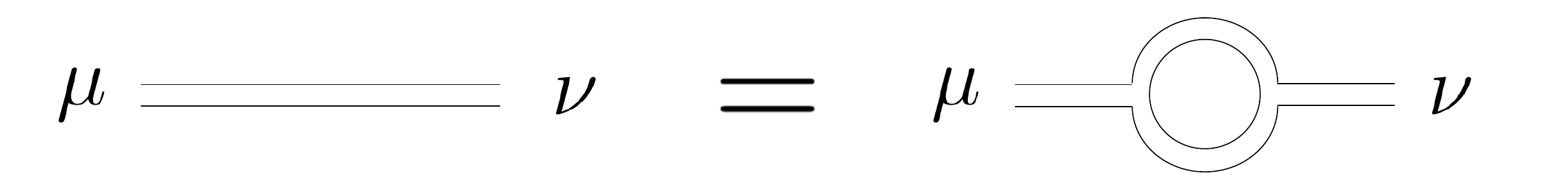}}} : \qquad 
g_{\mu \nu} = {C_{\mu \rho}}^{\sigma} {C_{\nu \sigma}}^{\rho}. 
\end{align}

From the data of $C_{\mu \nu \rho}$ and $g^{\mu \nu}$, 
one can introduce an algebra ${\cal C} = \oplus_{\mu=1}^N \C \phi_{\mu}$ 
as 
\begin{align}
\phi_{\mu} \phi_{\nu} = C_{\mu \nu \rho} \phi_{\rho}. 
\end{align}
The fusion condition (\ref{eq:fusion_cond}) means ${\cal C}$ is associative 
$(\phi_{\mu} \phi_{\nu}) \phi_{\rho} = \phi_{\mu} (\phi_{\nu} \phi_{\rho})$. 
We define a bilinear form by $Q(\phi_{\mu}, \phi_{\nu}) = g_{\mu \nu}$. 
Existence of inverse of $g_{\mu \nu}$ ensures that $Q$ is non-degenerate and 
the algebra ${\cal C}$ is semi simple. 
The cyclicity condition of $C_{\mu \nu \rho}$ 
leads to the Frobenius condition 
$Q(\phi_{\mu} \phi_{\nu}, \phi_{\rho}) = Q(\phi_{\mu}, \phi_{\nu} \phi_{\rho})$, 
i.e.,\ ${\cal C}$ is a semi simple Frobenius algebra. 

One can show that all the physical observables 
constructed from the data $C_{\mu \nu \rho}$ and $g^{\mu \nu}$ 
depend only on the 
center of ${\cal C}$, 
$Z({\cal C}) = \{\phi \in {\cal C} | \phi \phi' = \phi' \phi , \forall \phi' \in {\cal C} \}$.~\cite{FHK}
In other words, 
the Fukuma-Hosono-Kawai state sum construction 
describes 2d oriented TFTs 
which are equivariant to commutative semisimple Frobenius algebras. 
For example, for a matrix algebra ${\cal C} = {\rm Mat}(\C^N)$ with 
$Q(A,B) := {\rm Tr} A B$, 
the center is trivial: $Z({\rm Mat}(\C^N)) = \C 1_{N \times N}$.



\subsubsection{$G$-equivariant state sum construction}
The state sum construction of $G$-equivariant closed TFTs
(both oriented and unoriented)
can be formulated in a way analogous to  
the Fukuma-Hosono-Kawai construction of 2d oriented TFTs.~\cite{Turaev}
In the following, we 
will discuss this within the context of TFTs describing SPT phases.   

As before,
let $G$ be a symmetry group 
which possibly includes orientation-reversing symmetries. 
We specify orientation-preserving elements by subgroup $G_0 \subset G$. 
We fix a group 2-cocycle $b(g,h) \in Z^2(G,U(1)_{\phi})$ and 
assume $[b(g,h)] \in H^2(G,U(1)_{\phi})$ is nontrivial. 
Let $V$ be a $b$-projective $N$-dimensional irrep.\ and $V^*$ be its dual. 
Recall that $V$ represents the ``bond Hilbert space'' in MPSs.
$V$ also play an analogous role in the state sum construction,
which will be developed in the following.
The $G$ symmetry is projectively represented in the bond Hilbert space as 
\begin{align}
&\hat g (\ket{i}) 
= \ket{j} [V_g]_{ji}, 
\quad V_g V_h = b(g,h) V_{gh}, 
\quad 
\ket{i}, \ket{j} \in V, 
\end{align}
in the same way as Sec.\ \ref{Fixed point MPSs}. 

As in the Fukuma-Hosono-Kawai construction, we need 
the input data -- the Frobenius algebra -- to boot-strap  
a $G$-equivariant TFT. 
To describe SPT phases (i.e., invertible TFTs), 
we take the matrix algebra of $V$ 
as the algebra ${\cal C}$,
${\cal C} := {\rm End}(V) \cong V \otimes V^*$.
The bilinear non-degenerate form is defined 
by the matrix trace $Q(X,Y) = N {\rm tr} (X Y) = N \sum_{ij} X_{ij} Y_{ji}$. 

A canonical basis of ${\cal C}$ can  be given as 
\begin{align}
\left\{ E_{ij} = \ket{i} \bra{j} \right\}. 
\end{align}
In this basis, $C_{ij, kl, mn} = C(E_{ij}, E_{kl}, E_{mn})$ and $g = g^{ij,kl} E_{ij} \otimes E_{kl}$ 
are given by  
\begin{align}
C_{ij,kl,mn}& = 
N \ 
\vcenter{\hbox{\includegraphics[width=0.15\linewidth]{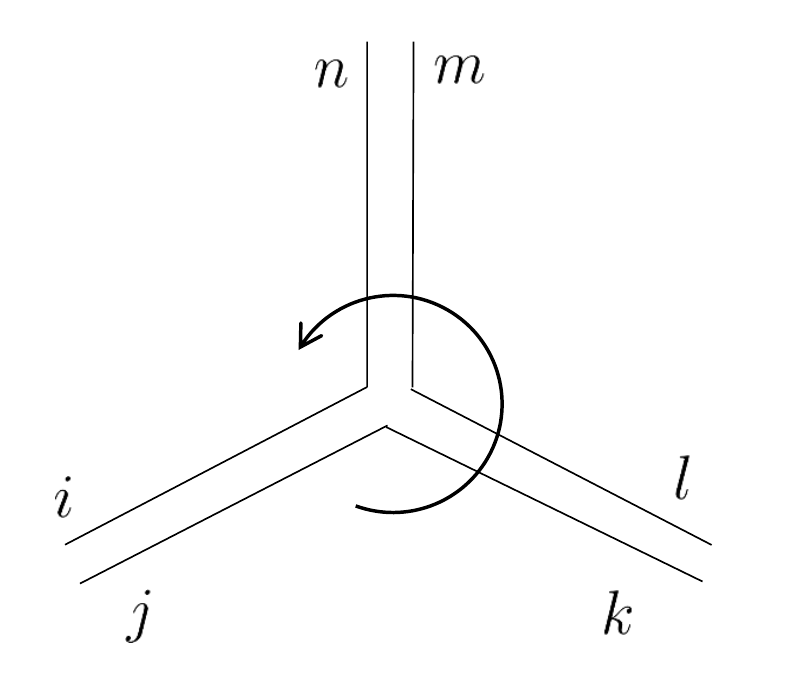}}} \ 
= N \delta_{ml} \delta_{kj} \delta_{in}, 
\\
g^{ij,kl} &= 
\frac{1}{N} \ \vcenter{\hbox{\includegraphics[width=0.15\linewidth]{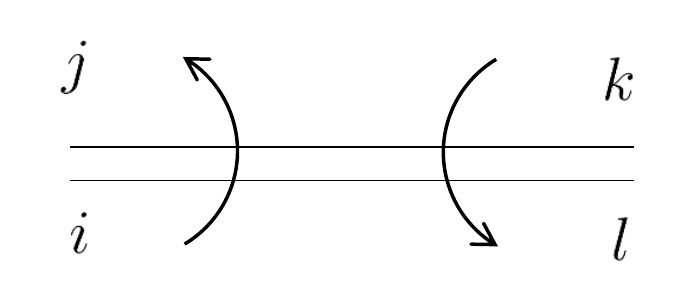}}} \
=\frac{1}{N} \delta^{il} \delta^{kj}. 
\end{align}
One can show 
$Q(E_{ij},E_{kl}) = N g_{ij,kl} = N \delta_{il} \delta_{kj}$, 
${C_{ij,kl}}^{mn} = \delta_{kj} \delta_i^m \delta_l^n$, 
$E_{ij} E_{kl} = {C_{ij,kl}}^{mn} E_{mn}$, 
and can check 
(\ref{eq:fusion_cond}), 
and (\ref{eq:bubble_cond}). 

Form the construction, 
the algebra ${\cal C}$ has $G$ action 
\begin{align}
&\hat g (X) = V_g X V^{\dag}_g, \quad X \in {\cal C}, g \in G_0, \\
&\hat P (X) = V_P X^T V^{\dag}_P, \quad X \in {\cal C}, P \notin G_0. 
\end{align}
Here, observe that the orientation-reversing symmetry $P \notin G_0$ exchanges left and right. 
This $G$ action can be 
used to 
to incorporate the background $G_0$ gauge field 
in the networks of the state sum construction.
We introduce a symmetry twisted metric 
by
\begin{align}
[T_g]^{ij,kl} 
:= g^{ij,pq} Q\big( E_{pq}, \hat g(E_{rs}) \big) g^{rs,kl} 
= \frac{1}{N} [V_g]_{il} [V^{\dag}_g]_{kj}
= \vcenter{\hbox{\includegraphics[width=0.25\linewidth]{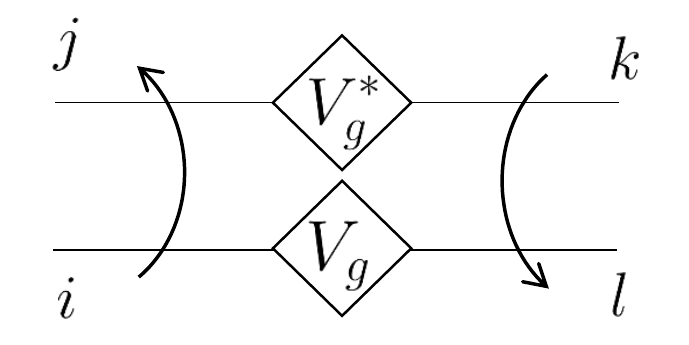}}} 
\quad (g \in G_0). 
\end{align}
We replace $g^{ij,kl}$ by $[T_g]^{ij,kl}$ on a nontrivial 1-cycle of the triangulation. 
On the other hand, an orientation reversing symmetry $g \notin G_0$ induces 
the exchange of indices $i$ and $j$. 
We introduce the orientation reversing twisted metric~\cite{Karimipour-Mostafazadeh} by 
\begin{align}
[T_P]^{ij,kl} 
:= g^{ij,pq} Q\big( E_{pq}, \hat P(E_{rs}) \big) g^{rs,kl} 
= \frac{1}{N} [V_P]_{ik} [V^{\dag}_P]_{lj} 
= \vcenter{\hbox{\includegraphics[width=0.3\linewidth]{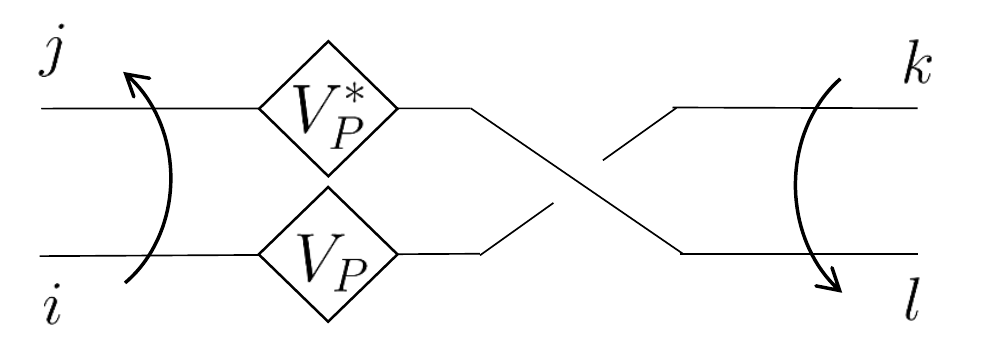}}} 
\quad  (P \notin G_0). 
\end{align}

\subsubsection{Partition functions}

Let us now construct, by using the state sum,
the partition functions on $T^2$, the Klein bottle, and $\mathbb{R}P^2$
(with symmetry twist).  
We will show that these match precisely with the
topological invariants discussed and constructed by using MPSs
in Sec.\ \ref{Topological invariants}.

\paragraph{Partition function on $T^2$ with twist}
A background $G_0$ gauge field 
on a torus $T^2$ is specified by 
two commuting elements $g,h \in G_0, [g,h] = 0, g,h \in G_0$. 
From the twisted metrics $T_g, T_h$ 
we have the torus partition function with twist 
\begin{align}
Z_{T^2}(g,h) 
&= \vcenter{\hbox{\includegraphics[width=0.2\linewidth]{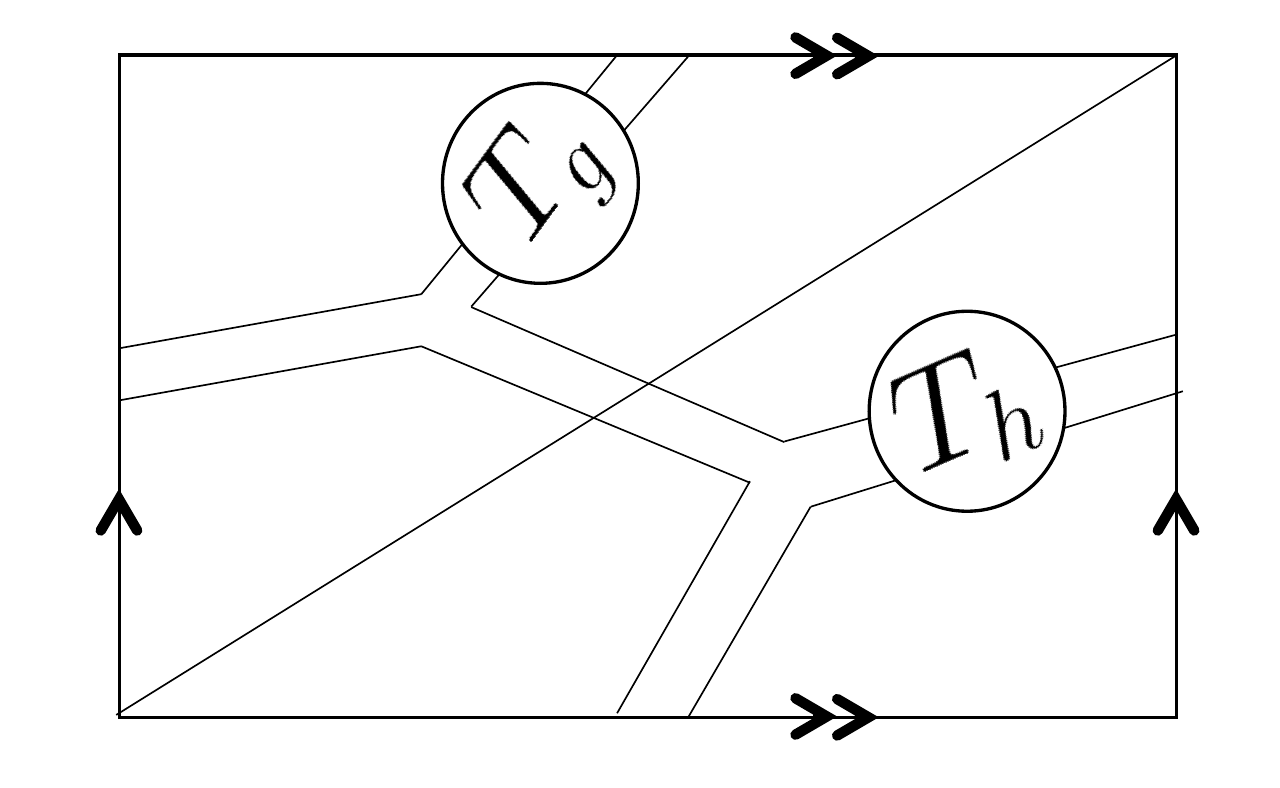}}} 
= {C_{ij,kl}}^{mn} C_{mn,pq,rs} [T_g]^{ij,pq} [T_h]^{rs,kl} \\ 
&= \frac{1}{N} {\rm tr}(V_g V_h V^{\dag}_g V^{\dag}_h) = \epsilon(g,h), \ \ 
(g,h \in G_0, h g h^{-1} = g). 
\end{align}
This is the discrete torsion phase (\ref{eq:T^2inv}), 
a topological invariant that characterizes $H^2(G,U(1)_{\phi})$. 

\paragraph{Partition function on the Klein bottle with twist}

Similar to the torus partition function with twist, 
the Klein bottle partition function with twist is computed in 
the state sum construction. 
Let $P \notin G_0$ be an orientation reversing symmetry and 
$g \in G_0$ be a orientation preserving symmetry. 
We have 
\begin{align}
Z_{KB}(P;g) 
&= \vcenter{\hbox{\includegraphics[width=0.2\linewidth]{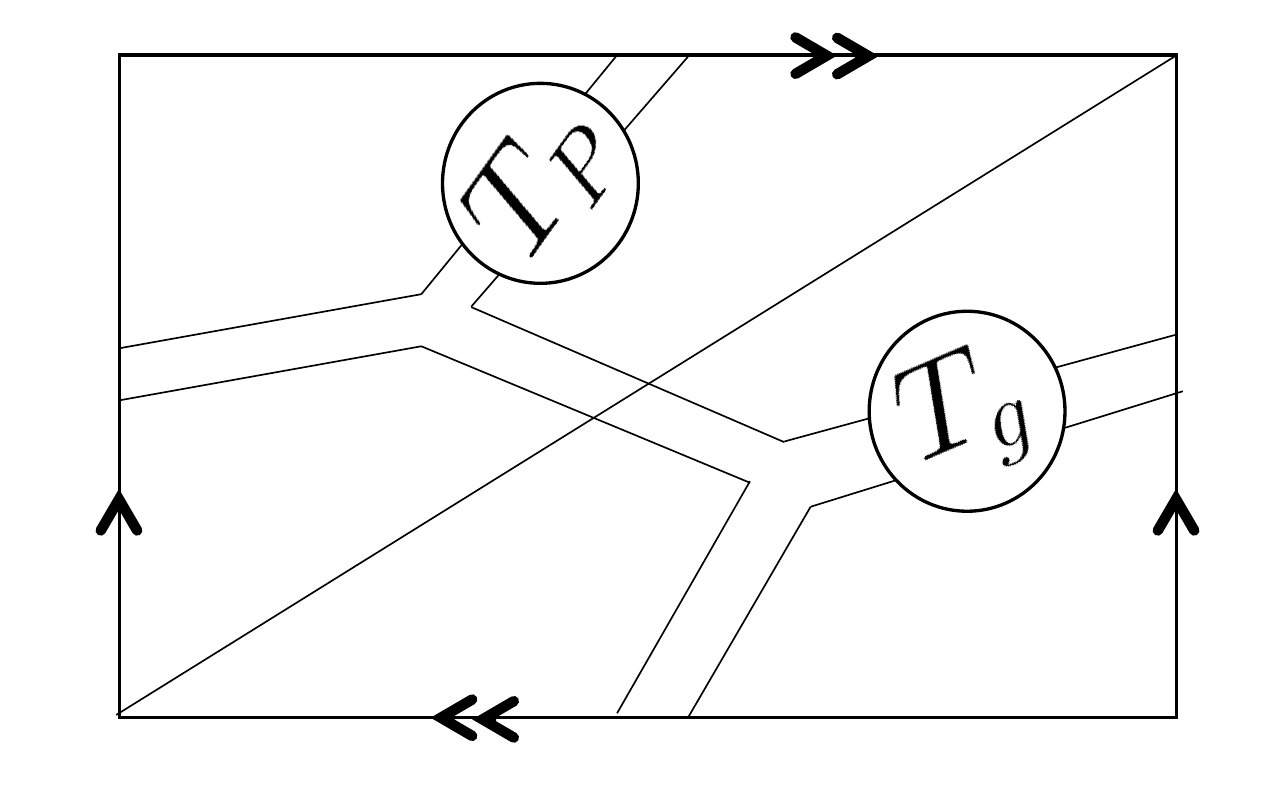}}} 
= {C_{ij,kl}}^{mn} C_{mn,pq,rs} [T_P]^{ij,pq} [T_g]^{rs,kl} \\ 
&= \frac{1}{N} {\rm tr}(V_P V_g^T V^{\dag}_P V^{\dag}_g) = \kappa(P;g) , \ \ 
(g \in G_0, P \notin G_0, P g^{-1} P^{-1} = g). 
\end{align}
Here, $\kappa(P;g)$ is the Klein bottle invariant of $H^2(G,U(1)_{\phi})$ 
introduced in (\ref{eq:KBinv}).

\paragraph{Partition function on $\R P^2$}
By using the orientation-reversing symmetry $P \in G_0$, 
we can construct the partition funciton on 
the real projective plane $\R P^2$ as 
\begin{align}
Z_{\R P^2}(P) 
&= \vcenter{\hbox{\includegraphics[width=0.2\linewidth]{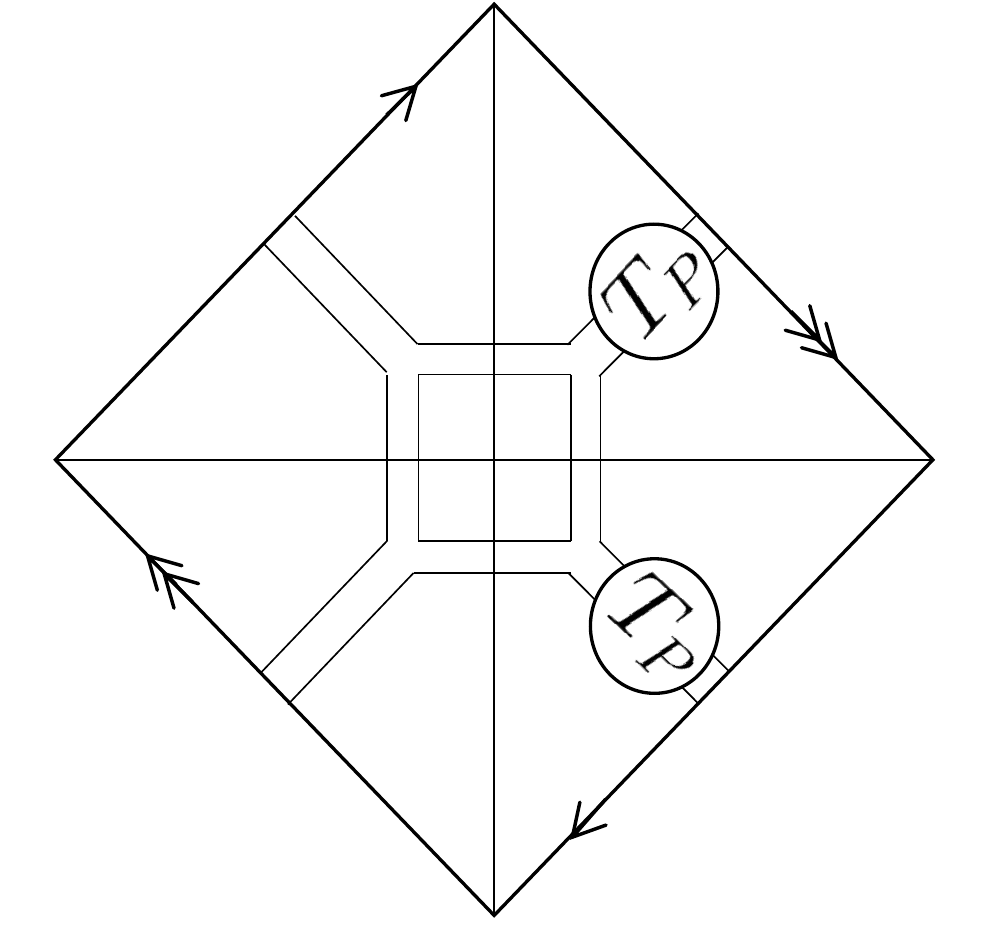}}} 
= \vcenter{\hbox{\includegraphics[width=0.2\linewidth]{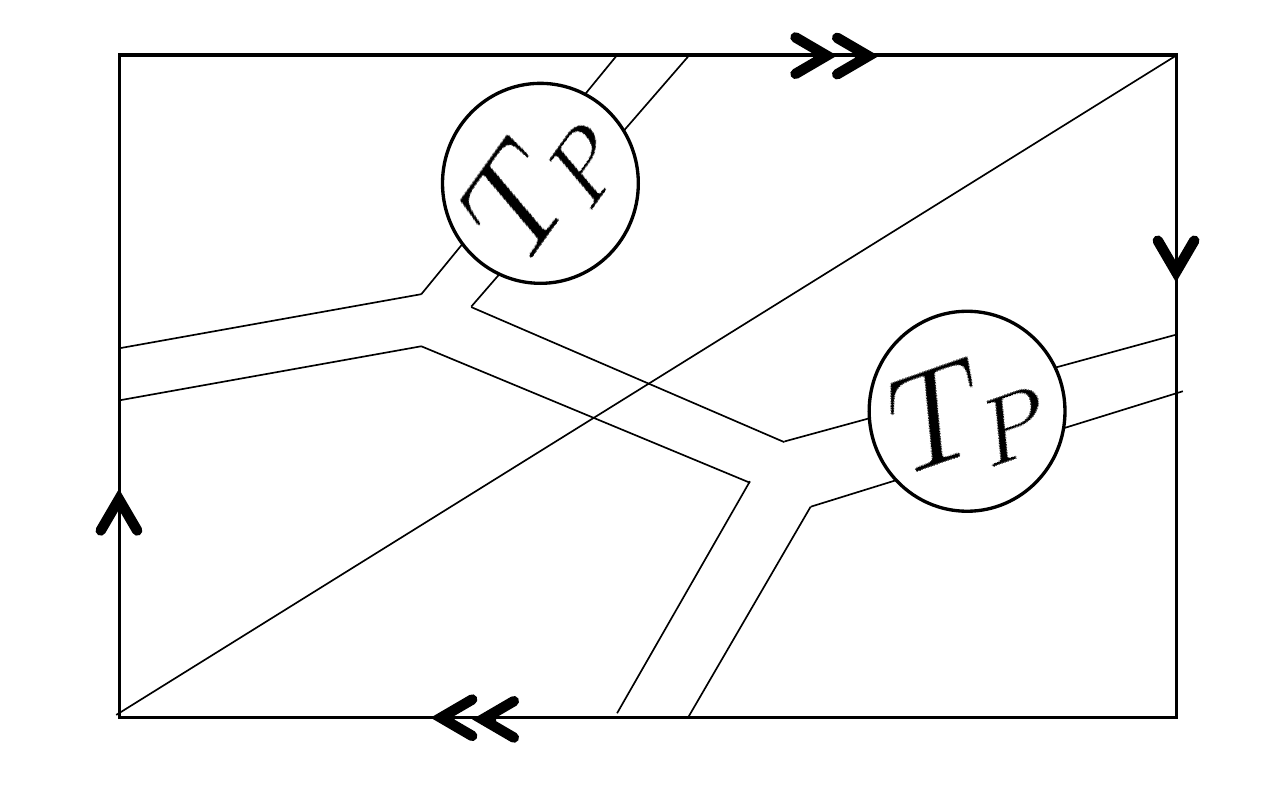}}} \\
&= {C_{ij,kl}}^{mn} C_{mn,pq,rs} [T_P]^{ij,pq} [T_P]^{rs,kl} 
= \frac{1}{N^2} {\rm tr}(V_P V_P^*) {\rm tr}(V_P V_P^{\dag}) = \theta(P), 
\quad
(P \notin G_0, P^2 = 1). 
\end{align}
This is the cross cap invariant (\ref{eq:Corsscapinv}).

\subsubsection{Cobordisms}

In addition to the closed surfaces considered above, 
we can also consider surfaces with boundaries by using 
the state sum construction. 
From the generalities of TFTs, 
a surface with boundary represents a state of the Hilbert space. 
Here, we will construct various states that can be obtained 
by considering state sum with open boundary/boundaries.
For our TFTs that describe SPT phases, 
the physical Hilbert space ${\cal C}$ is spanned by a 
basis of algebra $\{E_{ij}\}_{i,j = 1}^N$.

\paragraph{Disc (cap state)}

By the path integral on the disc,
we define a state associated to the disc (the cap state). 
The cap state is the vacuum state on untwisted sector.
By triangulating the disc, the path-integral can be evaluated 
explicitly as
\begin{align}
1_{\cal C}
&= \vcenter{\hbox{\includegraphics[width=0.15\linewidth]{figs/tft/disc}}} 
= \vcenter{\hbox{\includegraphics[width=0.15\linewidth]{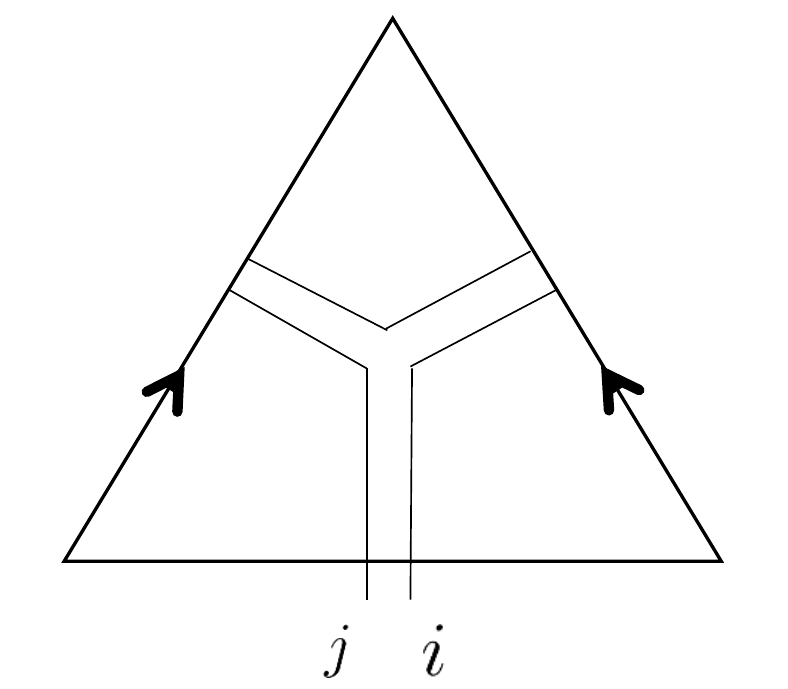}}} 
= \vcenter{\hbox{\includegraphics[width=0.1\linewidth]{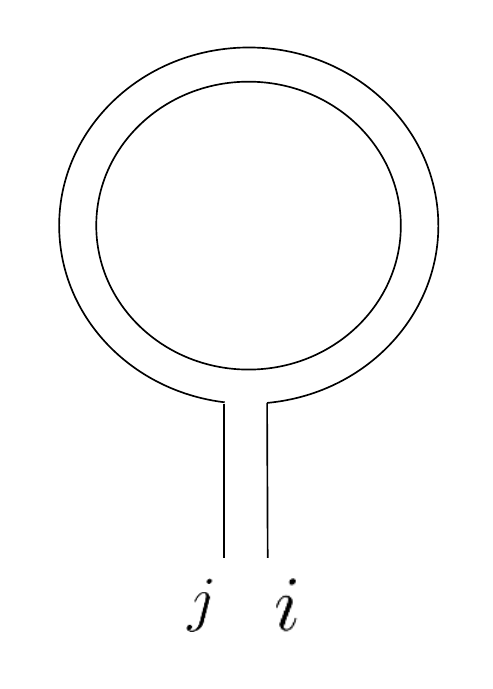}}} 
\nonumber \\
&
= {C_{kl,mn}}^{mn} g^{kl,ij} E_{ij} 
= \delta^{ij} E_{ij}
= \sum_i \ket{i} \bra{i}. 
\end{align}
This is nothing but the simple and fixed point MPS representation of the ground state of SPT phases 
$\ket{\Psi_1} = {\rm tr}(A_{ij}) \ket{i^L} \otimes \ket{j^R} = \frac{1}{\sqrt{N}} \sum_{i} \ket{i^L} \otimes \ket{i^R}$ 
introduced in Sec.\ \ref{Fixed point MPSs} up to a normalization.

\paragraph{M\"obius strip (cross cap state)}
By the path integral on the M\"obius strip,
we define a state associated to the M\"obius strip (the cross cap state). 
By triangulating the M\"obius strip, the path-integral can be evaluated 
explicitly as
\begin{align}
\theta_P
&= \vcenter{\hbox{\includegraphics[width=0.15\linewidth]{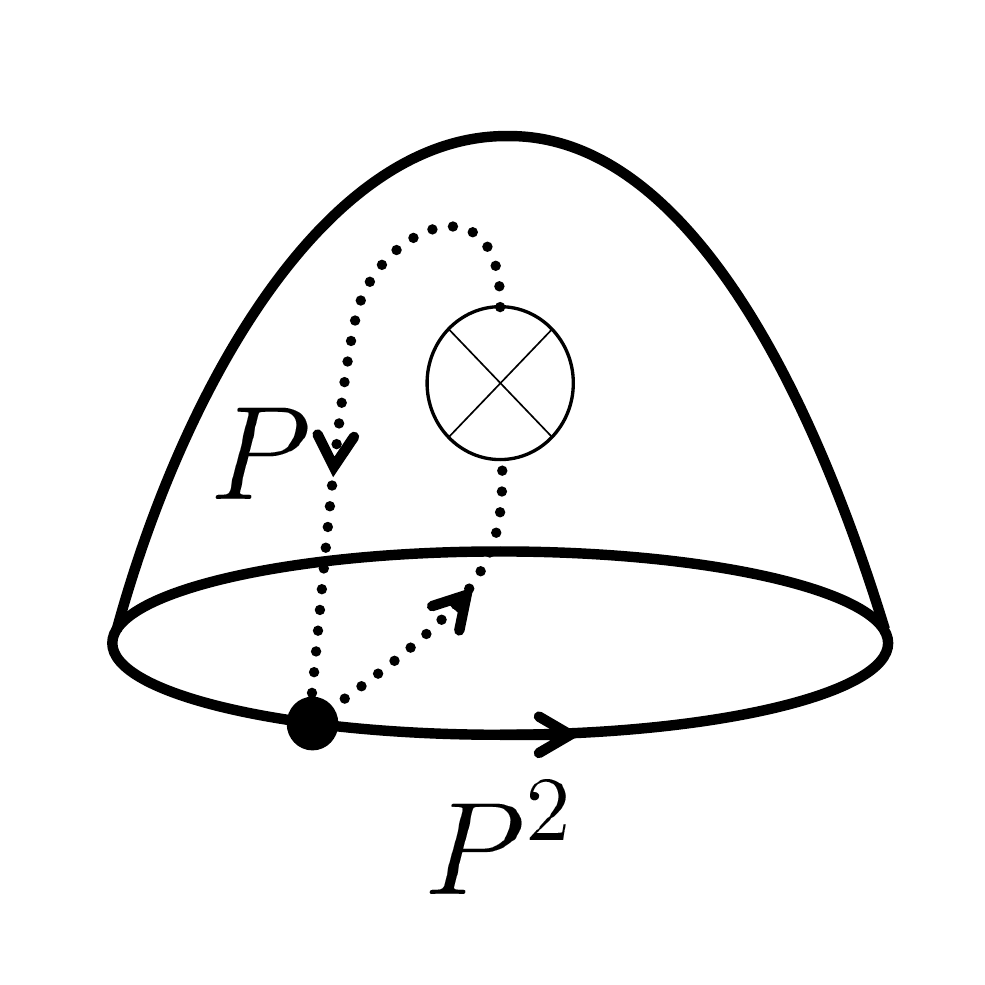}}} 
= \vcenter{\hbox{\includegraphics[width=0.15\linewidth]{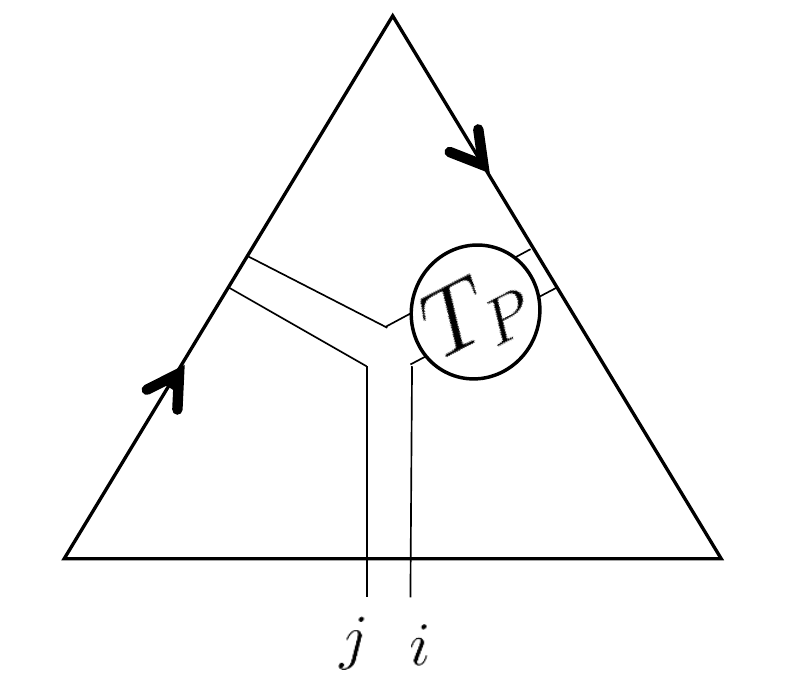}}} 
= \vcenter{\hbox{\includegraphics[width=0.1\linewidth]{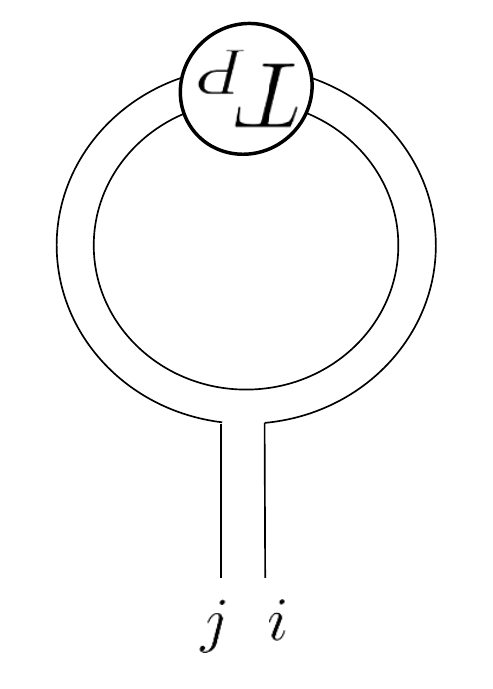}}} 
\nonumber \\
&
= {C_{mn,pq}}^{ij} [T_P]^{mn,pq} E_{ij} 
= \frac{1}{N} [V_P V_P^*]_{ij} E_{ij} 
= \frac{1}{N} b(P,P) \sum_{ij} [V_{P^2}]_{ij} \ket{i} \bra{j}, 
\quad
(P \notin G_0). 
\end{align}

\paragraph{Coform $\Delta_g$}

For a cylinder with two outgoing circles, following the axiom of TFTs,
we associate a coform $\Delta_g$. 
By triangulating the cylinder, 
we have 
\begin{align}
\Delta_g
&= \vcenter{\hbox{\includegraphics[width=0.2\linewidth]{figs/tft/coform}}} 
= \vcenter{\hbox{\includegraphics[width=0.2\linewidth]{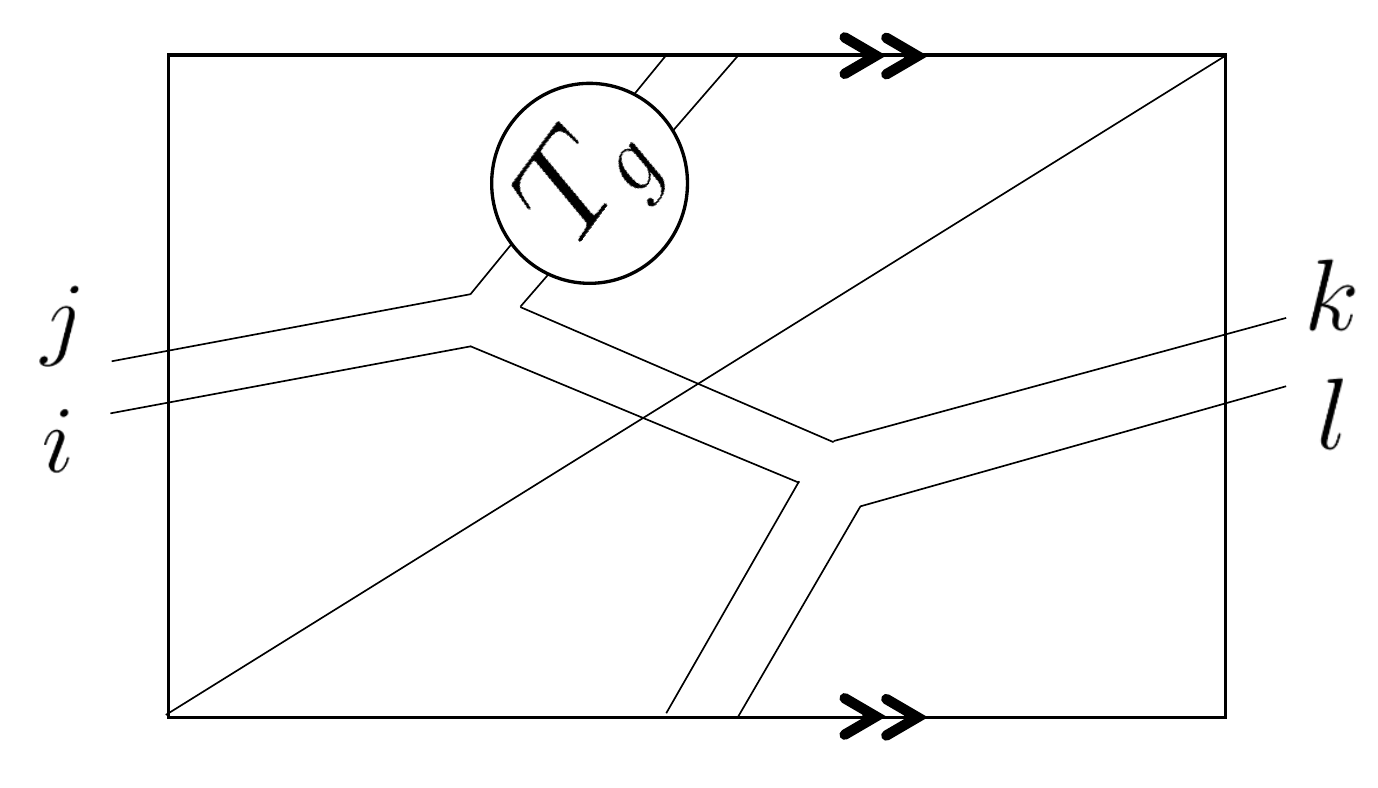}}}
\nonumber \\
&= {C^{ij}}_{mn,pq} {C^{kl,mn}}_{rs} [T_g]^{pq,rs} E_{ij} \otimes E_{kl} 
= \frac{1}{N^2} \sum_{ijkl} [V^{\dag}_g]_{ij} [V_g]_{kl} E_{ij} \otimes E_{kl}, 
\quad (g \in G_0). 
\end{align}
From this, we can read off the twisted ground state $\ell_g$ as 
\begin{align}
\ell_g = \frac{1}{N} \sum_{ij} [V_g]_{ij} \ket{i} \bra{j}.
\end{align}
This is the same as the fixed point MPS ground state with twist up to a normalization.

\paragraph{Cylinder $\alpha_g$}
Since the Hilbert space with twist ${\cal C}_h (h \in G_0)$ 
is defined on the circle with $h$-flux, 
we have to associate cylinder with twist by $T_h$. 
For orientation preserving action $g \in G_0$, $\alpha_g$ reads 
\begin{align}
\alpha_{g \in G_0} |_{{\cal C}_h} 
&= \vcenter{\hbox{\includegraphics[width=0.15\linewidth]{figs/tft/Cylinder}}} 
= \vcenter{\hbox{\includegraphics[width=0.1\linewidth]{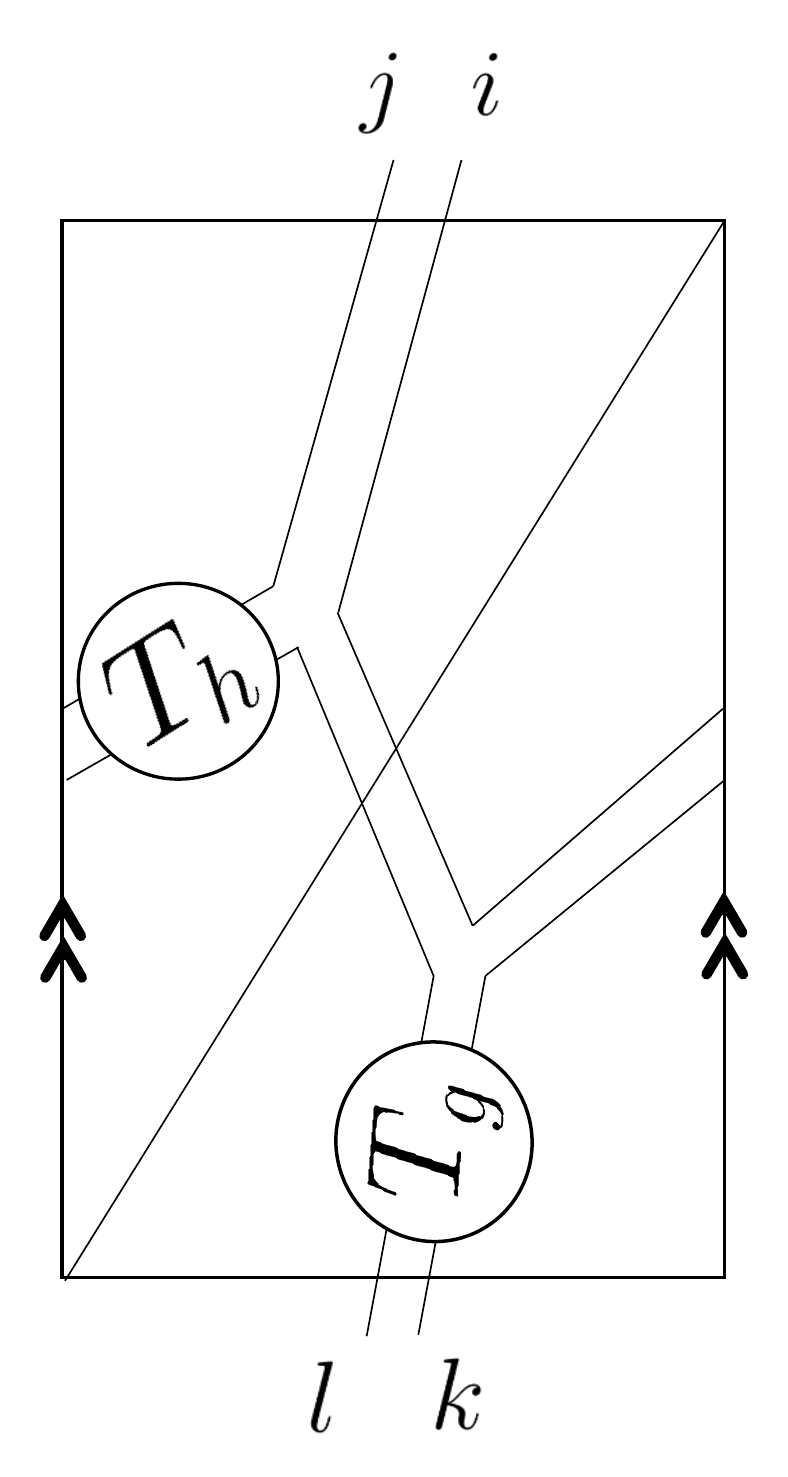}}} 
\nonumber \\
&= {C_{ij,mn}}^{pq} C_{pq,rs,tu} [T_h]^{tu,mn} [T_g]^{kl,rs} E^{ij} \otimes E_{kl} 
= \frac{1}{N} [V^*_h]_{ij} E^{ij} \otimes [V_g V_h V_g^{\dag}]_{kl} E_{kl} , 
\quad (g \in G_0). 
\end{align}
Here, $E^{ij}$ is dual basis of $E_{ij}$. 

For an orientation reversing action $g \notin G_0$, $\alpha_g$ reads 
\begin{align}
\alpha_{g \notin G_0} |_{{\cal C}_h} 
&= \vcenter{\hbox{\includegraphics[width=0.15\linewidth]{figs/tft/cylinder_P}}} 
= \vcenter{\hbox{\includegraphics[width=0.1\linewidth]{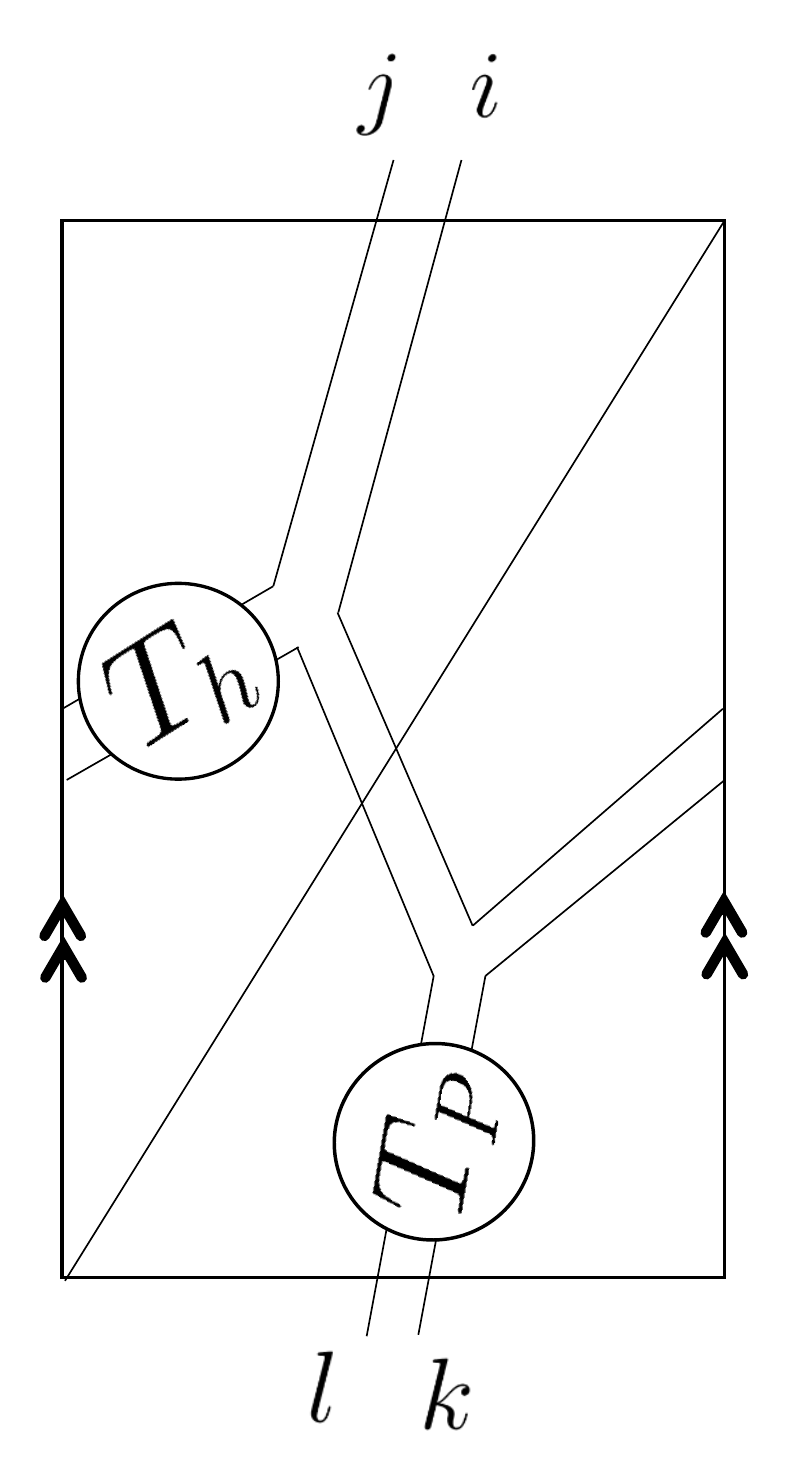}}} 
\nonumber \\
&= {C_{ij,mn}}^{pq} C_{pq,rs,tu} [T_h]^{tu,mn} [T_P]^{kl,rs} E^{ij} \otimes E_{kl} 
= \frac{1}{N} [V^*_h]_{ij} E^{ij} \otimes [V_P V^T_h V_P^{\dag}]_{kl} E_{kl} , 
\quad (g \in G_0). 
\end{align}

\paragraph{Fusion}

The sphere with three punctures (the pants diagram) 
describes a
fusion process ${\cal C}_g \otimes {\cal C}_h \to {\cal C}_{gh}$ ($g,h \in G_0$).
The path integral can be evaluated as 
\begin{align}
&\vcenter{\hbox{\includegraphics[width=0.15\linewidth]{figs/tft/pants}}} 
= \vcenter{\hbox{\includegraphics[width=0.25\linewidth]{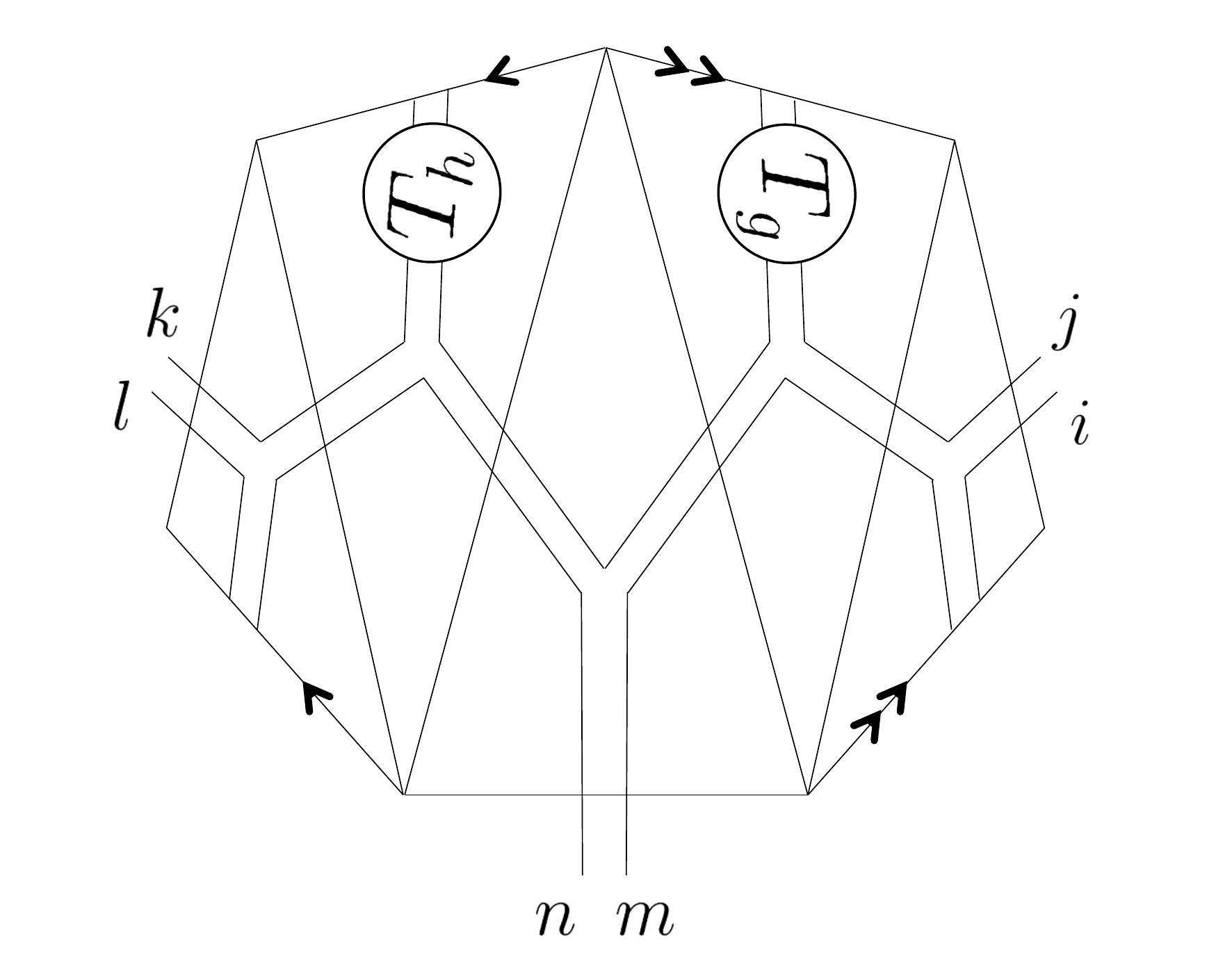}}}
\nonumber \\
&= 
{C_{kl,ab}}^{cd} {C_{ef,cd}}^{gh} [T_h]^{ef,ab} 
{C_{pq,ij}}^{rs} {C_{rs,tu}}^{vw} [T_g]^{pq,tu} 
{C_{vw,gh}}^{mn}
E^{ij} \otimes E^{kl} \otimes E_{mn}
\nonumber \\
&= \frac{1}{N^2} [V^*_g]_{ij} E^{ij} \otimes [V^*_h]_{kl} E^{kl} \otimes [V_g V_h]_{mn} E_{mn}. 
\end{align}

\section{Conclusion}

Tensor networks methods have have been employed as an efficient way to represent
correlated, entangled, many-body ground states.
In particular, they are expected to provide a powerful framework to study 
gapped quantum many-body systems with (symmetry-protected) topological order. 
On the other hand, 
topological quantum field theories
have been playing an important guiding
role in topological phases of matter. 
Indeed, one definition of a topological phase of matter is simply that it is 
described by a TFT. 
These two descriptions (methods) are complementary to each other: 
The tensor network methods in general 
can provide a powerful practical (numerical) framework   
to study a given lattice model.  
Within the tensor network framework, it is important to 
develop a methodology to diagnose topological properties of 
a given (ground state) many-body wave function.
E.g., to develop a method to extract topological invariants from
a given many-body wave function (in the tensor network representation).  
On the other hand, 
TFTs allow us to work directly in the topological limit (the limit of zero-correlation 
length), and hence provide a systematic and abstract (axiomatic) framework
to,  e.g., systematically classify possible topological phases of matter.

In this paper,  focusing on $(1+1)$d bosonic SPT phases,  
we bridge MPTs and TFTs in $(1+1)$d.
In particular,
we discuss $(1+1)$d $G$-equivariant (possibly unoriented) TFTs, 
which are TFTs coupled with a background gauge field. 

Our results are briefly summarized as follows: 

-- In Sec.\ \ref{Classification and topological invariants of MPSs},
we summarized the construction of SPT invariants in terms of 
MPS networks.~\cite{PollmannTurner2012}
By expressing those by a reduced density matrix, 
MPS networks representing SPT 
topological invariants can be identified 
with path integrals on manifold with a twist. 
We showed the partial inversion and 
the adjacent partial transpose leads to 
partition function on the real projective plane.  

-- In Sec.\ \ref{$G$-equivariant Topological field theory} 
we reviewed 
$(1+1)$d $G$-equivariant open and closed TFTs by Moore-Segal~\cite{Moore-Segal}
which,
in addition to closed chains in $(1+1)$d closed TFTs, 
have $(1+0)$d open chains as an object.
We established a 
fixed point MPS representation of 
$(1+1)$d $G$-equivariant open and closed TFTs
(when TFTs are invertible). 
A concrete connection between the MPS and TFT descriptions 
is summarized in Table \ref{tab:functor_ori_tft} and \ref{Tab:Functor_Open}. 
In particular, 
we noted, for example, 
that the classification of the $G$-equivariant closed unoriented simple TFTs
is given by the second group cohomology, 
which precisely is the known classification of $(1+1)$d SPT phases (without
orientation-reversing symmetry). 
We also noted that semisimple TFTs correspond to a combination of 
symmetry breaking and symmetry fractionalization discussed in the MPS context.  
Furthermore, for $G$-equivariant open TFTs, 
the category of boundary conditions is equivalent to 
the known boundary degrees of freedom in $(1+1)$d SPT phases.  

-- In Sec.\ \ref{Fukuma-Hosono-Kawai state sum construction}, 
we presented a state sum construction for 
$G$-equivariant unoriented closed TFT for $(1+1)$d bosonic SPT phases. 
The symmetry twisted metrics play roles of nontrivial holonomy. 
Partition functions and correlation functions can be 
calculated in a unified framework.
In particular, we showed that the partition functions 
on the torus, the real projective plane, and the Klein bottle
match precisely with the SPT invariants constructed from the MPS method.

There is a number of natural extensions of the current paper:   
For example, it is natural to speculate that we can make a precise dictionary between
higher-dimensional TFTs
and higher-dimensional tensor networks, such as 
projected entangled pair states (PEPS).  
Another interesting direction is 
to consider topological phases of fermions, and 
their descriptions in terms of (fermionic) tensor networks, and 
spin TFTs. 
(For recent works addressing these issues,
see Refs.\ \cite{barrett2015two, novak2015state}
(the state sum construction of (1+1)d oriented spin TFTs), 
Refs.\ \cite{gaiotto2015spin, bhardwaj2016state} 
($(1+1)$d and $(2+1)d$ oriented equivariant spin TFTs and the state sum construction),
and
Ref.\  \cite{Shapourian-Shiozaki-Ryu}.

\paragraph{Note added.} 
After completing this work, we became aware of an independent work \cite{Kapustin-Turzillo-You}, 
which established the connection between the state sum construction of $(1+1)$d $G$-equivariant TQFT and MPS representations. 

%
%

\acknowledgments
We thank 
Gil Young Cho, 
Kiyonori Gomi, 
Andreas W.\ W.\ Ludwig, 
Kantaro Ohmori, 
Hassan Shapourian, 
Tadashi Takayanagi, 
Apoorv Tiwari, 
Keisuke Totsuka, 
Alex Turzillo,
Juven C. Wang, 
Xueda Wen, 
and 
Peng Ye
for useful discussion.
Especially, K.S.\ is grateful to Takahiro Morimoto for pointing out 
the equivalence between 
a partial transposition and crosscap. 
This work was supported in part
by the National Science Foundation grant DMR-1455296,
and by Alfred P. Sloan foundation. 
K.S.\ is supported by JSPS Postdoctoral Fellowship for Research Abroad.

\appendix

\makeatletter
\renewcommand{\theequation}{%
\Alph{section}.\arabic{equation}}
\@addtoreset{equation}{section}
\makeatother

\section{Group cohomology}
\label{Group cohomology}

Let $G$ be a group and $\phi : G \to \Z_2$ be a homomorphism which specifies orientation preserving symmetries. 
Let $U(1)_{\phi}$ be a $G$-left module defined by 
\begin{align}
g \cdot z = \left\{ \begin{array}{ll}
z & (\phi(g) = 1) \\
z^* & (\phi(g) = -1) \\
\end{array} \right. 
\label{eq:app_g_action}
\end{align}
where $z \in U(1)$. 
The cochain complex $C^*(G,U(1)_{\phi})$ is defined by the differential $\delta : C^n(G,U(1)_{\phi}) \to C^{n+1}(G,U(1)_{\phi})$ as 
\begin{align}
\delta c_n(g_1, \dots g_{n+1}) &= 
c_n^{\phi(g_1)}(g_2, \dots, g_{n+1}) 
c_n^{-1}(g_1g_2, g_3, \dots, g_{n+1}) 
\cdots
\nonumber \\
&
\quad
\cdots
c_n(g_1, g_2 g_3, g_4, \dots, g_{n+1}) \cdots 
c_n^{(-1)^n}(g_1, \dots, g_{n} g_{n+1}) c_n^{(-1)^{n+1}}(g_1, \dots, g_n). 
\end{align}
For our purposes, only the cases of $n=1, 2$ are needed, 
\begin{align}
\delta c_1(g_1,g_2) 
= (g_1 \cdot c_1)(g_2) c^{-1}_1(g_1 g_2) c_1(g_1) = c_1^{\phi(g_1)}(g_2) c^{-1}_1(g_1 g_2) c_1(g_1), 
\end{align}
\begin{align}
\delta c_2(g_1,g_2,g_3) 
&= (g_1 \cdot c_2)(g_2,g_3) c^{-1}_2(g_1g_2, g_3) c_2(g_1, g_2 g_3) c^{-1}_2(g_1,g_2) \\
&= c_2^{\phi(g_1)}(g_2,g_3) c^{-1}_2(g_1g_2, g_3) c_2(g_1, g_2 g_3) c^{-1}_2(g_1,g_2), 
\end{align}
where $c_n : G^n \to U(1)$. 
The group cohomology (with $U(1)_{\phi}$ coefficient) is defined by 
\begin{align}
H^n(G,U(1)_{\phi}) 
= Z^2(G,U(1)_{\phi})/B^n(G,U(1)_{\phi}). 
\end{align}
For a trivial $G$-module $U(1)$, we have 
\begin{align}
H^n(G,U(1)) \cong H^{n+1}(G, \Z). 
\end{align}

\section{Projective representation}
Once a 2-group cocycle $b(g,h) \in Z^2(G,U(1))$ is given, 
a factor group of a $b$-projective representation is determined as 
\begin{align}
V_g V_h = b(g,h) V_{gh}. 
\end{align}
In the same manner as the ordinary linear representation, 
there may be multiple irreducible $b$-projective representations. 
The following quantity 
\begin{align}
\sum_{V \in b\mbox{-irreps.}} ({\rm dim}V)^2 = const.
\end{align}
does not depend on $b(g,h) \in Z^2(G,U(1))$. 
For example, for $G=\Z_n \times \Z_n = \braket{ \sigma_1,\sigma_2 | \sigma_1^n = \sigma_2^n=1}$, the second group cohomology is $H^2(\Z_n \times \Z_n,U(1)) = \Z_n$, 
and there are $n^2$ 
1-dimensional linear irreps.\ and only one $n$-dimensional nontrivial projective irreps., thus, the above identity holds as 
\begin{align}
\underbrace{1^2 +  \cdots +1^2}_{n^2} = n^2
\end{align} 
The trivial linear irreps.\ are constructed as 
\begin{align}
V_{\sigma_1} = e^{2 p \pi i /n}, && V_{\sigma_2} = e^{2 q \pi i/n}, && (p,q = 0, \dots, n-1). 
\end{align}
On the other hand,
a nontrivial projective irreps.\ belonging to $1 \in H^2(\Z_n \times \Z_n,U(1))$ is given by 
\begin{align}
V_{\sigma_1} = \begin{pmatrix}
1 & & & \\
& \omega & & \\
& & \cdots &  \\
& & & \omega^{n-1} \\
\end{pmatrix}, && 
V_{\sigma_2} = \begin{pmatrix}
& & & 1 \\
1 & & & \\
& \cdots & & \\
& & 1 & \\
\end{pmatrix}, && \omega = e^{2 \pi i/n}, 
\end{align}
which satisfies $V_{\sigma_1} V_{\sigma_2} = \omega V_{\sigma_2} V_{\sigma_1}$. 
In general, for commuting elements $g,h \in G$, we have 
\begin{align}
V_g V_h = \epsilon(g,h) V_h V_g, && (gh=hg). 
\end{align}

An example of the existence of multiple projective irreps.\ is 
a dihedral group $D_4 = \braket{ c_4,\sigma| c^4_4=\sigma^2=1, \sigma c_4 \sigma= c_4^{-1} }$ 
of which the second group cohomology is $H^2(D_4,U(1)) = \Z_2$. 
For $1 \in H^2(D_4,U(1))$, there are two inequivalent irreps.\ $V_g, W_g$ as 
\begin{align}
\left\{ \begin{array}{l}
V_{c_4} = e^{-\frac{\pi}{4} i \sigma_y} \\
V_{\sigma} = \sigma_z \\
\end{array}\right. && 
\left\{ \begin{array}{l}
W_{c_4} = e^{-\frac{3\pi}{4} i \sigma_y} \\
W_{\sigma} = \sigma_z \\
\end{array}\right. 
\end{align}
Here $\sigma_i(i=x,y,z)$ are the Pauli matrices.

\section{Orbifolding: Dijkgraaf-Witten theory in (1+1)d}
\label{Orbifolding: Dijkgraaf-Witten theory in (1+1)d}

In Sec.~\ref{Topological invariants}, we have discussed 
gauging of symmetry $G$. 
The partition function $Z_M(P)$ on a 2-space $M$ with various background $G$-field $P$ 
gives the topological invariants for (1+1)d bosonic SPT phase.
In other words, what we have discussed are the response theory of SPT phases. 
One can make one further step and consider orbifolding
by summing over all possible flat background $G$-field.
(I.e., we are promoting the (flat) $G$-field to dynamical entities.)
This procedure leads to the so-called 
the Dijkgraaf-Witten theory 
\cite{dijkgraaf1990topological, Moore-Segal, Levin2012,2012PhRvB..85x5132R, 2013PhRvB..88g5125S}
in (1+1)d. 
Here we show the partition functions of the orbifolded theories on some spacetimes: 
\begin{align}
Z_{\rm orb}(T^2) &= \frac{1}{|G|} \sum_{g,h \in G_0;[g,h] = 0} \epsilon(g,h) = \mbox{number\ of\ } b \mbox{-irreps.}, 
\nonumber \\
Z_{\rm orb}(KB) &= \frac{1}{|G|} \sum_{g \notin G_0, h \in G_0; g h^{-1} g^{-1}=h} \kappa(g;h), 
\nonumber \\
Z_{\rm orb}(\R P^2) & = \frac{1}{|G|} \sum_{g \notin G_0;g^2=1} \theta(g). 
\end{align}

\section{Algebraic relations in equivariant open and closed TFTs}

In this section, we summarize the 
algebraic relations which are followed from 
the $G$-equivariant cobordism category.~\cite{Moore_lecture, Moore-Segal, Kapustin-Turzillo, tagami2012unoriented, sweet2013equivariant}
In the following picture, dotted lines without specifying a group element represent trivial holonomies.

\subsection{Closed TFT}
\label{app:Closed TFT}

\begin{align}
&\vcenter{\hbox{\includegraphics[width=0.6\linewidth]{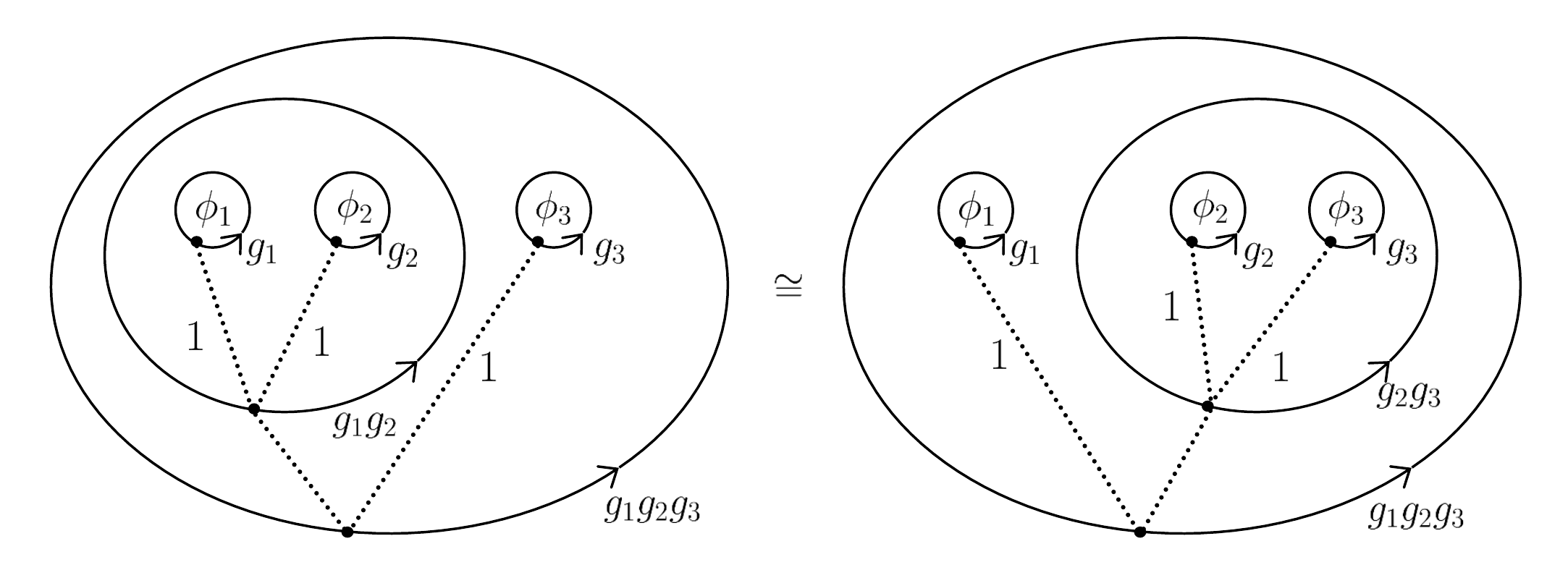}}} 
&& \Rightarrow && 
(\phi_1 \phi_2) \phi_3 = \phi_1 (\phi_2 \phi_3), \ \  (\phi_i \in {\cal C}_{g_i}, i=1,2,3), 
\end{align}
\begin{align}
&\vcenter{\hbox{\includegraphics[width=0.4\linewidth]{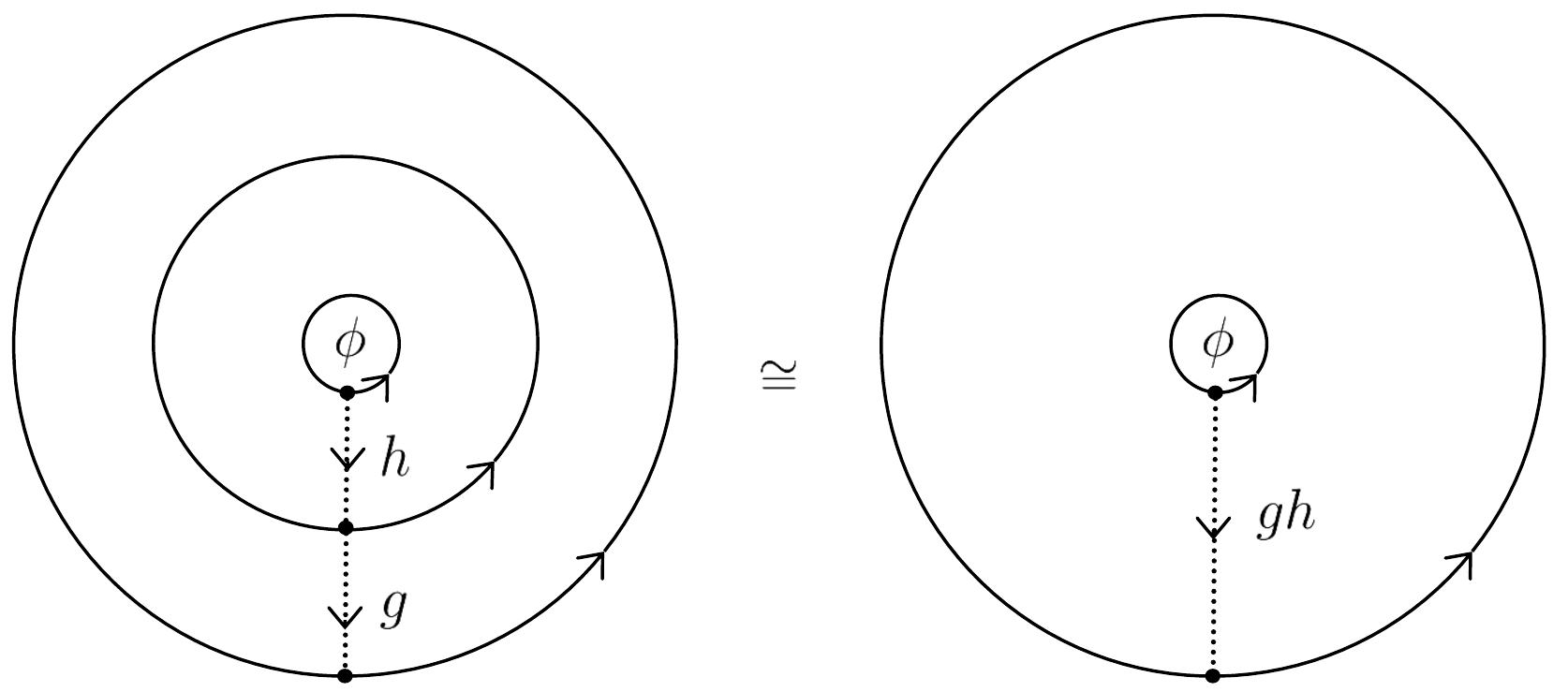}}} 
&& \Rightarrow &&
\alpha_g \circ \alpha_h = \alpha_{gh}, \ \ g,h \in G, 
\end{align}
\begin{align}
&\vcenter{\hbox{\includegraphics[width=0.6\linewidth]{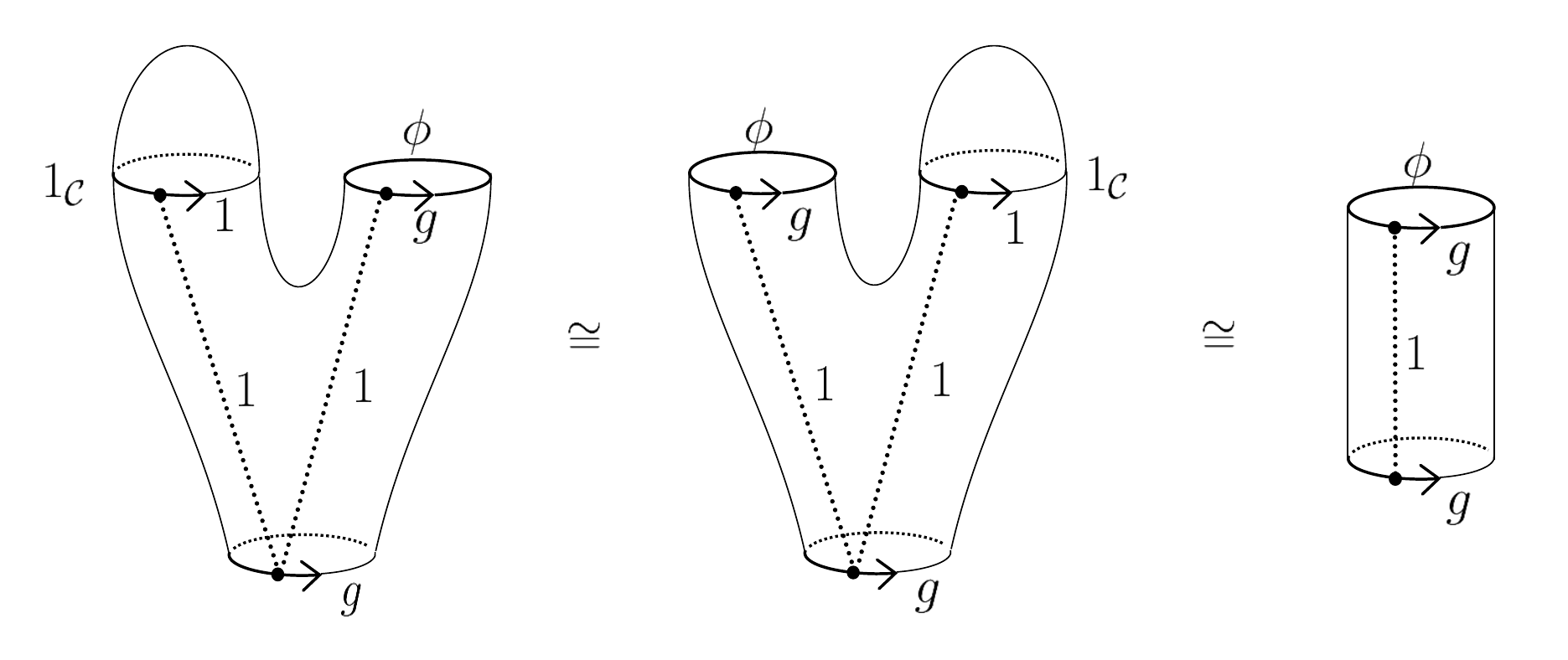}}} 
&& \Rightarrow && 
1_{\cal C} \phi = \phi 1_{\cal C} = \phi, \ \  (\phi \in {\cal C}_g), 
\end{align}
\begin{align}
&\vcenter{\hbox{\includegraphics[width=0.4\linewidth]{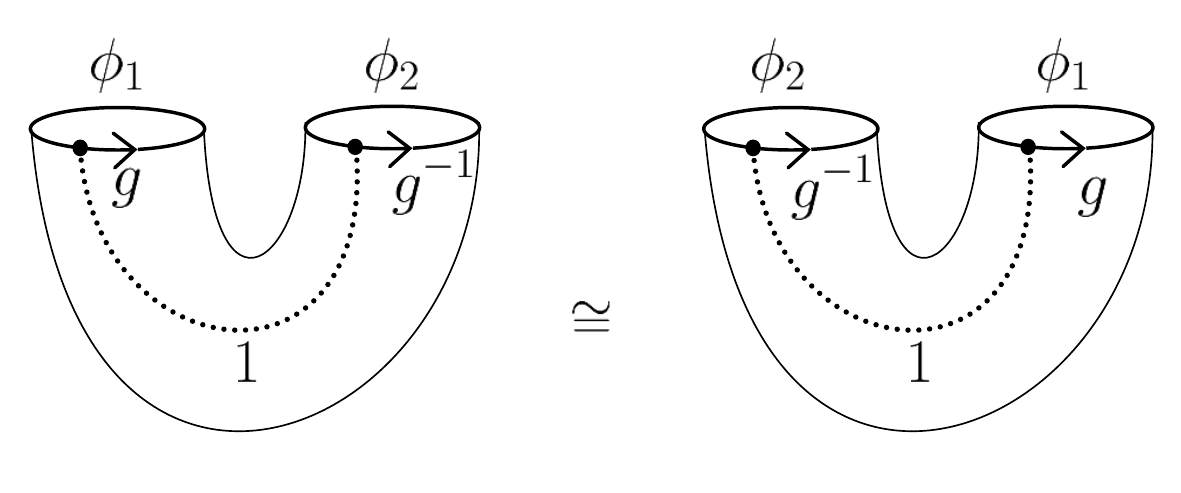}}} 
&& \Rightarrow && 
\theta_{\cal C}(\phi_1 \phi_2) = \theta_{\cal C}(\phi_2 \phi_1), \ \ (\phi_1 \in {\cal C}_g, \phi_2 \in {\cal C}_{g^{-1}}), 
\end{align}
\begin{align}
&\vcenter{\hbox{\includegraphics[width=0.4\linewidth]{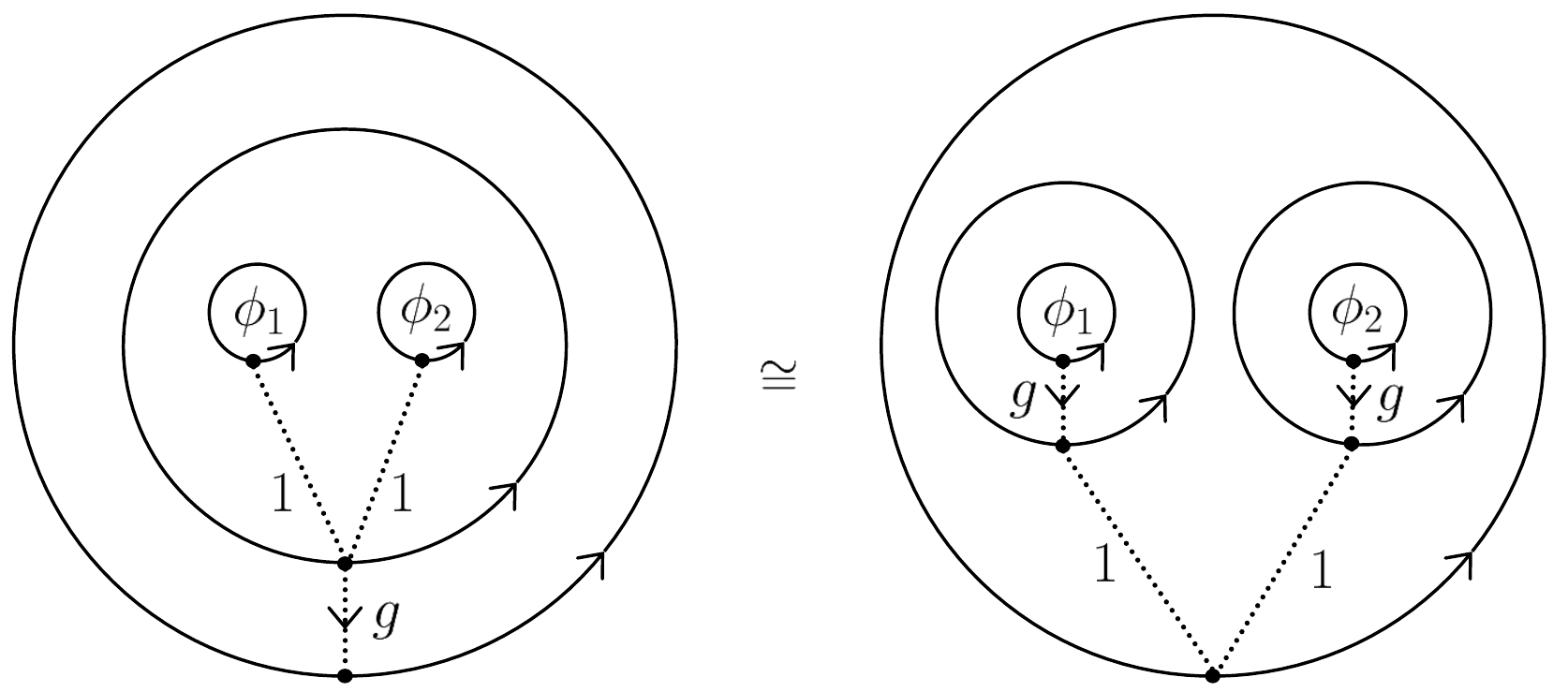}}} 
&& \Rightarrow && 
\alpha_g(\phi_1 \phi_2) = \alpha_g(\phi_1) \alpha_g(\phi_2), \ \ (g \in G_0), 
\end{align}
\begin{align}
&\vcenter{\hbox{\includegraphics[width=0.4\linewidth]{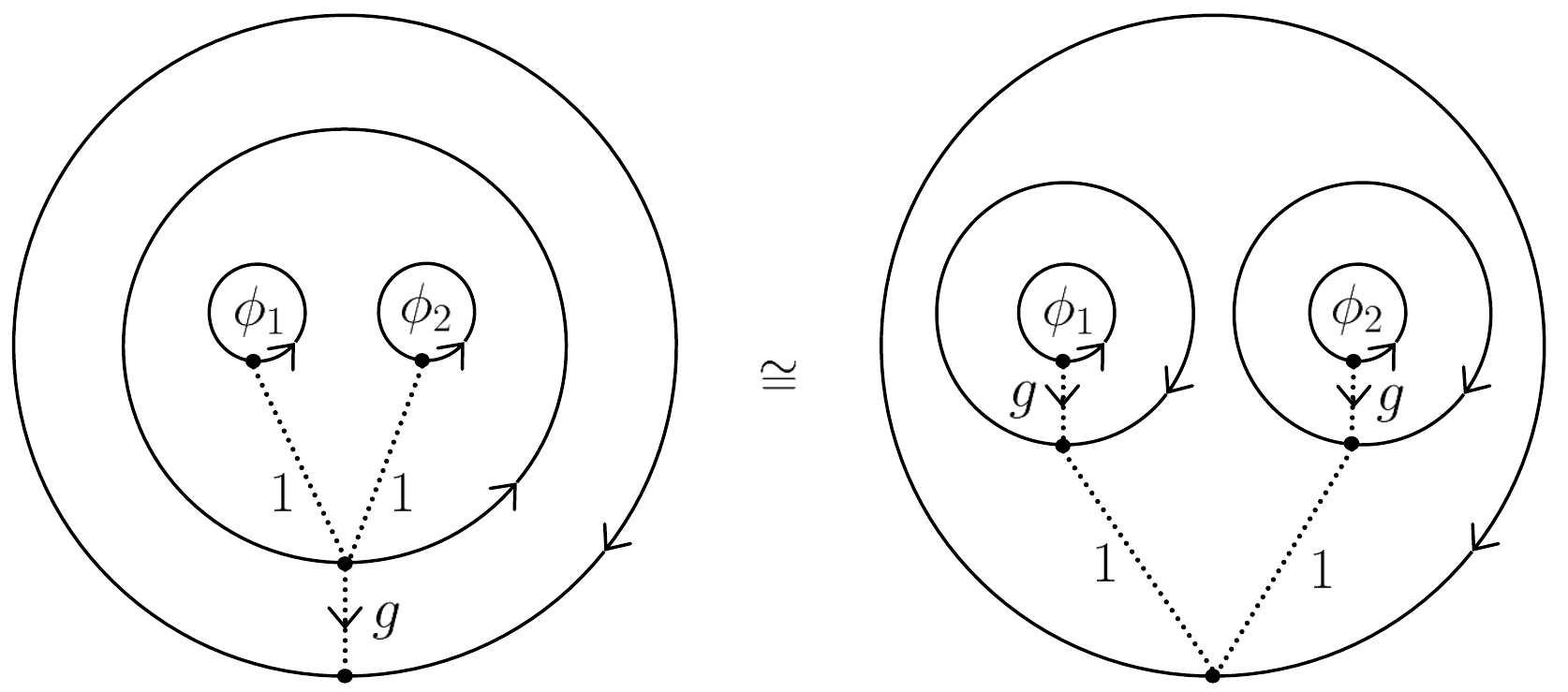}}} 
&& \Rightarrow && 
\alpha_g(\phi_1 \phi_2) = \alpha_g(\phi_2) \alpha_g(\phi_1),\ \ (g \notin G_0), 
\end{align}
\begin{align}
&\vcenter{\hbox{\includegraphics[width=0.4\linewidth]{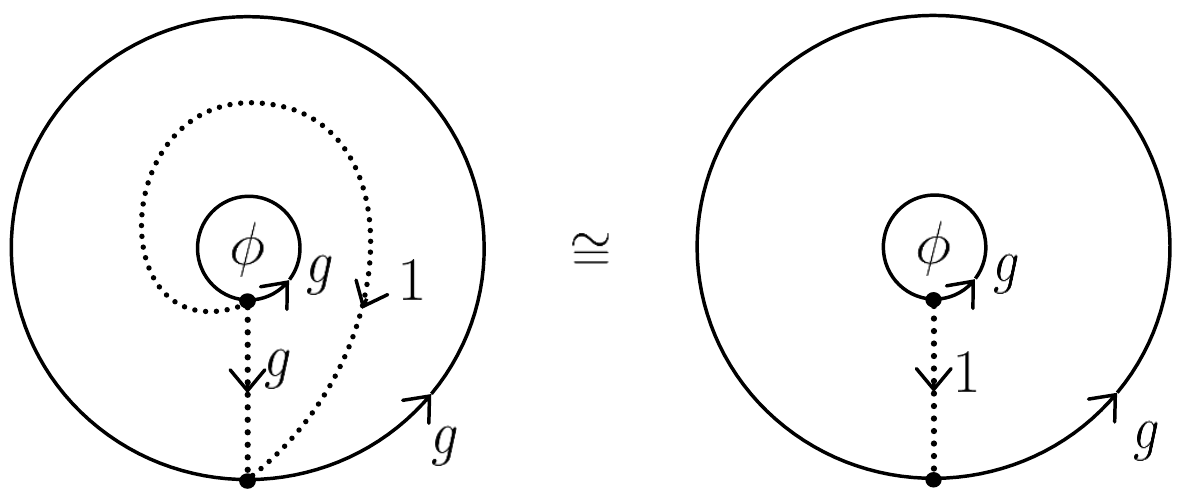}}} 
&& \Rightarrow && 
\alpha_g(\phi) = \alpha_1(\phi) = \phi, \ \ (\phi \in {\cal C}_g), 
\end{align}
\begin{align}
&\vcenter{\hbox{\includegraphics[width=0.4\linewidth]{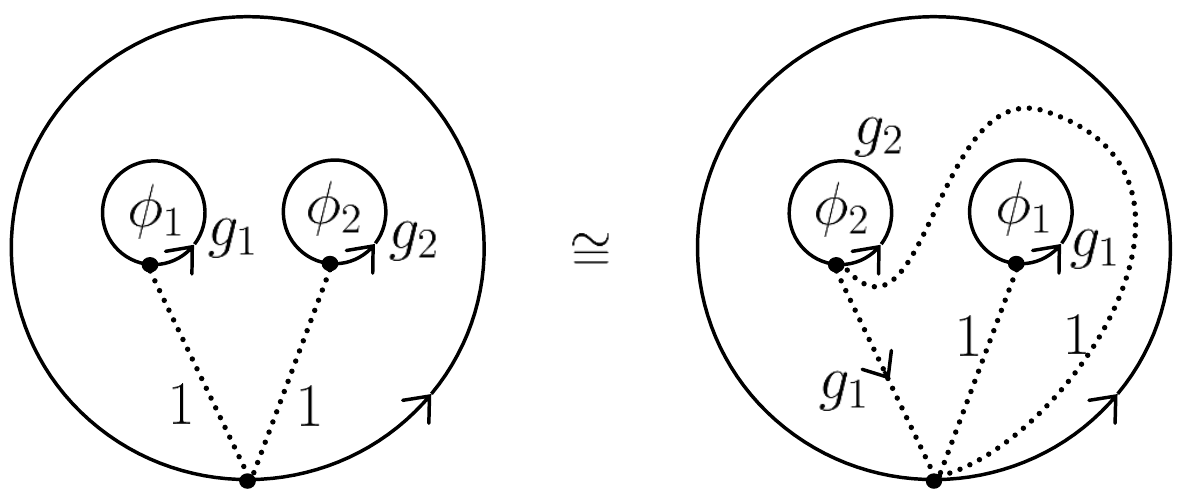}}} 
&& \Rightarrow && 
\phi_1 \phi_2 = \alpha_{g_1}(\phi_2) \phi_1, \ \ (\phi_1 \in {\cal C}_{g_1}, \phi_2 \in {\cal C}_{g_2}), 
\end{align}
\begin{align}
&\vcenter{\hbox{\includegraphics[width=0.4\linewidth]{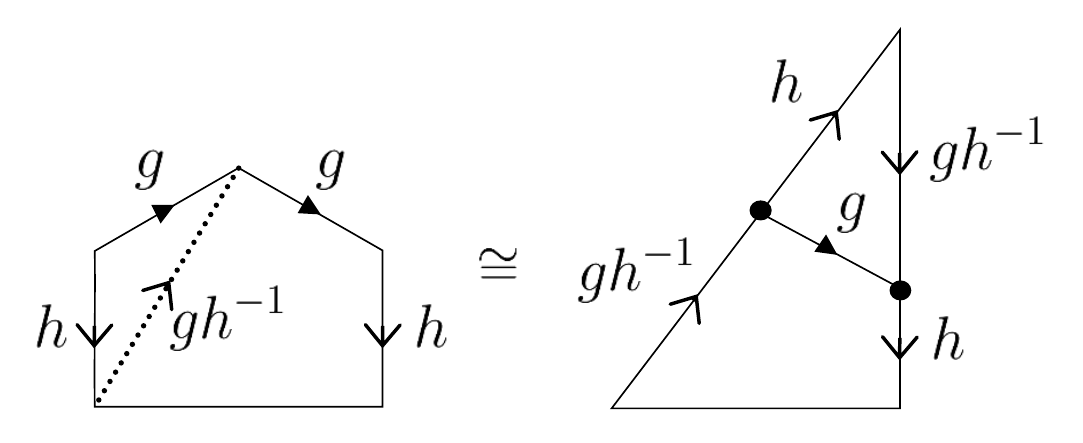}}} 
&& \Rightarrow && 
\alpha_{h} (\theta_g) = \theta_{h g h^{-1}}, \ \ (g \notin G_0, h \in G_0), 
\end{align}
\begin{align}
&\vcenter{\hbox{\includegraphics[width=0.4\linewidth]{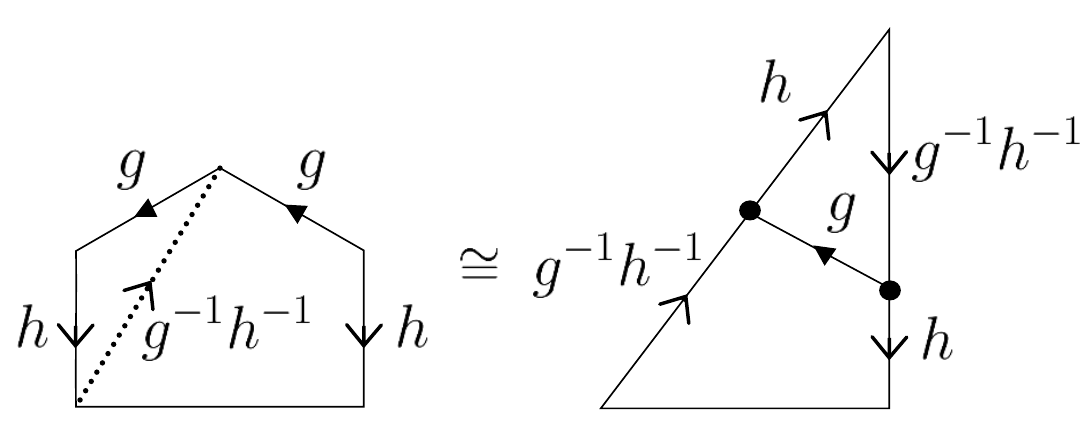}}} 
&& \Rightarrow && 
\alpha_{h} (\theta_g) = \theta_{h g^{-1} h^{-1}}, \ \ (g,h \notin G_0), 
\end{align}
\begin{align}
&\vcenter{\hbox{\includegraphics[width=0.6\linewidth]{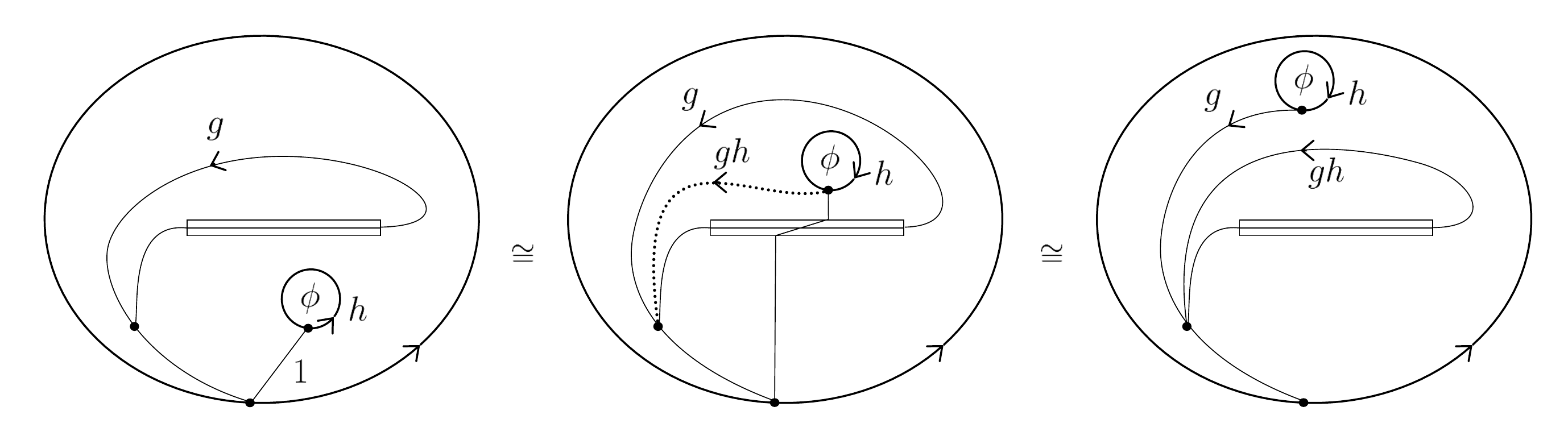}}} \nonumber \\
& \Rightarrow \ \ 
\theta_g \phi = \alpha_g(\phi) \theta_{gh}, \ \ (g \notin G_0, h \in G_0, \phi \in {\cal C}_h), 
\end{align}
\begin{align}
&\vcenter{\hbox{\includegraphics[width=\linewidth]{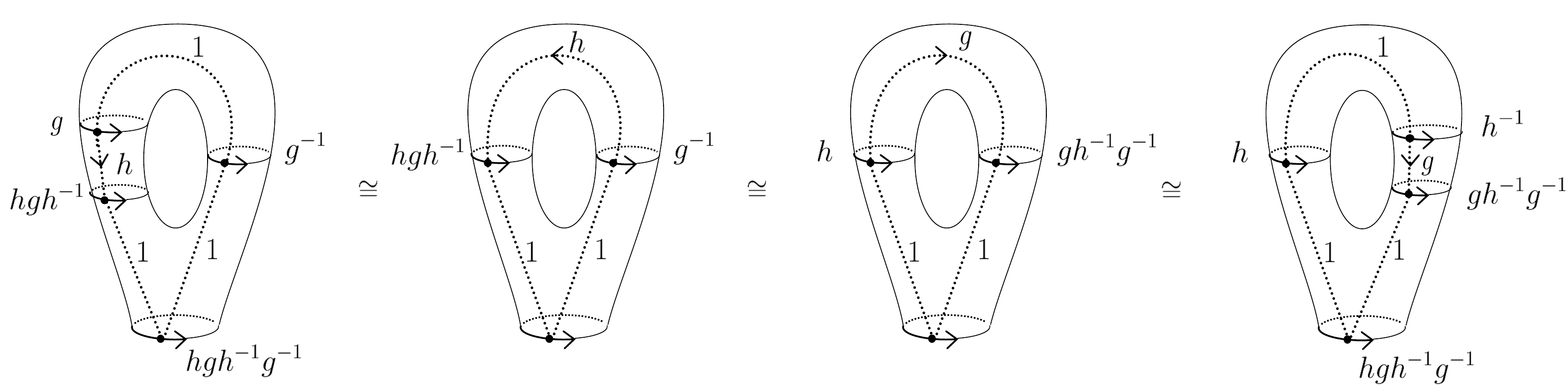}}} \nonumber \\
& \Rightarrow \ \ \sum_i \alpha_h (\xi^g_i) \xi^{g^{-1}}_i =\sum_i \xi_i^{h} \alpha_g(\xi^{h^{-1}}_i), 
\end{align}
\begin{align}
&\vcenter{\hbox{\includegraphics[width=\linewidth]{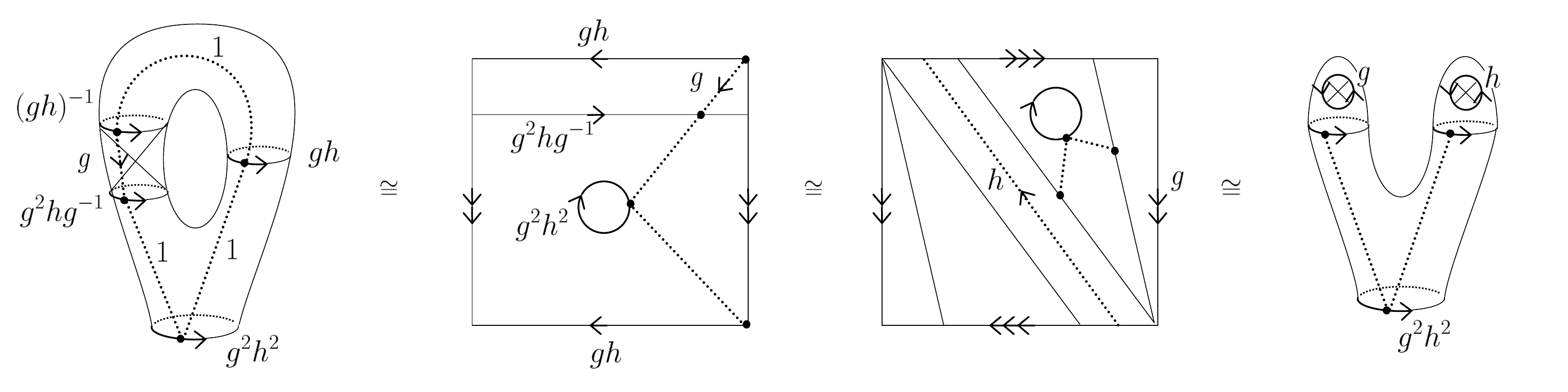}}} \nonumber \\
& \Rightarrow \ \ \sum_i \alpha_{g}(\xi^{(gh)^{-1}}_i) \xi_i^{gh} = \theta_g \theta_h, g,h \notin G_0, 
\end{align}

\subsection{Open TFT}
\label{app:open TFT}
\begin{align}
\vcenter{\hbox{\includegraphics[width=0.4\linewidth]{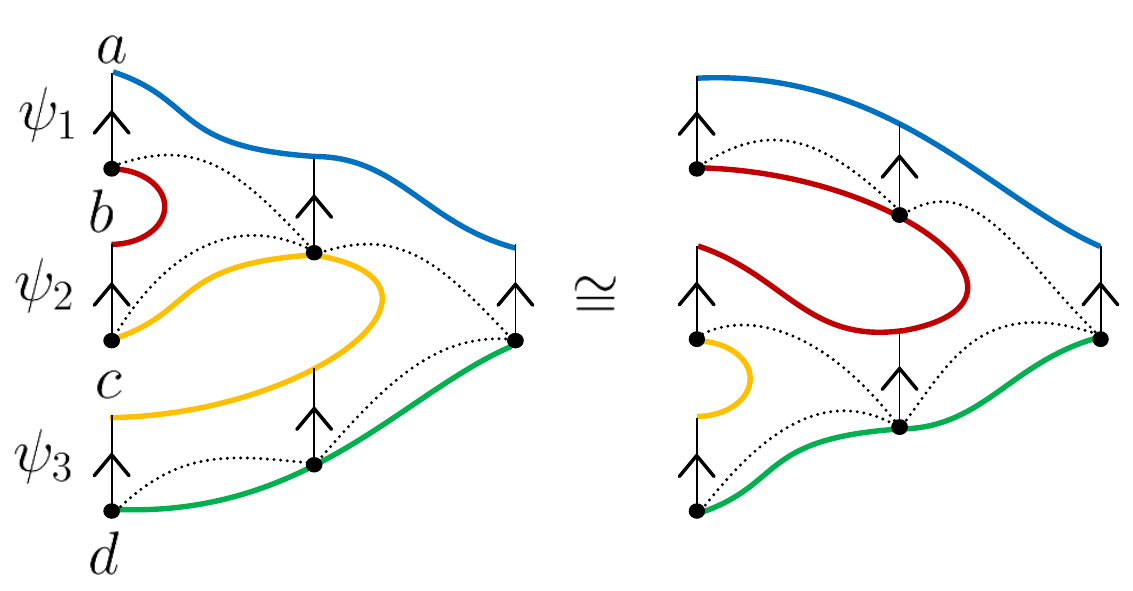}}} 
&& \Rightarrow && 
(\psi_1 \psi_2) \psi_3 = \psi_1 (\psi_2 \psi_3), \ \ 
\psi_1 \in {\cal O}_{ab},  \psi_2 \in {\cal O}_{bc}, \psi_3 \in {\cal O}_{cd}, 
\end{align}
\begin{align}
\vcenter{\hbox{\includegraphics[width=0.4\linewidth]{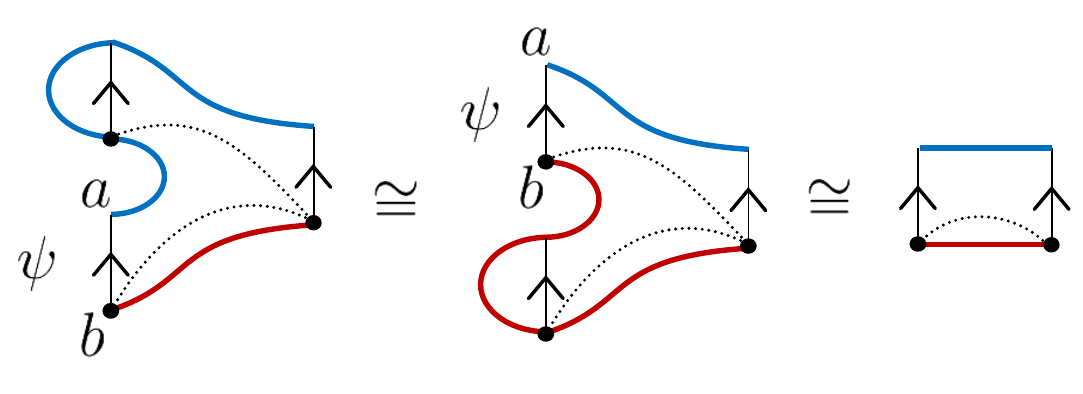}}} 
&& \Rightarrow && 
1_a \psi = \psi 1_b = \psi, \ \ \psi \in {\cal O}_{ab},  
\end{align}
\begin{align}
\vcenter{\hbox{\includegraphics[width=0.4\linewidth]{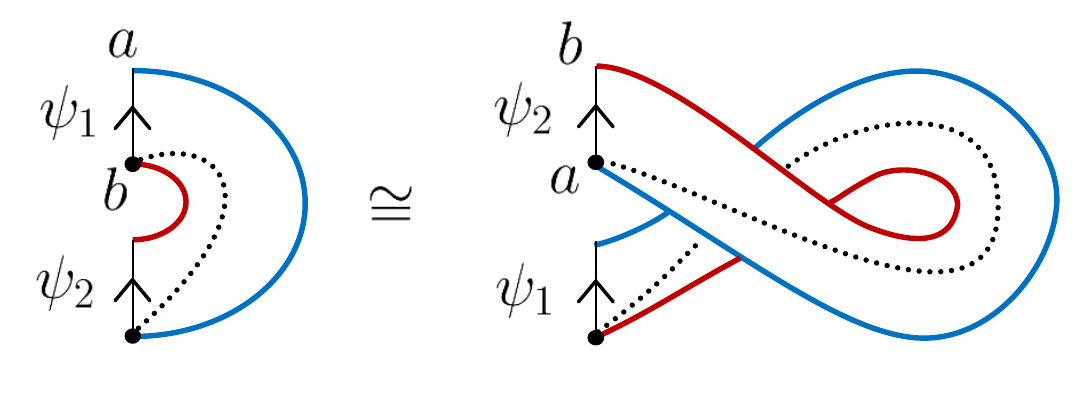}}} 
&& \Rightarrow && 
\theta_a( \psi_1 \psi_2 ) = \theta_b( \psi_2 \psi_1), \ \ 
\psi_1 \in {\cal O}_{ab}, \psi_2 \in {\cal O}_{ba}, 
\end{align}
\begin{align}
\left\{\begin{array}{l}
\vcenter{\hbox{\includegraphics[width=0.25\linewidth]{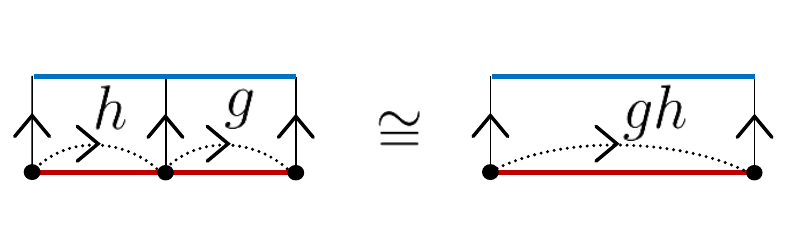}}} \\
\vcenter{\hbox{\includegraphics[width=0.3\linewidth]{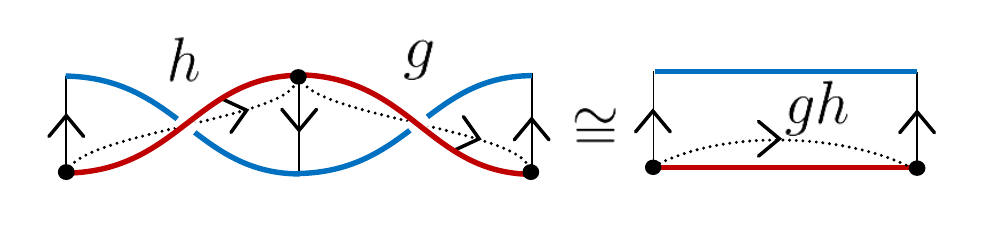}}} \\
\vcenter{\hbox{\includegraphics[width=0.3\linewidth]{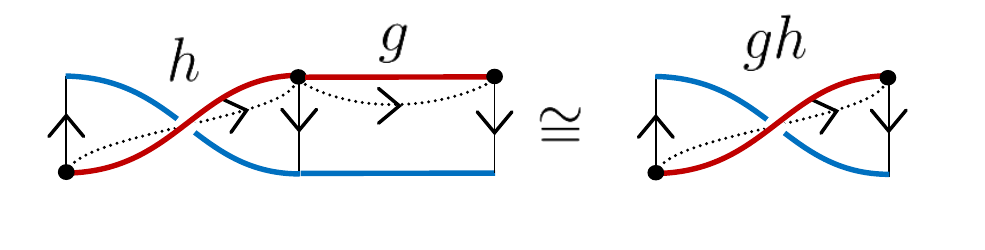}}} \\
\vcenter{\hbox{\includegraphics[width=0.3\linewidth]{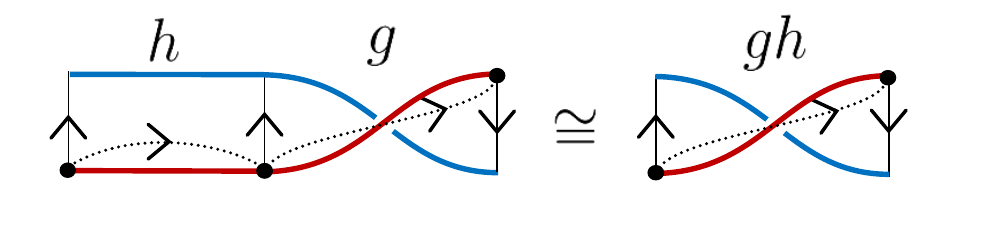}}} \\
\end{array}\right.
&& \Rightarrow && 
\rho_g \circ \rho_h = \rho_{gh}, \ \ g,h \in G,  
\end{align}
\begin{align}
\vcenter{\hbox{\includegraphics[width=0.4\linewidth]{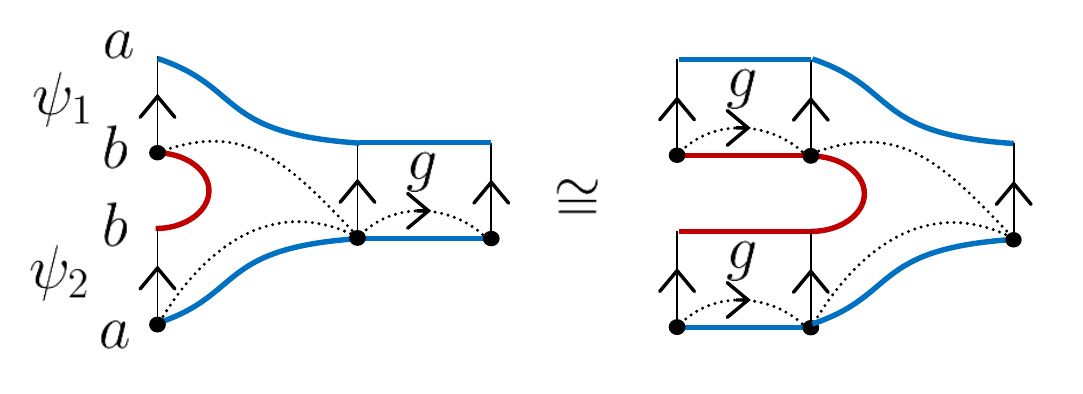}}} 
&& \Rightarrow && 
\rho_{g \in G_0}(\psi_1 \psi_2) = \rho_g(\psi_1) \rho_g(\psi_2), \ \ 
\psi_1 \in {\cal O}_{ab}, \psi_2 \in {\cal O}_{ba},  
\end{align}
\begin{align}
\vcenter{\hbox{\includegraphics[width=0.4\linewidth]{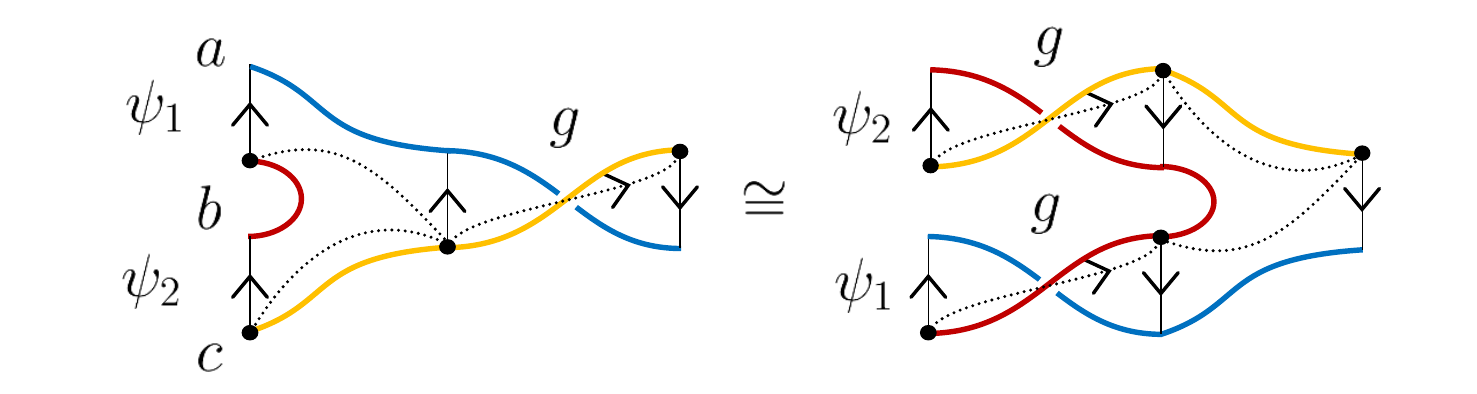}}}  
&& \Rightarrow && 
\rho_{g \in G_0}(\psi_1 \psi_2) = \rho_g(\psi_2) \rho_g(\psi_1), \ \ 
\psi_1 \in {\cal O}_{ab}, \psi_2 \in {\cal O}_{ba},  
\end{align}
\begin{align}
\vcenter{\hbox{\includegraphics[width=0.4\linewidth]{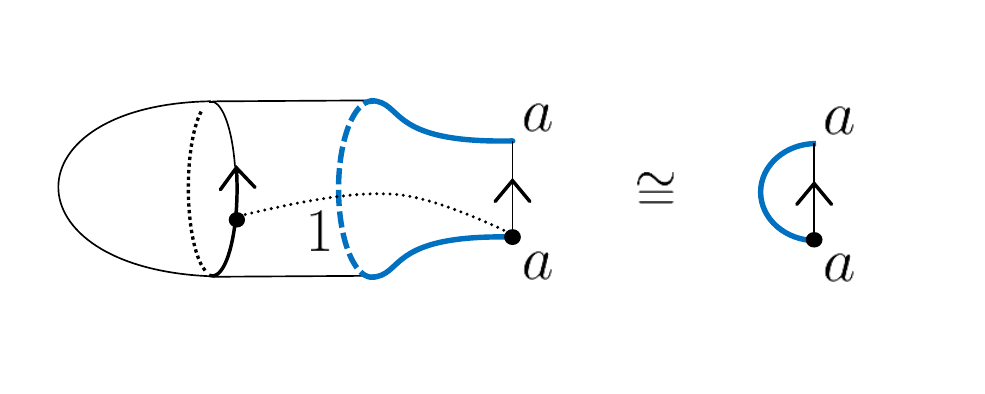}}} 
&& \Rightarrow && 
\imath_{1,a}(1_{\cal C}) = 1_a, 
\end{align}
\begin{align}
\vcenter{\hbox{\includegraphics[width=0.4\linewidth]{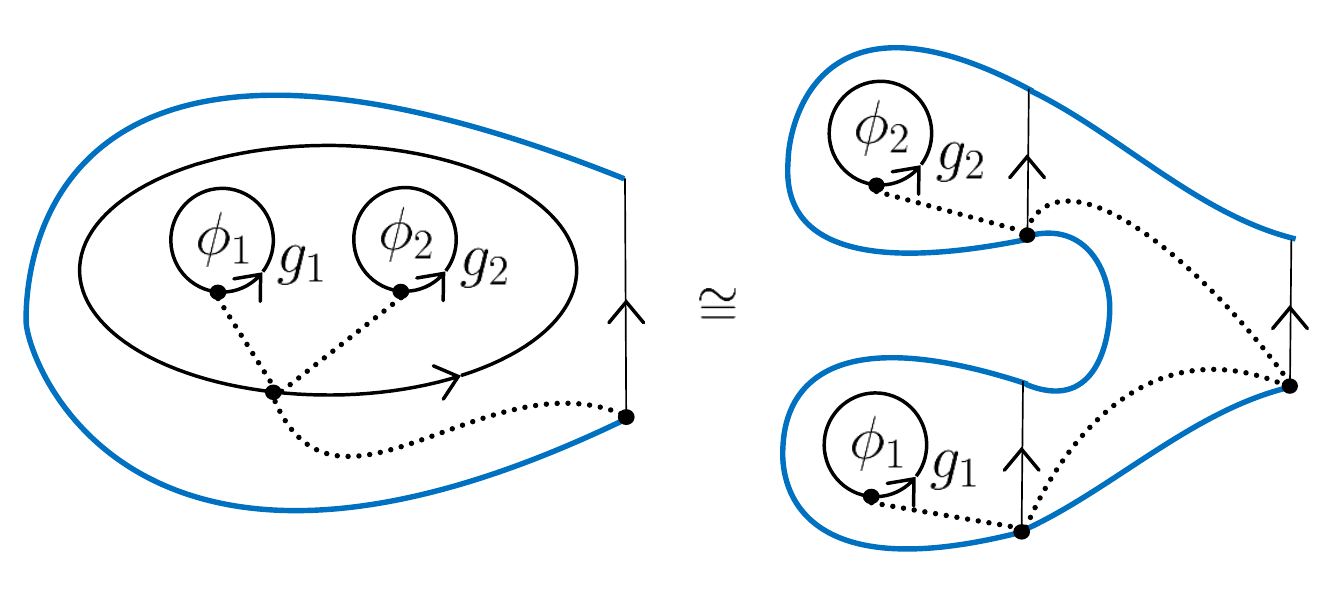}}} 
&& \Rightarrow && 
\imath_{g_1 g_2,a}(\phi_1 \phi_2) = \imath_{g_2,a}(\phi_2) \imath_{g_1,a}(\phi_1), \ \ 
\phi_1 \in {\cal C}_{g_1}, \phi_2 \in {\cal C}_{g_2},
\end{align}
\begin{align}
\vcenter{\hbox{\includegraphics[width=0.6\linewidth]{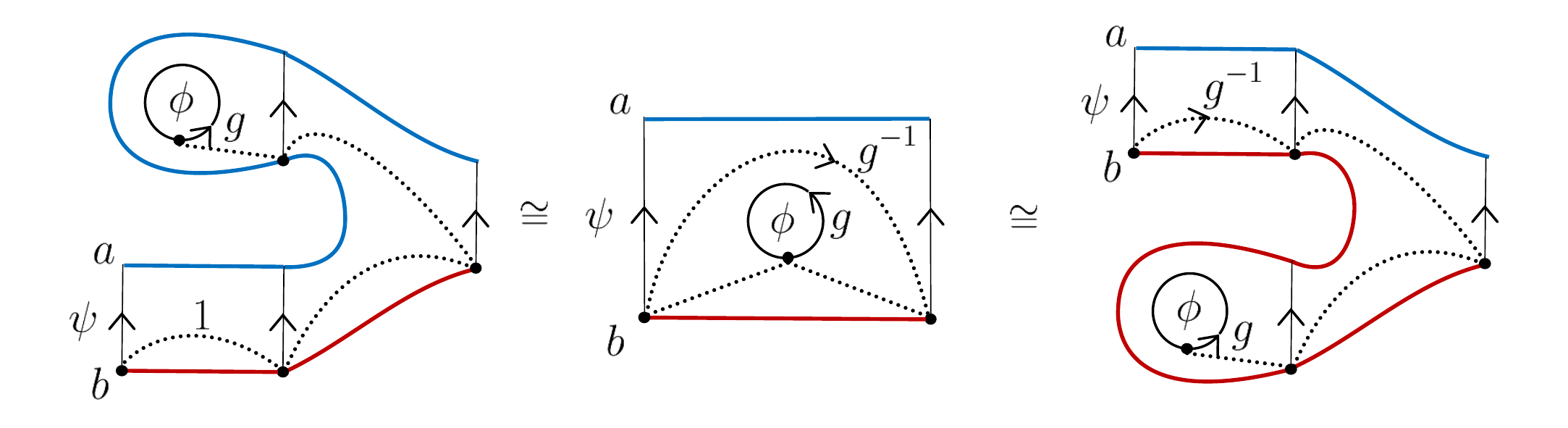}}} 
&& \Rightarrow && 
\imath_{g,a}(\phi) \psi = \rho_{g^{-1}}(\psi) \imath_{g,b}(\phi), \ \ 
\phi \in {\cal C}_{g}, \psi \in {\cal O}_{ab},
\end{align}
\begin{align}
\vcenter{\hbox{\includegraphics[width=0.4\linewidth]{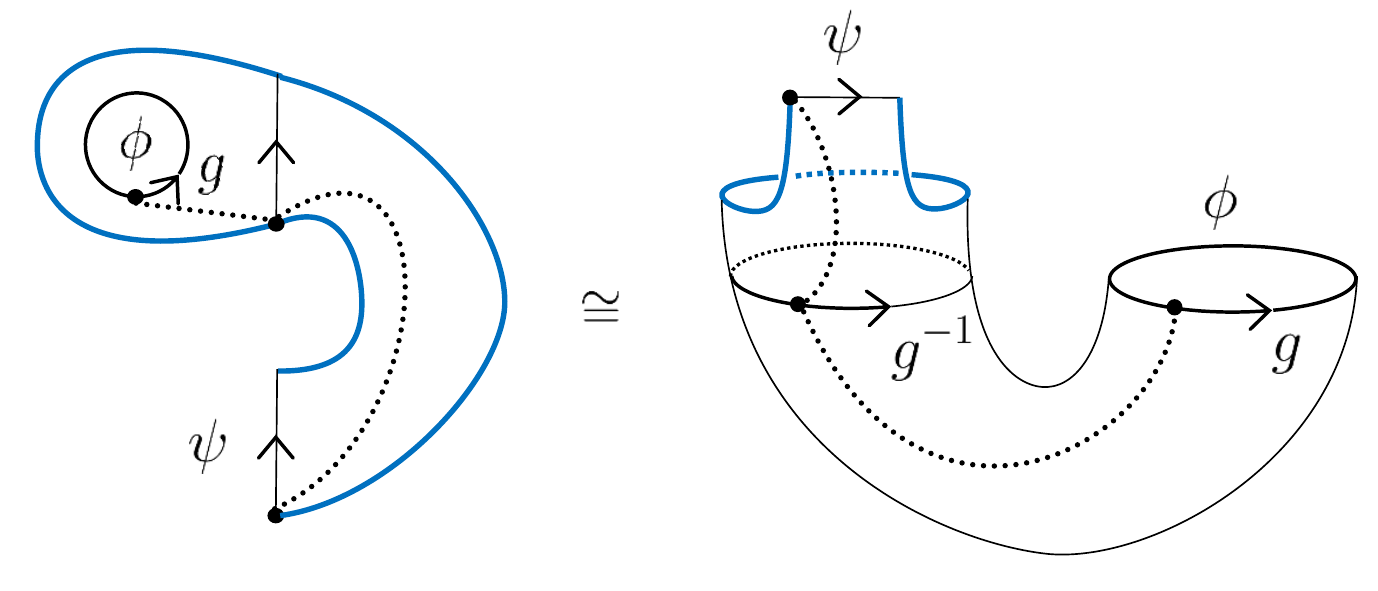}}} 
&& \Rightarrow && 
\theta_a( \imath_{g,a}(\phi) \psi ) = \theta_{\cal C} (\imath^{g^{-1},a}(\psi) \phi), \ \ 
\phi \in {\cal C}_{g}, \psi \in {\cal O}_{aa},
\end{align}
\begin{align}
&\vcenter{\hbox{\includegraphics[width=0.8\linewidth]{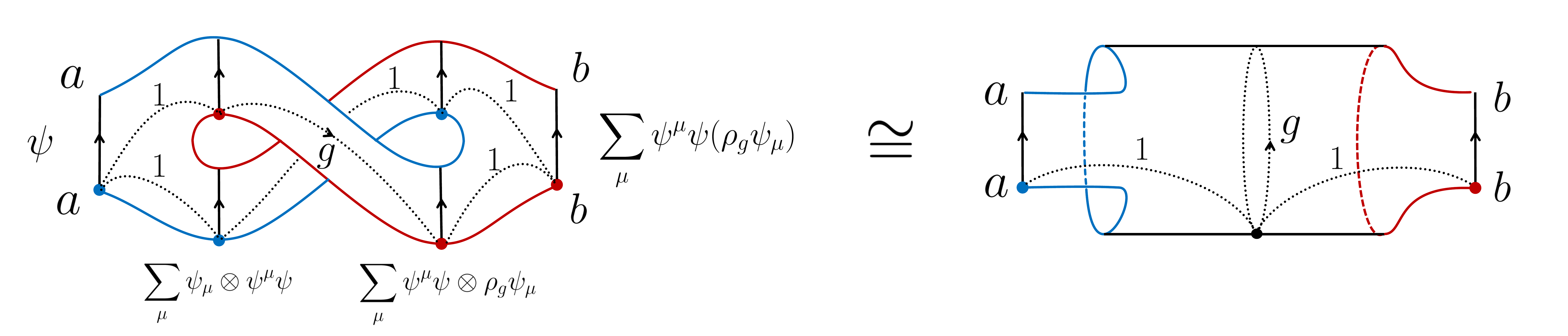}}}  \nonumber \\
&\Rightarrow \ \ \pi_{g,b}^a = \imath_{g,b} \circ \imath^{g,a}, \ \ 
\pi_{g,b}^a(\psi) = \sum_{\mu} \psi^{\mu} \psi (\rho_g \psi_{\mu}), \psi_{\mu} \in {\cal O}_{ab}, \psi^{\mu} \in {\cal O}_{ba},  
\end{align}
\begin{align}
&\vcenter{\hbox{\includegraphics[width=0.5\linewidth]{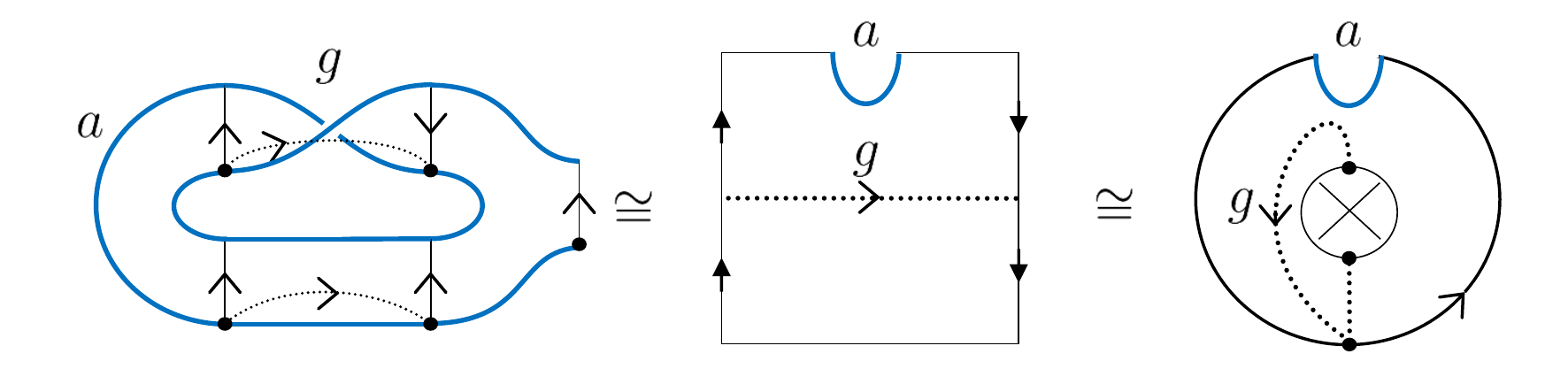}}} \nonumber \\
&\Rightarrow \ \ \sum_{\mu} (\rho_g \psi_{\mu} ) \psi^{\mu} = \imath_{g^2,a} (\theta_g), \ \ g \notin G_0. 
\end{align}

\bibliographystyle{JHEP}
\bibliography{ref}

\end{document}